\documentclass[useAMS,usenatbib]{mnras}
\usepackage[dvips]{graphicx}
\usepackage{amsmath}
\usepackage{natbib}
\usepackage{pifont}
\usepackage{txfonts}
\usepackage{mathrsfs}
\usepackage{amsfonts}
\usepackage{amssymb}
\usepackage{hyperref}
\usepackage{epsfig}
\usepackage{threeparttable}
\usepackage{xfrac}

\title[Variability in young VLMs]{Variability in young very low mass stars: Two surprises from spectrophotometric monitoring}

\author[]{
      I. Bozhinova$^{1}$, A. Scholz$^{1}$, J. Eisl{\"o}ffel$^{2}$  \\
      $^{1}$SUPA, School of Physics and Astronomy, University of St Andrews, North Haugh, St Andrews,  KY16 9SS, UK \\
      $^{2}$Th{\"u}ringer Landessternwarte Tautenburg, Sternwarte 5, D-07778 Tautenburg, Germany
    }

\begin{document}    
    
\maketitle

  \begin{abstract}
      
We present simultaneous photometric and spectroscopic observations of seven young and highly variable
M dwarfs in star forming regions in Orion, conducted in 4 observing nights with FORS2 at ESO/VLT. All 
seven targets show significant photometric variability in the I-band, with amplitudes between 0.1-0.8\,mag, 
The spectra, however, remain remarkably constant, with spectral type changes less than 0.5 subtypes. Thus, 
the brightness changes are not caused by veiling that 'fills in' absorption features. Three objects in the 
$\sigma$ Ori cluster (age $\sim$3\,Myr) exhibit strong H$\alpha$ emission and  H$\alpha$ variability, in 
addition to the continuum variations. Their behaviour is mostly consistent with the presence of spots with 
temperature of $\sim300$\,K above the photosphere and filling factors between 0.2-0.4, in contrast to typical 
hot spots observed in more massive stars. The remaining targets near $\epsilon$\,Ori, likely to be older, show 
eclipse-like lightcurves, no significant $H\alpha$ activity and are better represented by variable 
extinction due to circumstellar material. Interestingly, two of them show no evidence of infrared excess 
emission. Our study shows that high-amplitude variability in young very low mass stars can be caused by different 
phenomena than in more massive T Tauri stars and can persist when the disk has disappeared and accretion 
has ceased.       
	
\end{abstract}
 

\section{Introduction}
	
Variability is a key characteristic of young stars. Apart from its relevance
in the history of the discovery of the first prototypical young stellar
objects (\citealt{1945ApJ...102..168J}), variability is important for three different reasons:

First, variability is a nuisance parameter that needs to be constrained before analysing 
the emission from young stars and their environment, e.g., when estimating masses 
from spectral types/luminosities (\citealt{2009MNRAS.398..873S}) or when 
fitting spectral energy distributions derived from multi-wavelength, single-epoch
data \citep{2014MNRAS.443.1587R}.
Second, observations of variability inform our understanding of the dynamics of
the environment of young stellar objects (YSOs). In particular, it can provide
insights into episodic accretion \citep{2014prpl.conf..387A}, 
instabilities in the magnetospheric accretion \citep{2007prpl.conf..479B} as well as changes in 
the disk structure \citep{2015AJ....150...32R}.
Third, variability provides spatial information about the inner disk and accretion zone
that is impossible to obtain with any other means. For a typical low-mass star, features
at the inner disk rim ($\sim 0.1$\,AU) in Keplerian rotation can cause brightness
changes with a period of $\sim$days, which can be easily detected in optical/infrared
lightcurves \citep{2014AJ....147...83S,2015A&A...577A..11M}. This scale corresponds to
milliarcsecs for the nearest young stars and can barely be resolved by infrared 
interferometry \citep{2010ARA&A..48..205D}. 

Here we focus mostly on this third aspect. A number of fundamental topics in current
research on YSOs rely on our understanding of the inner 1\,AU of the disks -- the physics
of accretion, the interaction between stellar magnetic fields and the disks, the launching
of the wind, the control of angular momentum, and, last but not least, the interaction
between newly formed planets and the accretion disk. For very low mass stars (M dwarfs,
$M<0.3\,M_{\odot}$), the most common type of star in the Galaxy, monitoring studies are a 
unique tool for obtaining a close-up view of the central circumstellar zone during the 
era of planet formation.

In the canonical view, optical/near-IR variability in YSOs on timescales of days is mostly 
caused by a) cool spots on the stellar surface induced by stellar magnetic fields, b) hot 
spots close to the surface formed by the accretion shockfront, and c) variable extinction or
obscuration along the line of sight due to inhomogeneities in the inner disk (e.g., \citealt{1995A&A...299...89B,
1994AJ....108.1906H,1996A&A...310..143F,2001AJ....121.3160C,2009MNRAS.398..873S}. 
All three causes make the star redder as it gets fainter (or conversely bluer as it 
gets brighter) and should have minimum effect in the red part of the spectrum. Therefore, 
the wavelength domain around 1\,$\mu m$ is often used to determine stellar luminosities
and spectral types (e.g., \citealt{1995ApJS..101..117K,1999ApJ...525..466L}). That this is an oversimplification
is rather obvious, as there is solid evidence for significant and variable excess emission at 
these wavelengths, with unclear origin \citep{1999A&A...352..517F,2014MNRAS.442.1586D,2011ApJ...730...73F}.
\citealt{2013ApJ...767..112I} use shock accretion models to reproduce UV and optical spectra along with 1\,$\mu m$ veiling for a set of 21 low-mass T Tau stars. They suggest an explanation where multiple accretion column components with different energy densities and filling factor values can account for excess emission in both the blue and red (IR) parts of the spectrum. Even so, there are still some objects in their sample that the models fail to reproduce.   

In this paper, we analyse simultaneous broad-band photometry and low-resolution spectroscopy
of seven mid-M type objects in star forming regions in Orion, with the goal of constraining
the spatial structure of the accretion zone. All our targets are known to exhibit strong long-term
variability in the I-band ($\sim 0.8\,\mu m$) from previous studies 
(\citealt{2004A&A...419..249S,2005A&A...429.1007S}, hereafter SE04 and SE05). Their lightcurves 
are comparable to the prototypical T Tauri-type variations seen in YSOs. In line with the previously 
published work by our group, this paper extends similar studies for solar-mass T Tauri stars to 
very low mass stars with masses close to the substellar boundary.	

\section{Targets}

This study targets seven highly variable objects in open clusters in Orion. Based on their spectral type and/or magnitudes all seven are objects near the bottom of the stellar mass function; for convenience we will refer to them as very low mass (VLM) objects/stars in the following, loosely defined as having masses of $<$\,0.4M$_{\odot}$. Three of them are in the area of the well-known $\sigma$\,Ori cluster \citep{2008AJ....135.1616S}, the remaining four near $\epsilon$\,Ori. All seven have been identified from optical/near-infrared colour-magnitude diagrams as being redder than typical field stars and thus potentially young pre-main sequence objects.
Targets near $\sigma$\,Ori are likely members of this cluster, which would put their age at 2-4\,Myr \citep{2002A&A...384..937Z}. Objects 
near $\epsilon$\,Ori are likely to be slightly older with ages between 2 and 10 Myr \citep{1996PhDT........63W}. Table \ref{objects} 
summarises target names, coordinates and alternative ID-s and J-band photometry from 2MASS. 

The targets near $\sigma$\,Ori were first identified in SE04 as young very low mass (VLM) objects, using colour-magnitude diagrams and follow-up low-resolution spectroscopy. Their variability was studied in two I-band monitoring campaigns in January and December 2001. In these campaigns three objects stand out in both runs because of their high variability with amplitudes between 0.3 and 1.1 mag. Their lightcurves show evidence for periodicity on timescales of days, but significant irregular residuals remain after subtracting a sine curve. 
In addition, these three stars exhibit very strong H$\alpha$ emission as well as the lines from the Ca triplet, indicating ongoing accretion and thus youth.

Two of these objects, V2737 Ori and V2721 Ori, are later revisited in \cite{2009MNRAS.398..873S}. Near-infrared time series are taken in the J and K bands over eight nights in 2006. High level variability for these targets is observed again thereby confirming that their variability persists over timescales of years. Object V2737 Ori also showed high variability in an additional optical lightcurve taken in 2005. Colour plots provide insights into the most likely reasons for the variability. The causes are said to be inner disk edge inhomogeneity for one of the stars (V2721 Ori) and hot spots for the other (V2737 Ori). SEDs show excess flux in the infrared wavelength range (see Fig. 8 in their paper) which is an indication of the presence of disks, again confirming their young age.

The targets near $\epsilon$\,Ori have been first studied in SE05. About 30 objects with periodic variability were identified in that paper, but four stand out because they exhibit high-amplitude, partly irregular variability. Similar to the three objects near $\sigma$ Ori, they display a periodic component with superimposed variability on the scale of hours. So far, no spectra have been published for the $\epsilon$ Ori targets. 

\begin{figure*}
   
	\includegraphics[width=0.7\linewidth]{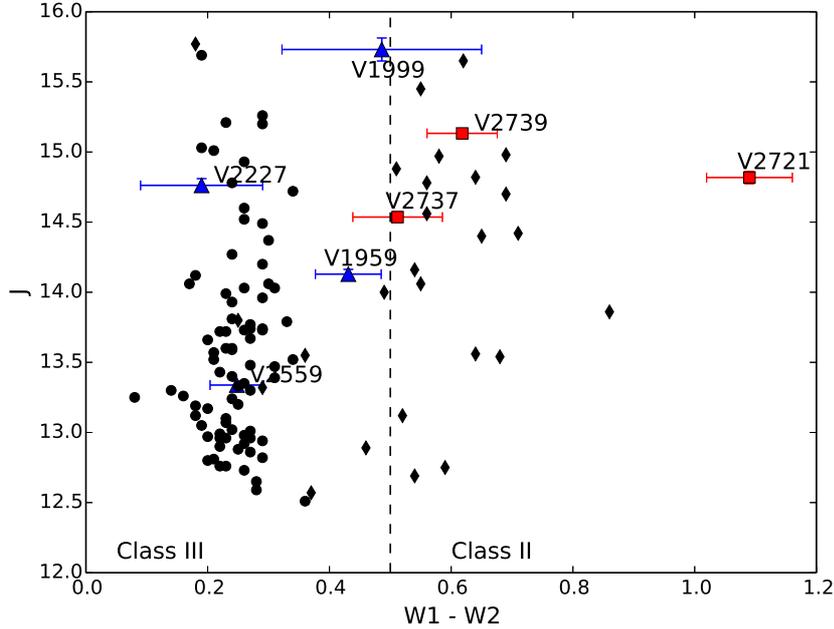} 
	
	\caption{Optical/mid-infrared colour-magnitude diagrams for observed objects (see Appendix \ref{appA} for SEDs). All targets near $\sigma$\,Ori (red squares) show a W1-W2 colour excess of more than 0.5. All stars near $\epsilon$\,Ori (blue triangles) have colours of less than 0.5. Black diamonds indicate class II brown dwarfs in Upper Scorpius \citep{2013MNRAS.429..903D}. Black circles correspond to class III objects from the same work. The vertical line at 0.5 is a visual guide for separating class II from class III objects.}
	\label{wise}	
	
\end{figure*}

To further investigate the possible presence of disks, we obtained the mid-infrared photometry from the WISE satellite \citep{2010AJ....140.1868W} for our targets. All 7 are detected in W1 (3.4$\,\mu m$) and W2 (4.5$\,\mu m$), but only 5 in W3 (12$\,\mu m$) and 1 in W4 (22$\,\mu m$). The J vs. $W1-W2$ colour magnitude diagram is shown in Figure. \ref{wise}. We compare the mid-infrared colours with published data for brown dwarf candidates in Upper Scorpius \citep{2013MNRAS.429..903D}. All objects near $\sigma$\,Ori (red squares) show a clear colour excess, consistent with the colours of class II objects (black diamonds) in \cite{2013MNRAS.429..903D}. Two of the stars near $\epsilon$\,Ori (V1999 and V1959) also display marginal colour excess, making them plausible disk candidates. In contrast, V2227 and V2559 lie within the class III population (black circles) from \cite{2013MNRAS.429..903D} and have colours consistent with a photosphere. In addition, we have constructed spectral energy distributions (SEDs) for each target based on 2MASS \citep{2006AJ....131.1163S}, Spitzer IRAC and MIPS (24$\,\mu m$) \citep{2004ApJS..154....1W} and WISE flux densities (where available). Data were obtained from the Spitzer SEIP source list catalogue on the NASA/IPAC Infrared Science Archive\footnote{http://irsa.ipac.caltech.edu/frontpage/}. The SEDs (Appendix \ref{appA}) agree with the predictions based on the WISE colour information. Three ($\sigma$\,Ori) have strong excess emission at all mid-infrared wavelengths, two more (V1999, V1959) have weak excess, the remaining two do not have measurable excess out to 10$\,\mu m$. Their upper limits at 22$\,\mu m$ still allow for excess emission at longer wavelengths.

\begin{table}
	
	\caption{Target names, coordinates, alternative ID-s, 2MASS J band magnitude \citep{2003AJ....126.1090C} and association.}
	\label{objects}
	
		\begin{threeparttable}
		\begin{tabular}{|c|c|c|c|c|c|c|}
	
			 \hline	
			{\bf Target} & {\bf RA} & {\bf DEC} & {\bf Other id } & {\bf J [mag]} & {\bf Region}\\ \hline
			V2737 Ori & 84.7042 & -2.3992 & \# 33\tnote{ a} & 14.54 & $\sigma$\,Ori \\
			V2721 Ori & 84.5542 & -2.2075 & \# 2\tnote{ a} & 14.82 & $\sigma$\,Ori\\
			V2739 Ori & 84.7792 & -2.2039 & \# 43\tnote{ a} & 15.13 & $\sigma$\,Ori\\
			V2227 Ori & 83.8042 & -0.6133 & \# 87\tnote{ b} & 14.76 & $\epsilon$\,Ori\\			
			V1999 Ori & 83.6625 & -0.9089 & \# 63\tnote{ b} & 15.73 & $\epsilon$\,Ori\\
			V1959 Ori & 83.6125 & -0.9061 & \# 51\tnote{ b} & 14.13 & $\epsilon$\,Ori\\
			V2559 Ori & 83.9083 & -0.8531 & \# 120\tnote{ b} & 13.34 & $\epsilon$\,Ori\\  \hline
	
		\end{tabular}
		
		\begin{tablenotes}
       		
       		\item[a] \citealt{2004A&A...419..249S}; Table 4., TLS Schmidt telescope, I band
			\item[b] \citealt{2005A&A...429.1007S}; Table 1., ESO/MPG WFI I band
			
     	\end{tablenotes}	
	
	\end{threeparttable}

\end{table} 
	
\section{Observations}
\label{obssect}
All data described hereafter have been obtained with the ESO-VLT/FORS2 instrument (FOcal Reducer and low dispersion Spectrograph) mounted on the 8.2 m UT1 (Antu) telescope, over the course of four consecutive nights, 28-31 Dec 2003, as part of ESO program 072.C-0197(A). The detector consists of a mosaic of two MIT CCDs, each with $4096 \times 2048$ pixels and 15 $\mu$m pixel size, with a total field of view of $6.8'\times 6.8'$. The seeing was $\sim0.6\arcsec$. Observing conditions were consistently good throughout the run.

For each object we obtained a series of spectra, accompanied with an acquisition image taken immediately prior to the spectrum. This gives us quasi-simultaneous (within 5-10\,min) photometry and spectroscopy. The images were taken in the I-band filter with an effective wavelength of 768 nm. The spectra were obtained in the LSS (long slit spectrum) mode with the 300I grism, which has central wavelength of 860 nm and wavelength range of 600-1100 nm. This setup yields a typical resolution of R = 660. Throughout the run we used a slit width of 1\,arcsec. During each night, we cycled through our seven targets 4-5 times. As part of each cycle, we also observed the A-star HD292956 for flux calibration. Two of our targets (V1999 and V1959) are covered in the same field. A typical length for one cycle is $\sim$1h 40min. This results in 4 epochs per object per night, each with photometry and spectroscopy. Each spectroscopy epoch consists of three spectra taken immediately after each other, which are co-added during data reduction. A few more additional photometry datapoints are available for each target. Table \ref{obs_log} contains the observing log for the four nights.
 
\begin{table}
 
	\caption{Target names, total number of images and spectra, exposure times}
	\label{obs_log}
	\centering 
 
	\begin{tabular}{|c|c|c|c|c|}
	
			 \hline	
			{\bf Target} & {\bf \# images} & {\bf t exp. (s)} & {\bf \# spectra } & {\bf t exp.(s)}  \\ \hline
			V2737 Ori & 26 & 2 & 16 & 50  \\
			V2721 Ori & 24 & 5 & 16 & 100 \\
			V2739 Ori & 25 & 5 & 16 & 160 \\
			V2227 Ori & 23 & 3 & 16 & 80 \\
			V1999 Ori & 27 & 10 & 17 & 300 \\
			V1959 Ori & 27 & 10 & 17 & 300  \\
			V2559 Ori & 25 & 1 & 17 & 30 \\  \hline
	
	\end{tabular}	
	
\end{table}

\section{Data reduction}	

\subsection{Photometry}
Standard image reduction, including bias and flat field correction, were done using IRAF routines in the {\it ccdred} package. We used twilight skyflats taken in the first night of the observing run for flatfielding. Aperture photometry on the aquisition images was carried out with the {\it daophot} package within IRAF. We measured magnitudes for the target and between 10 and 20 isolated, non-saturated, but well exposed comparison stars. The aperture was defined as 2-3 times the FWHM of the point spread function. The sky was measured in an annulus with a radius of 15 and a width 7 pix. 

For each field we carried out differential photometry to correct for changes in atmospheric transmission and systematics that affect all stars in the field in the same way. First, we averaged the lightcurves of all comparison stars, to create a preliminary 'master lightcurve'. This master lightcurve is subtracted from all lightcurves. We examined the root mean square (RMS) of the resulting lightcurves. All seven exhibit significant variations above the floor defined by the comparison stars. In two cases additional variable stars are found (RMS significantly larger than for the remaining objects), they are excluded from the sample of comparison stars. A small number of epochs (not more than one per field) produces outliers in the master lightcurve probably due to changing in atmospheric conditions not recognised during the run; these images were excluded from the analysis. The master lightcurve was re-calculated using the remaining set of comparison stars and images and again subtracted from the raw lightcurves.

\begin{figure*}
   
	\begin{tabular}{cc}
	    
	   \includegraphics[width=0.5\linewidth]{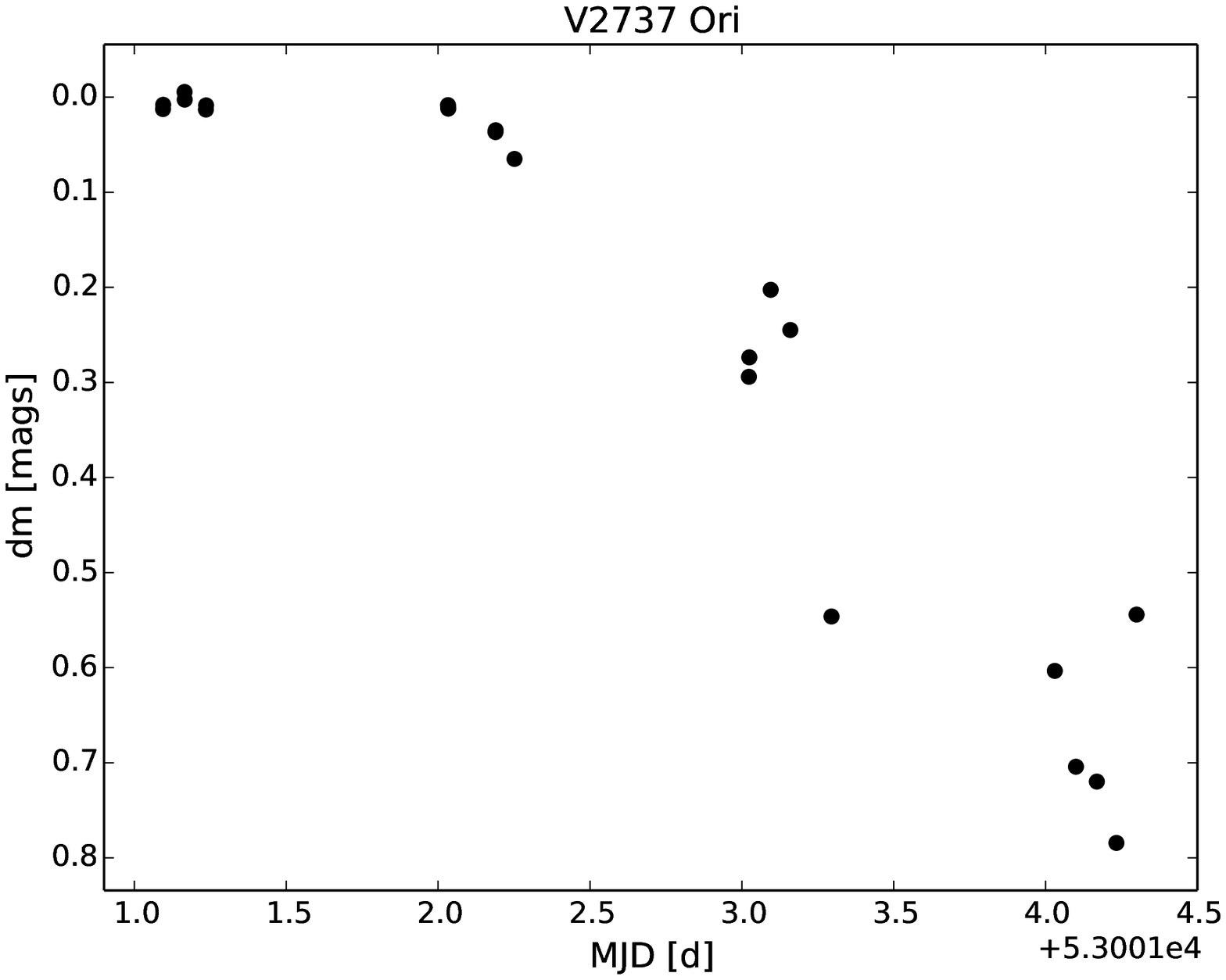}
	   \includegraphics[width=0.5\linewidth]{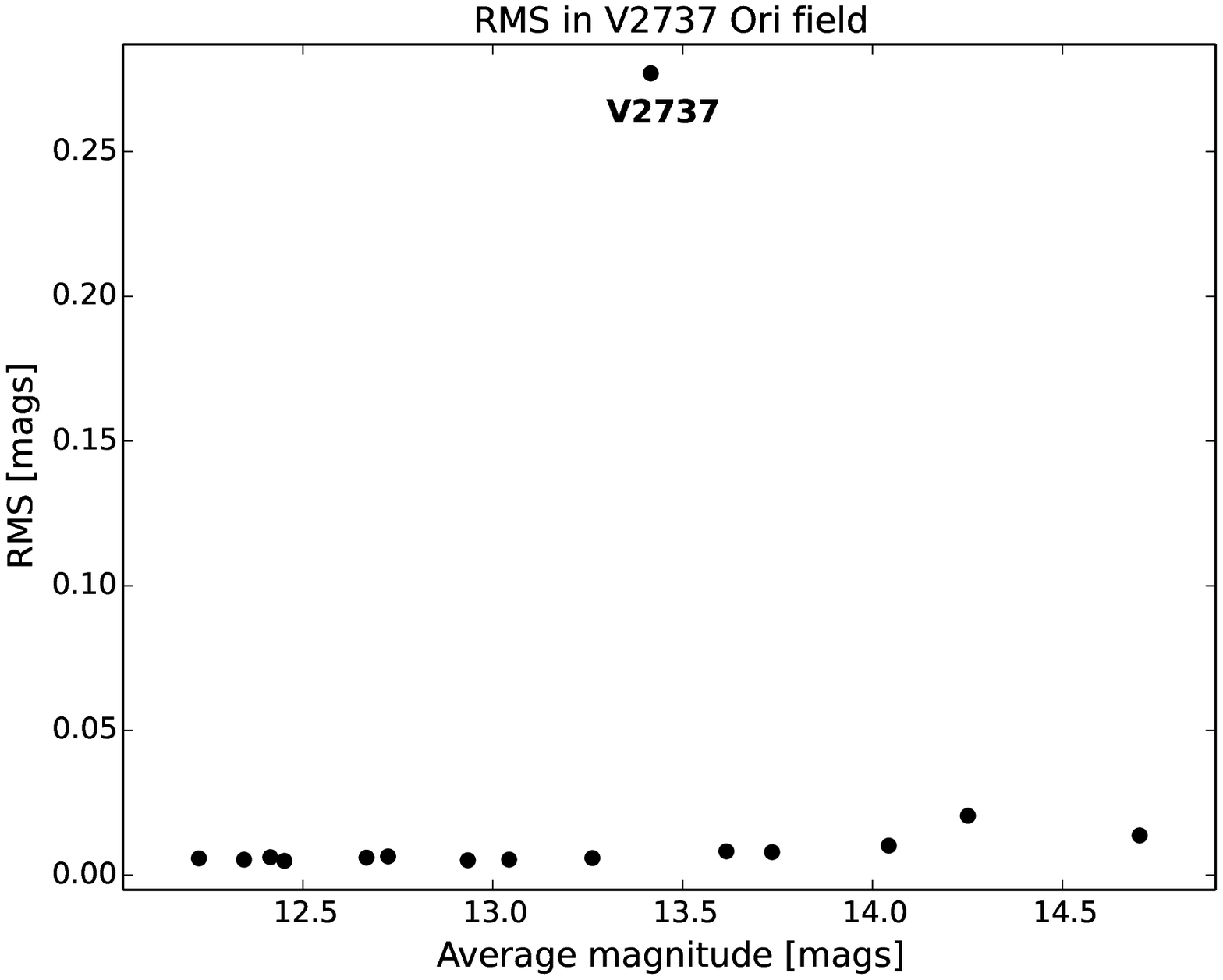} \\
	   
	    \includegraphics[width=0.5\linewidth]{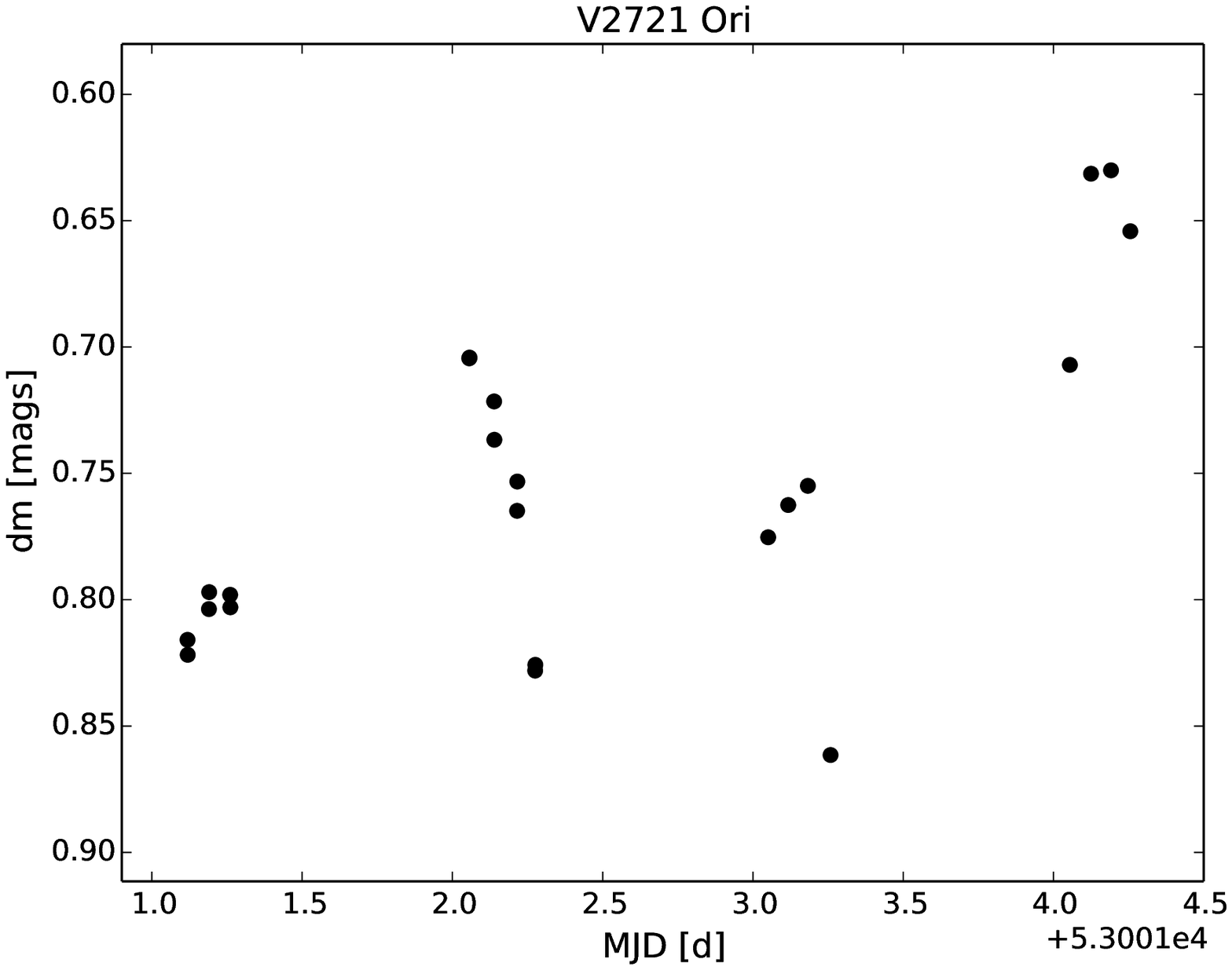}
	   \includegraphics[width=0.5\linewidth]{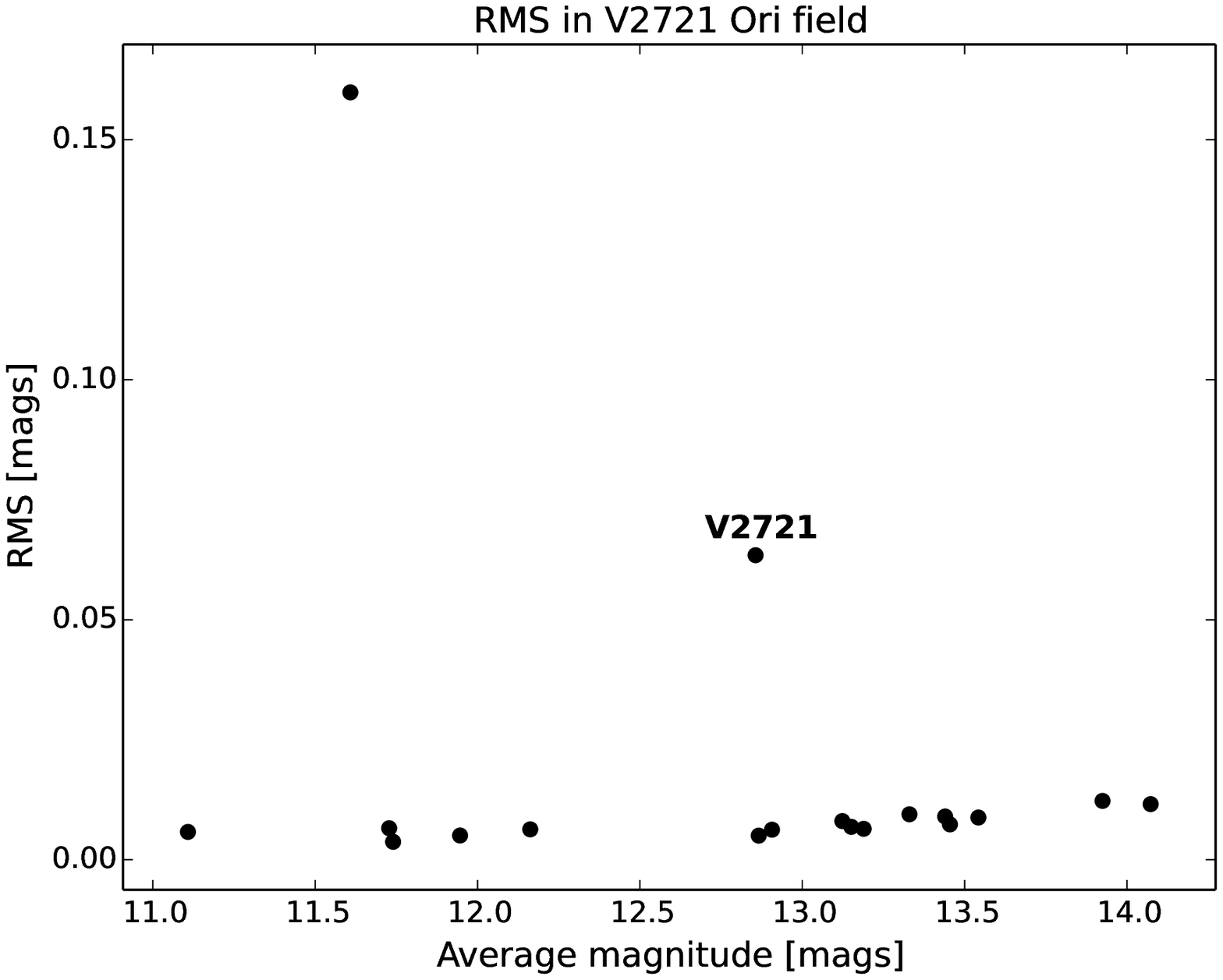} \\  
	   
	    \includegraphics[width=0.5\linewidth]{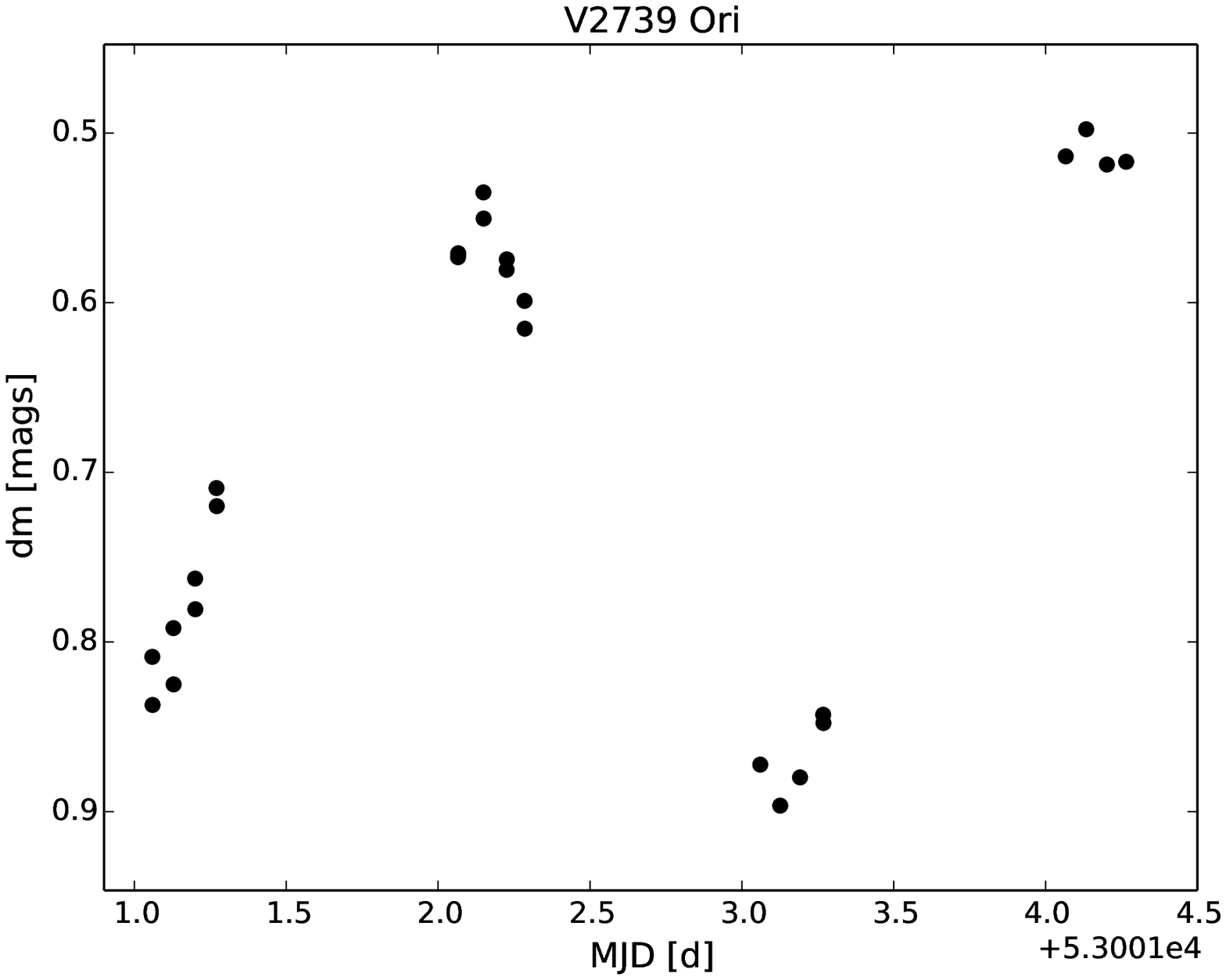}
	   \includegraphics[width=0.5\linewidth]{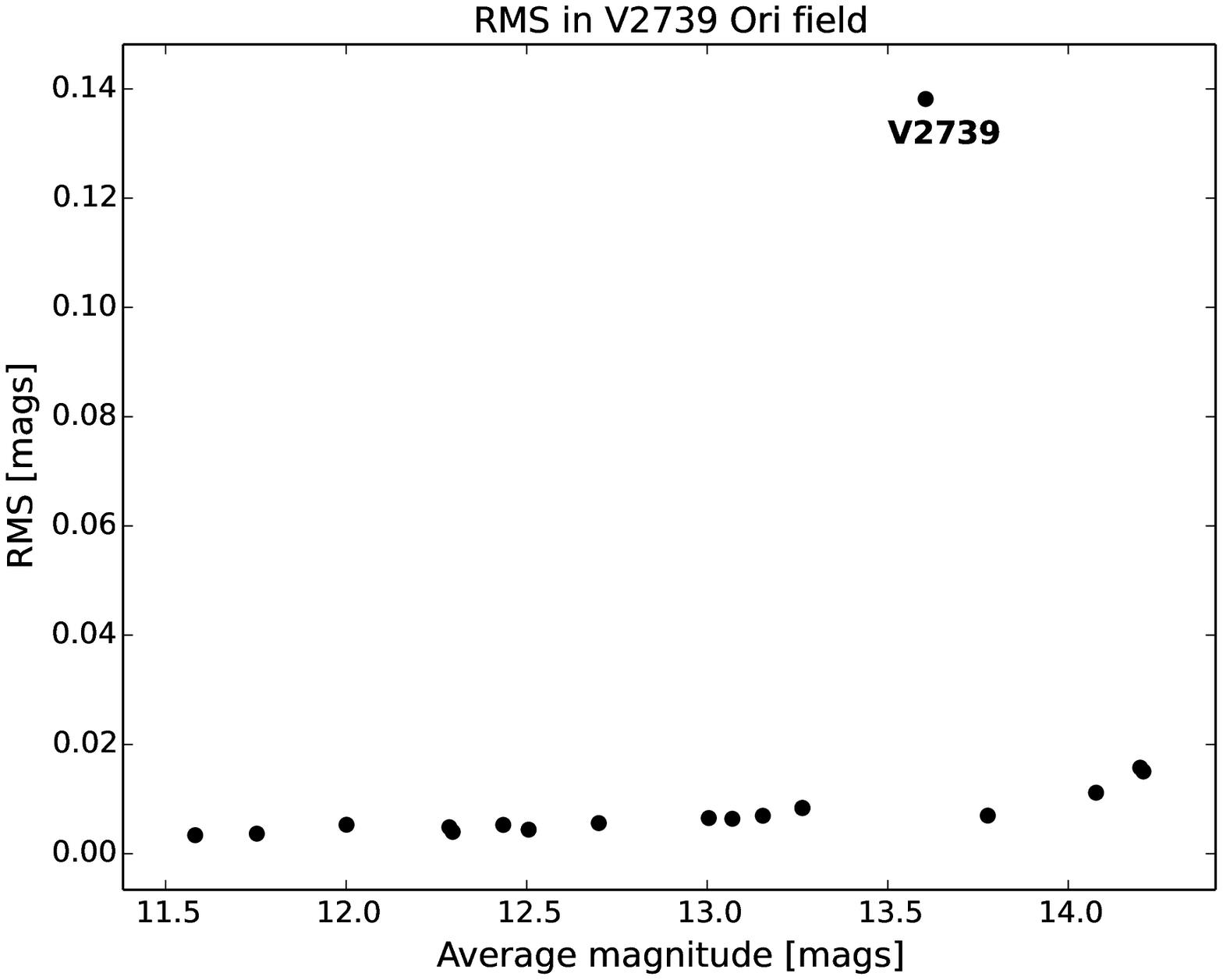} \\	 
	  
	\end{tabular}
	
	\caption{Observed lightcurves and Root Mean Square (RMS) plots for all 7 objects. Left column contains lightcurves, right column contains corresponding RMS plot for each target's field. The lightcurves for each object span 4 nights and shows relative magnitudes from differential photometry. We observe a spread of periodic, quasi-periodic and irregular behaviours. RMS plots indicate variable objects in each target's field.}
	\label{photometry}		
	
\end{figure*}

Figure \ref{photometry} presents the final differential light curves and RMS plots (RMS vs. average magnitude) for each of our seven objects. In these plots our targets are labelled. The typical photometric error (defined as the floor of the RMS of the comparison stars) is $\sim$ 1-3\%, significantly smaller than the amplitudes in the variable targets. Some of the target lightcurves exhibit a distinctive morphology -- V2737 shows a continuous downward trend, V2739 a periodic behaviour, and V2227 multiple short-period eclipses. The remaining lightcurves appear at least partially irregular. Although the sampling frequency of the current data set is less dense than in previous studies, the overall lightcurve morphology and variability time-scales are comparable to the previously published ones for all objects.

We calculate amplitudes for the lightcurves by taking the min-max difference for each star. To ensure no extreme outliers are included in this calculation we check for any 3-$\sigma$ outliers but find none. Table \ref{amplitudes} lists the derived amplitudes and compares them to previously published values (SE04, SE05). The amplitudes in SE04 and SE05 refer to the peak-to-peak values in the binned lightcurves.

\begin{table}
	
	\caption{Lightcurve amplitudes (A) derived in this work (see Sect 6.1) and previously published values (see text for details).}
	\label{amplitudes}
	\centering

	\begin{threeparttable}	
	
		\begin{tabular}{|c|c|c|c|c|c|}
	
			 \hline	
			{\bf Target} & {\bf Calculated A [mag]} & {\bf Published A [mag]} \\ \hline
			V2737 Ori & 0.79 & 1.114\tnote{ a} \\
			V2721 Ori & 0.24 & 0.547\tnote{ a} \\			
			V2739 Ori & 0.40 & 0.314\tnote{ a} \\
			V2227 Ori & 0.13 & 0.2\tnote{ b}  \\
			V1999 Ori & 0.54 & 0.426\tnote{ b} \\
			V1959 Ori & 0.24 & 0.073\tnote{ b}  \\
			V2559 Ori & 0.61 & 0.952\tnote{ b} \\  \hline
	
		\end{tabular}
	
		\begin{tablenotes}
       		
       		\item[a] \citealt{2004A&A...419..249S}; Table 4, I band
			\item[b] \citealt{2005A&A...429.1007S}; Table 1, I band

     	\end{tablenotes}	
	
	\end{threeparttable}

\end{table} 



\renewcommand{\thefigure}{\arabic{figure} (Cont.)}
\addtocounter{figure}{-1}

\begin{figure*}
   
	\begin{tabular}{cc}
	    
	\includegraphics[width=0.5\linewidth]{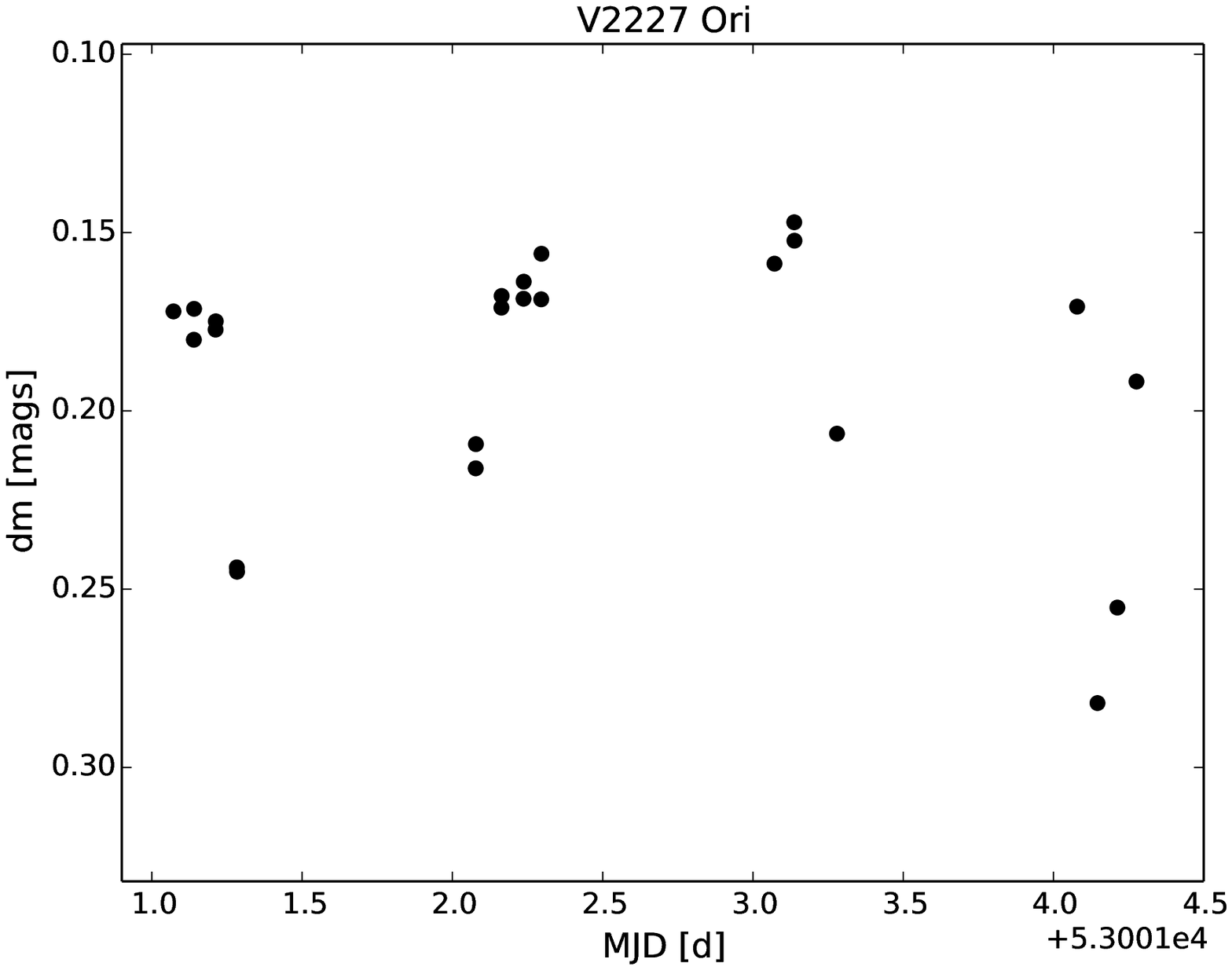}
	   \includegraphics[width=0.5\linewidth]{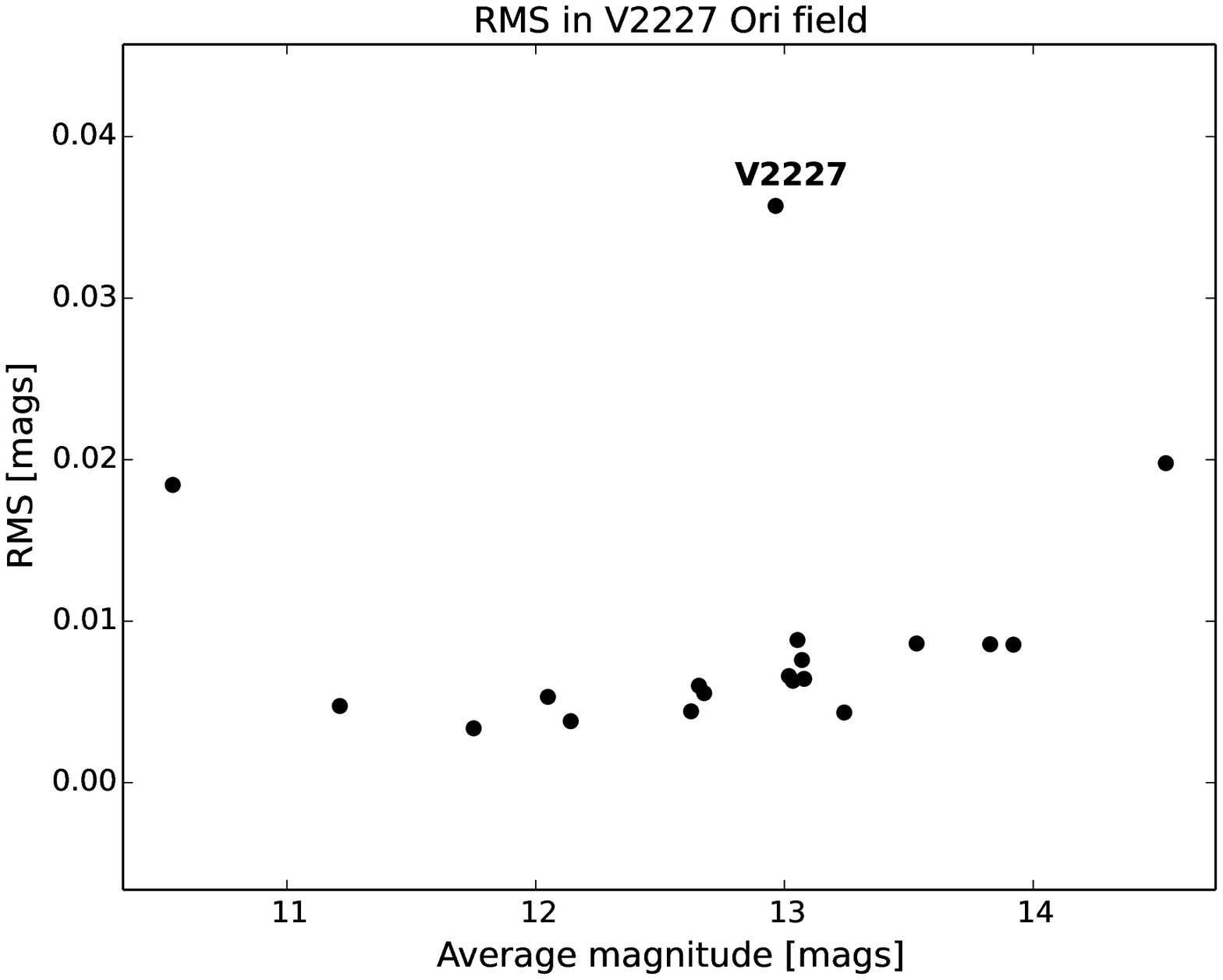} \\
	  
		\includegraphics[width=0.5\linewidth]{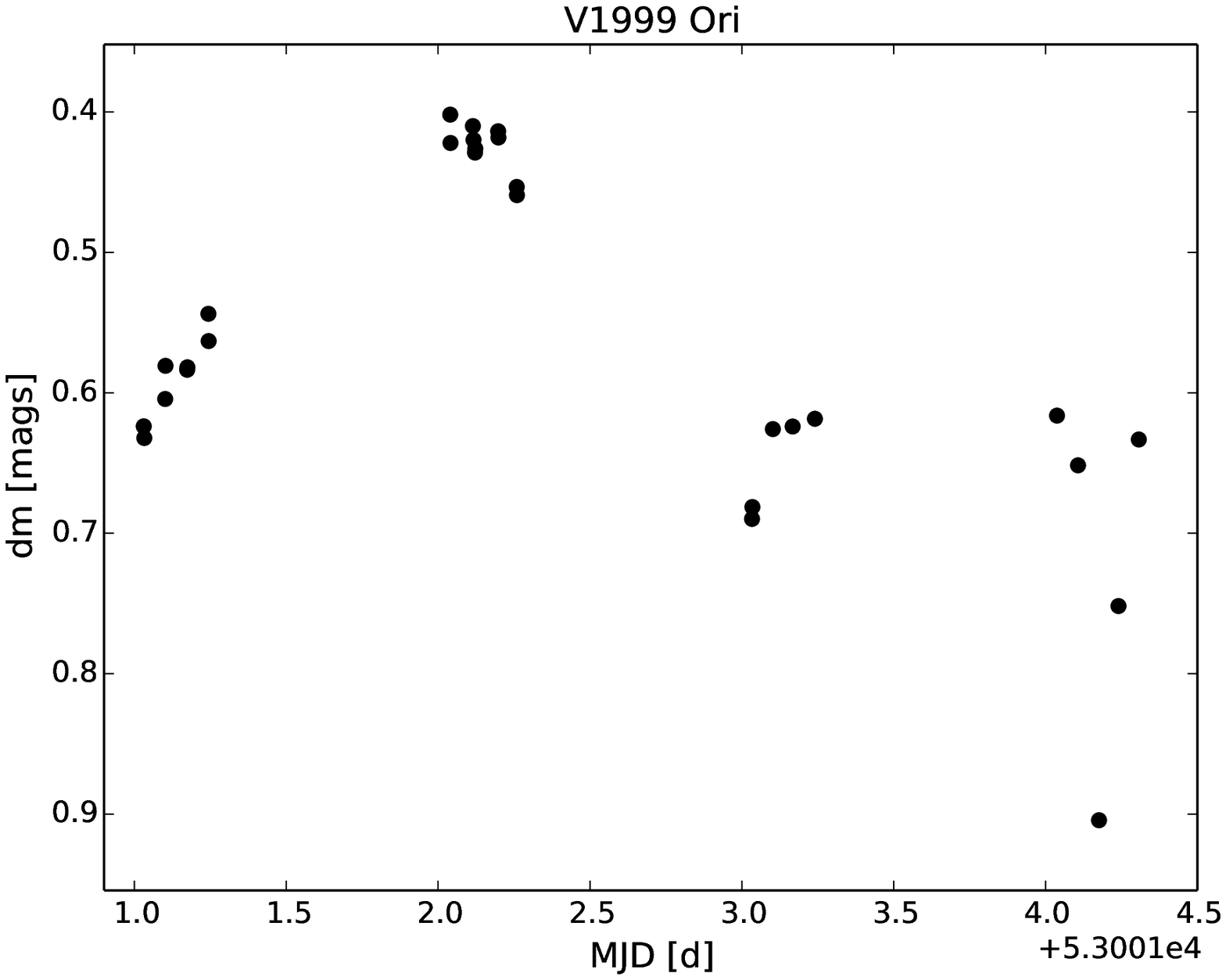}
	   \includegraphics[width=0.5\linewidth]{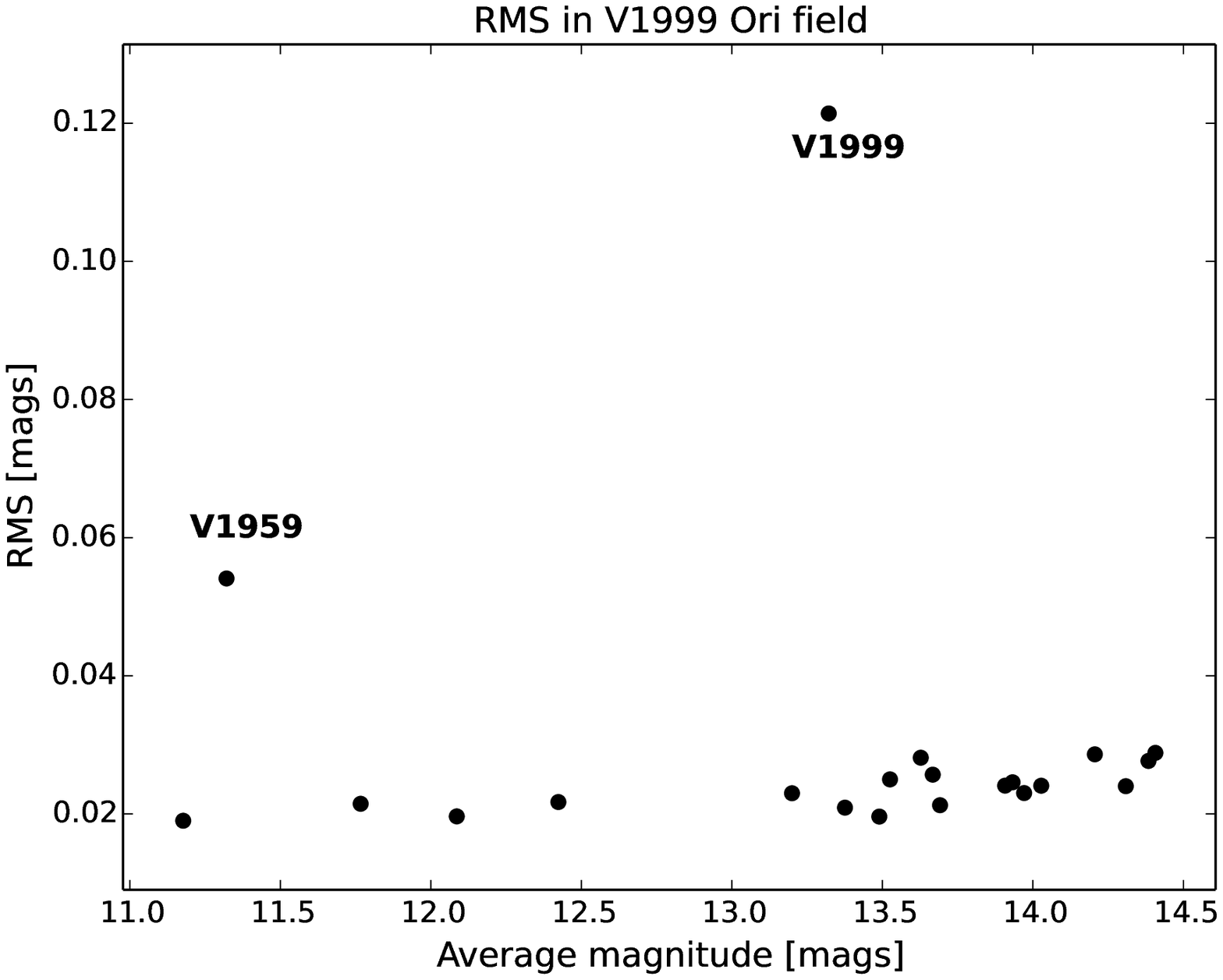} \\	  
	  
	    \includegraphics[width=0.5\linewidth]{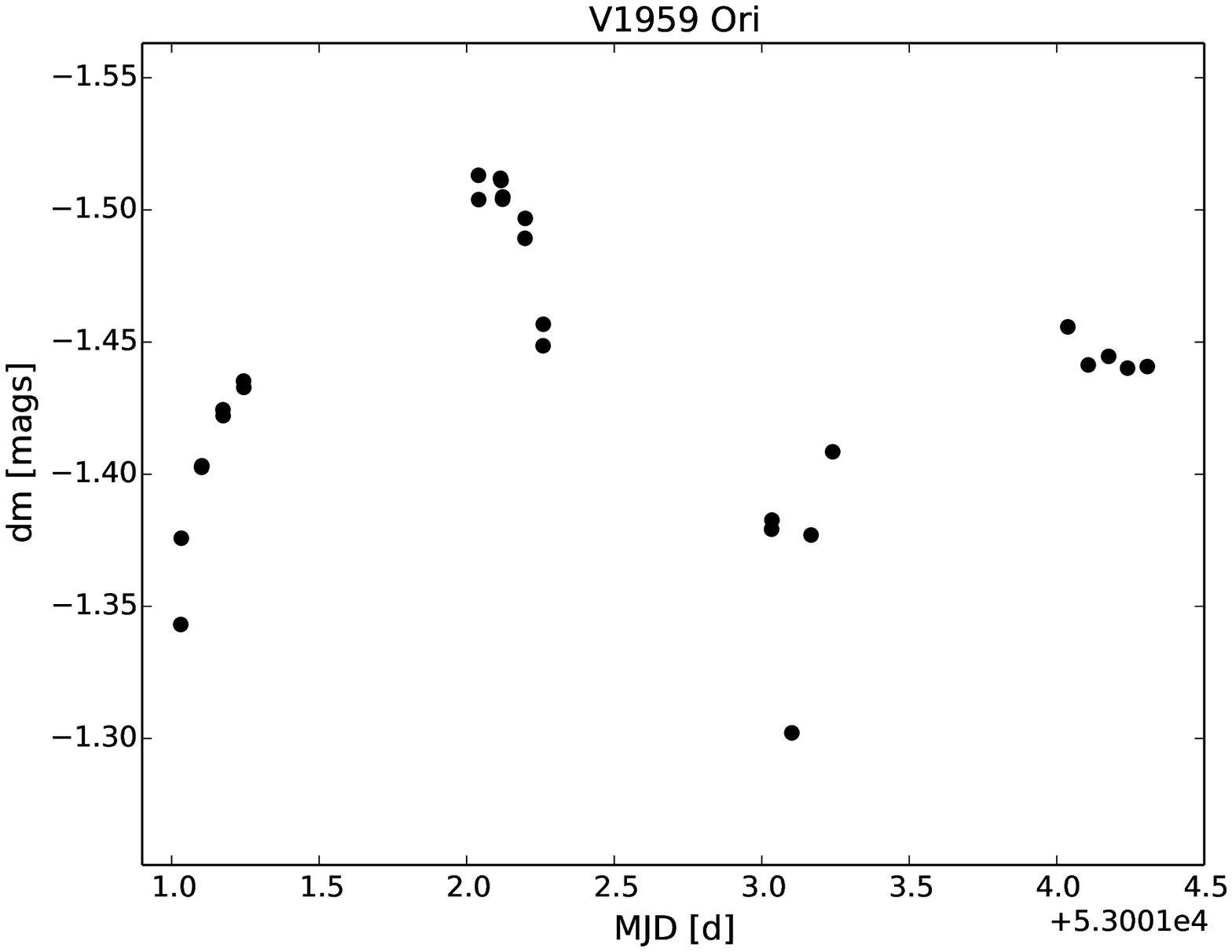}
	   \includegraphics[width=0.5\linewidth]{plots/phot/V1999_RMS.eps} \\

	\end{tabular}
	
	\caption{}	
	
\end{figure*}

\renewcommand{\thefigure}{\arabic{figure} (continued)}
\addtocounter{figure}{-1}

\begin{figure*}
   
	\begin{tabular}{cc}
	
	    \includegraphics[width=0.5\linewidth]{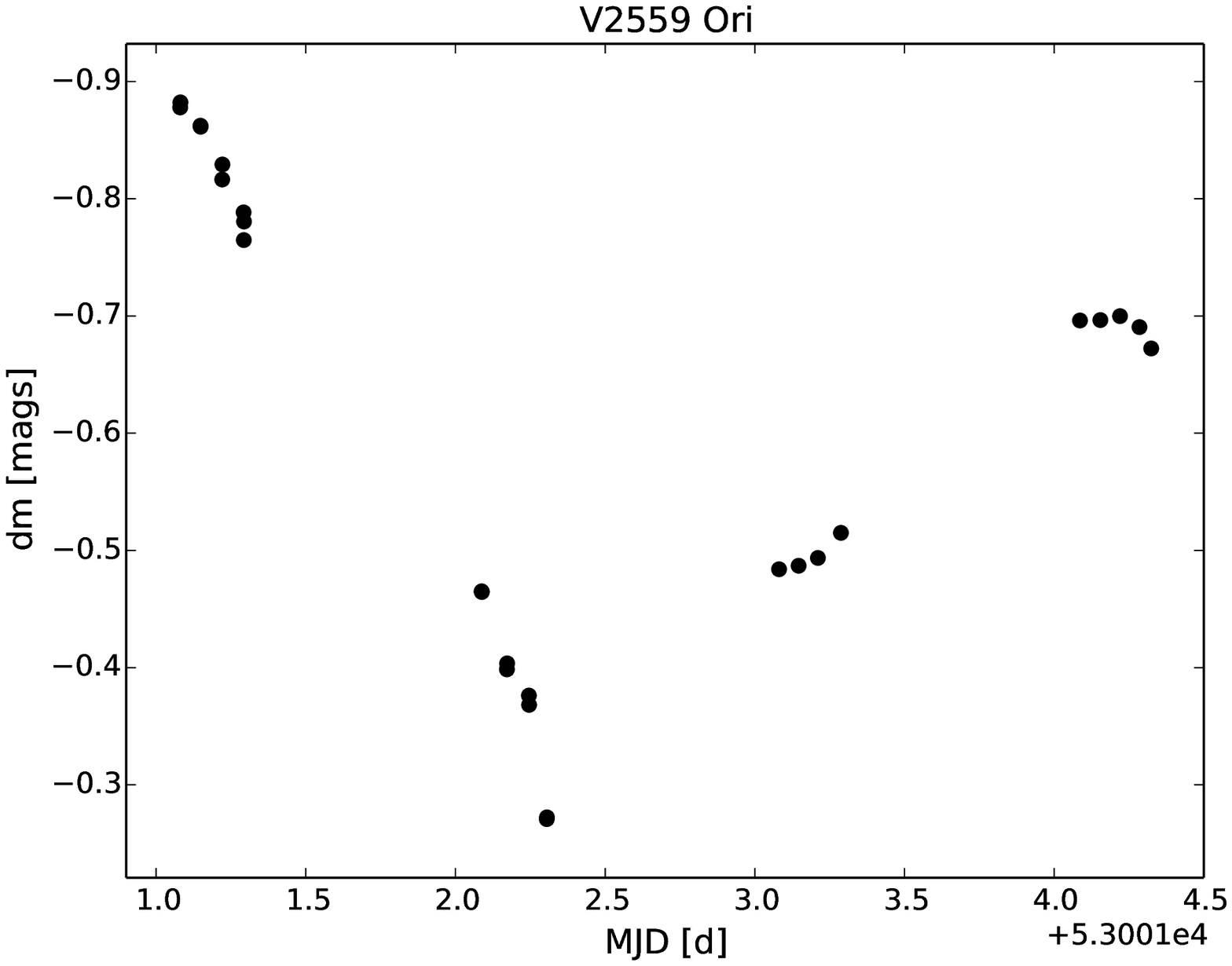}
	   \includegraphics[width=0.5\linewidth]{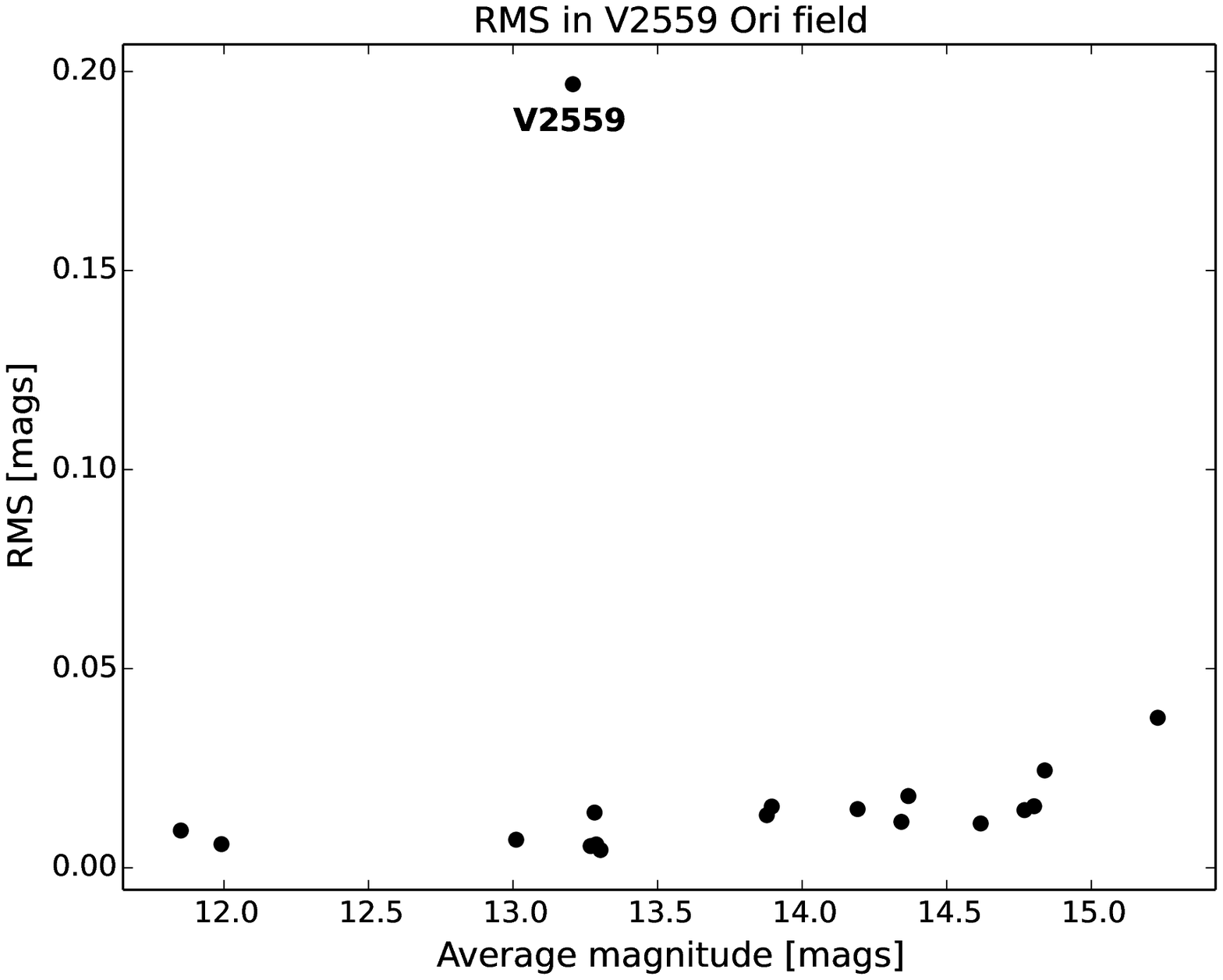} \\
	    
	\end{tabular}
	
	\caption{}	
	
\end{figure*}

\renewcommand{\thefigure}{\arabic{figure}}

\subsection{Spectroscopy}

\begin{figure*}
   
	\begin{tabular}{cc}
	    
	   \includegraphics[width=0.5\linewidth]{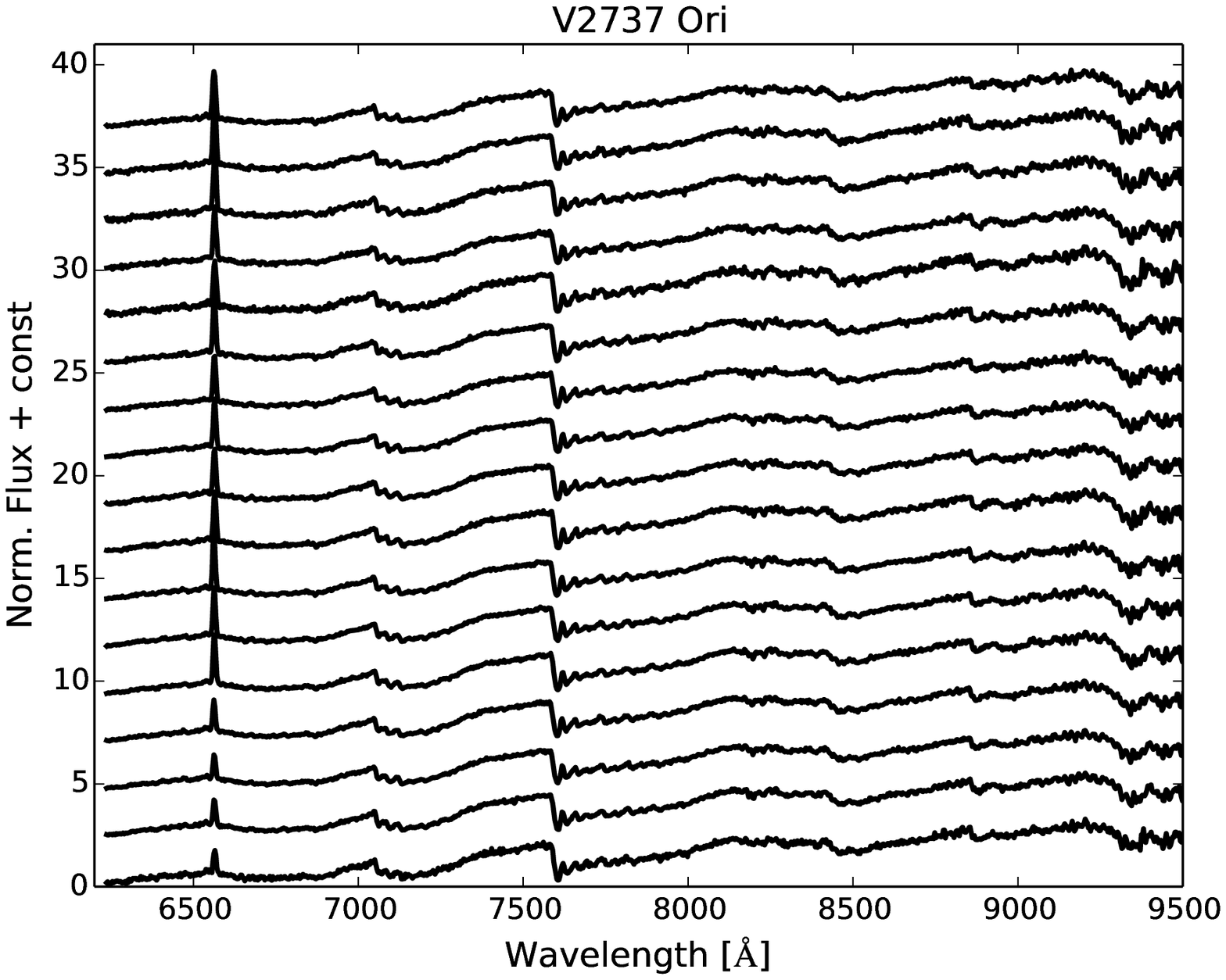}
	   \includegraphics[width=0.5\linewidth]{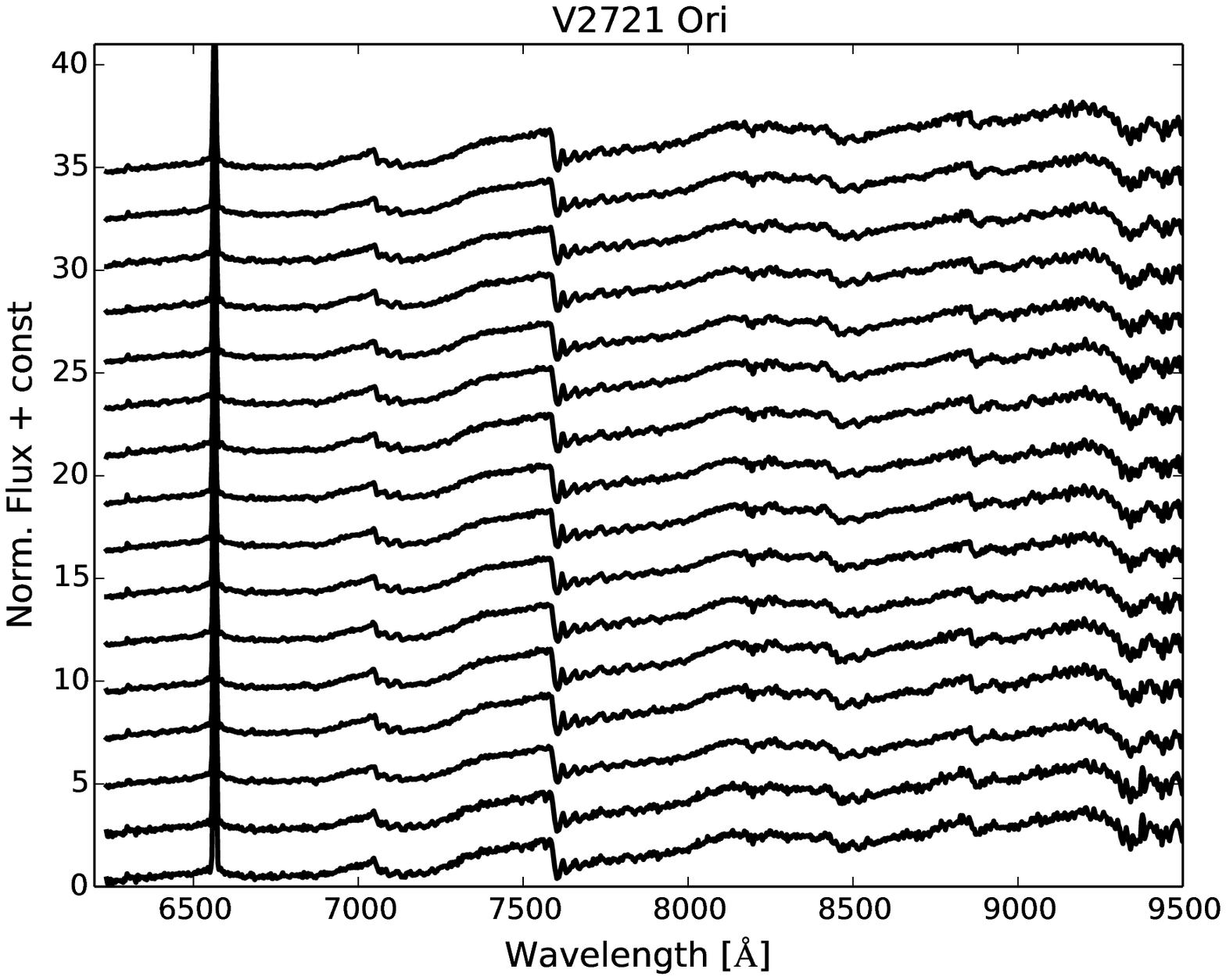} \\
	   
	    \includegraphics[width=0.5\linewidth]{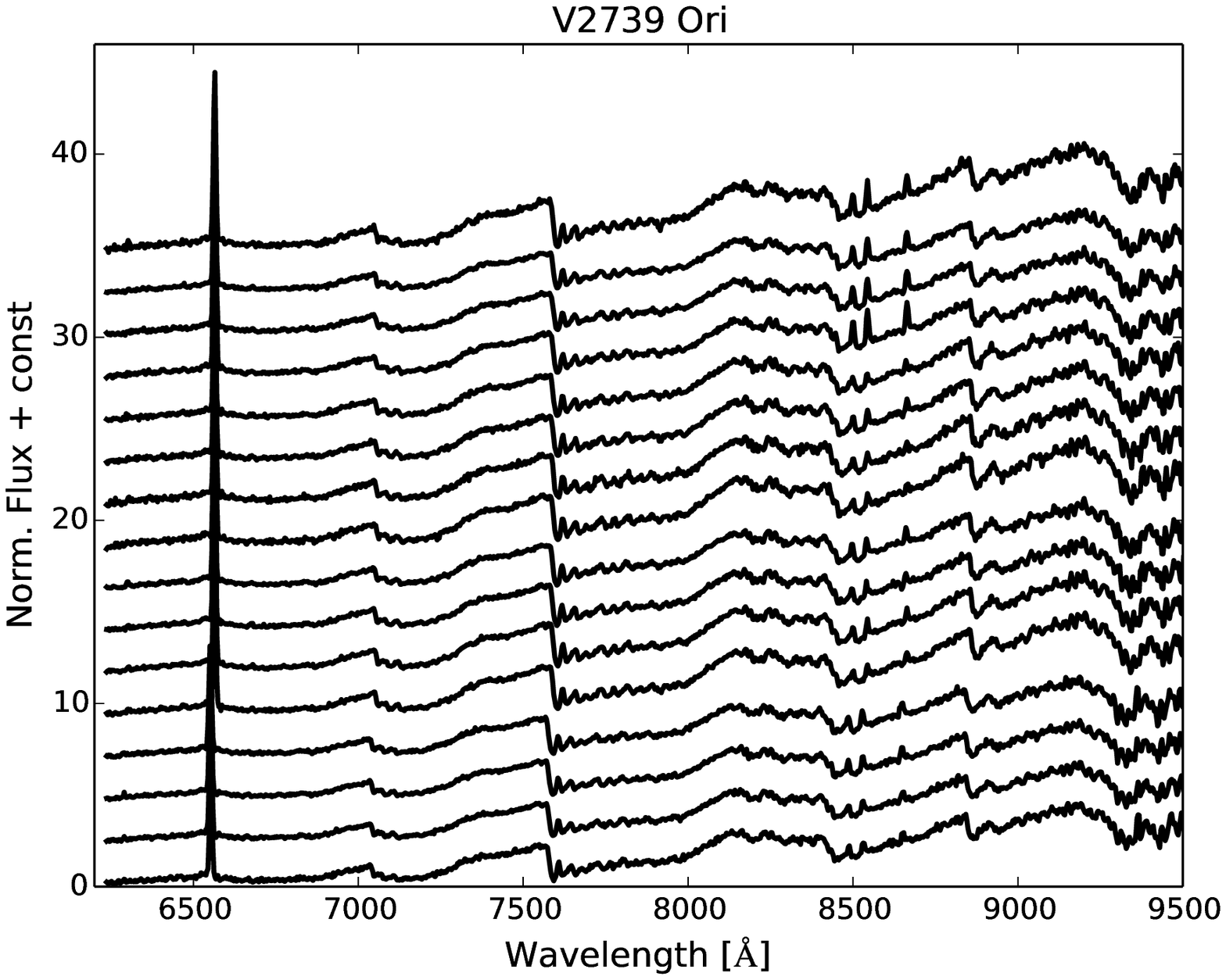}
	   \includegraphics[width=0.5\linewidth]{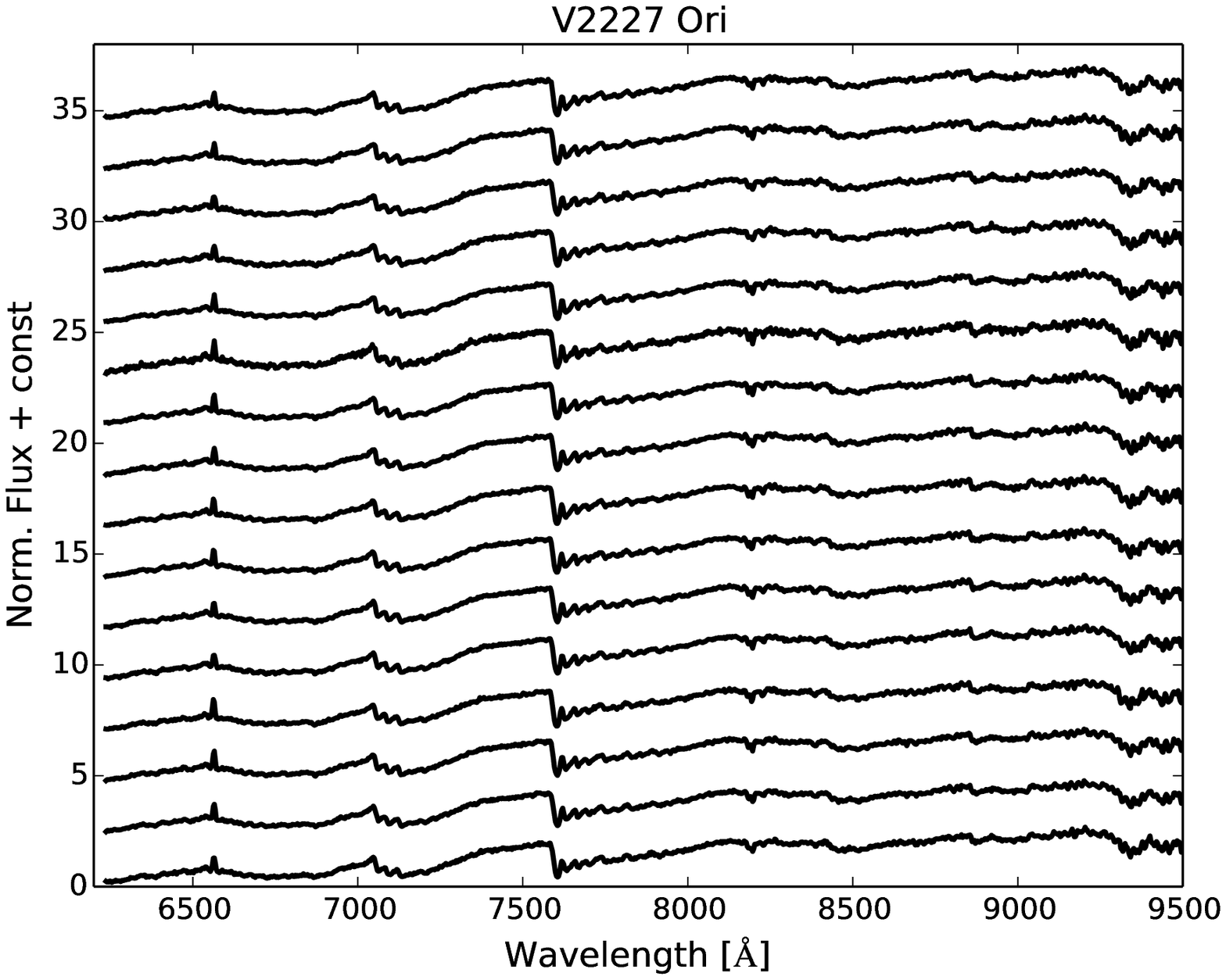} \\	 
	   
	    \includegraphics[width=0.5\linewidth]{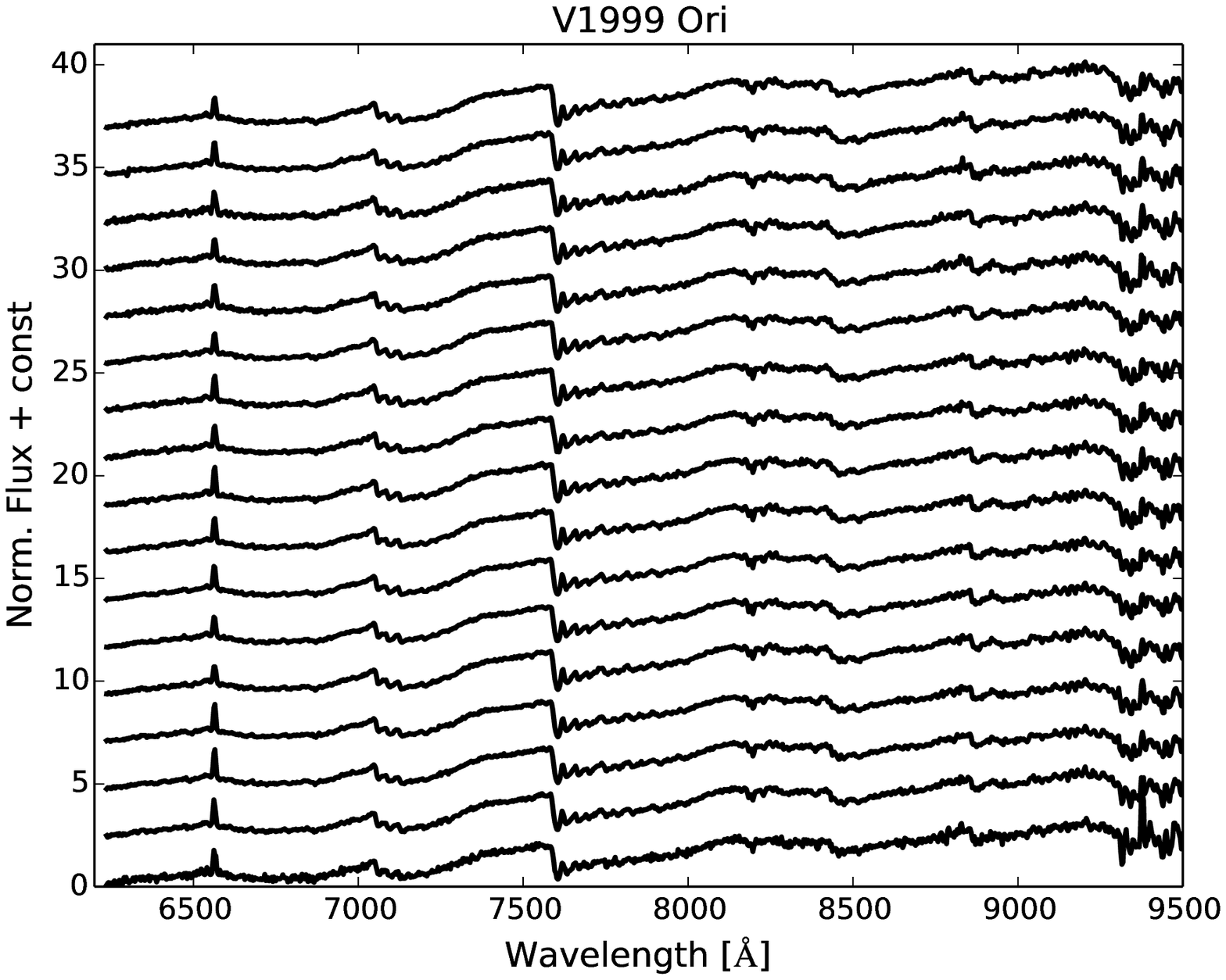}
	   \includegraphics[width=0.5\linewidth]{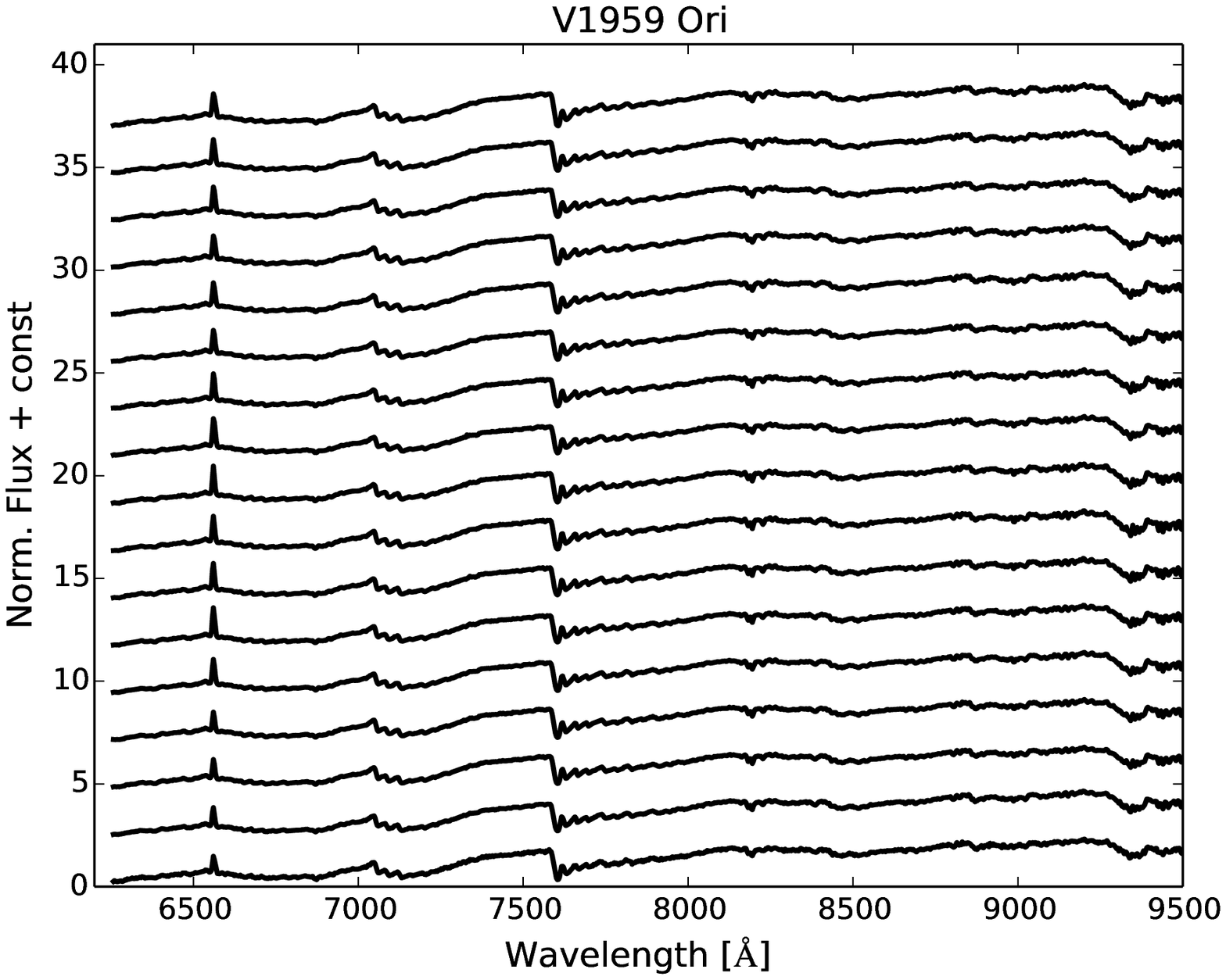} \\  
	  
	\end{tabular}
	
	\caption{Full sequence of spectra for all objects. The spectra are normalised at 7680$\AA$, then multiplied by a scale factor derived from the corresponding data point from the lightcurve (see Sect 4.1). }
	\label{allspecs}		
	
\end{figure*}

Standard spectroscopic data reduction was performed with IRAF routines from the {\it onedspec} and {\it twodspec} packages. Background was fit and subtracted during spectrum extraction. Wavelength calibration is done using a HeAr arclamp. The A-type star HD292956 was used for relative flux calibration. For each epoch, three spectra (see Sect \ref{obssect}) were co-added using a median filter. Figure \ref{allspecs} presents the complete sequences of calibrated spectra for our stars over the four nights of observations. Every spectrum is normalised to the central wavelength of the I-band filter (7680$\AA$). In the final figure (Fig. \ref{allspecs}) each spectrum is shifted by a constant for clarity.

\renewcommand{\thefigure}{\arabic{figure} (continued)}
\addtocounter{figure}{-1}

\begin{figure*}
   
	\begin{tabular}{c}
	
	    \includegraphics[width=0.5\linewidth]{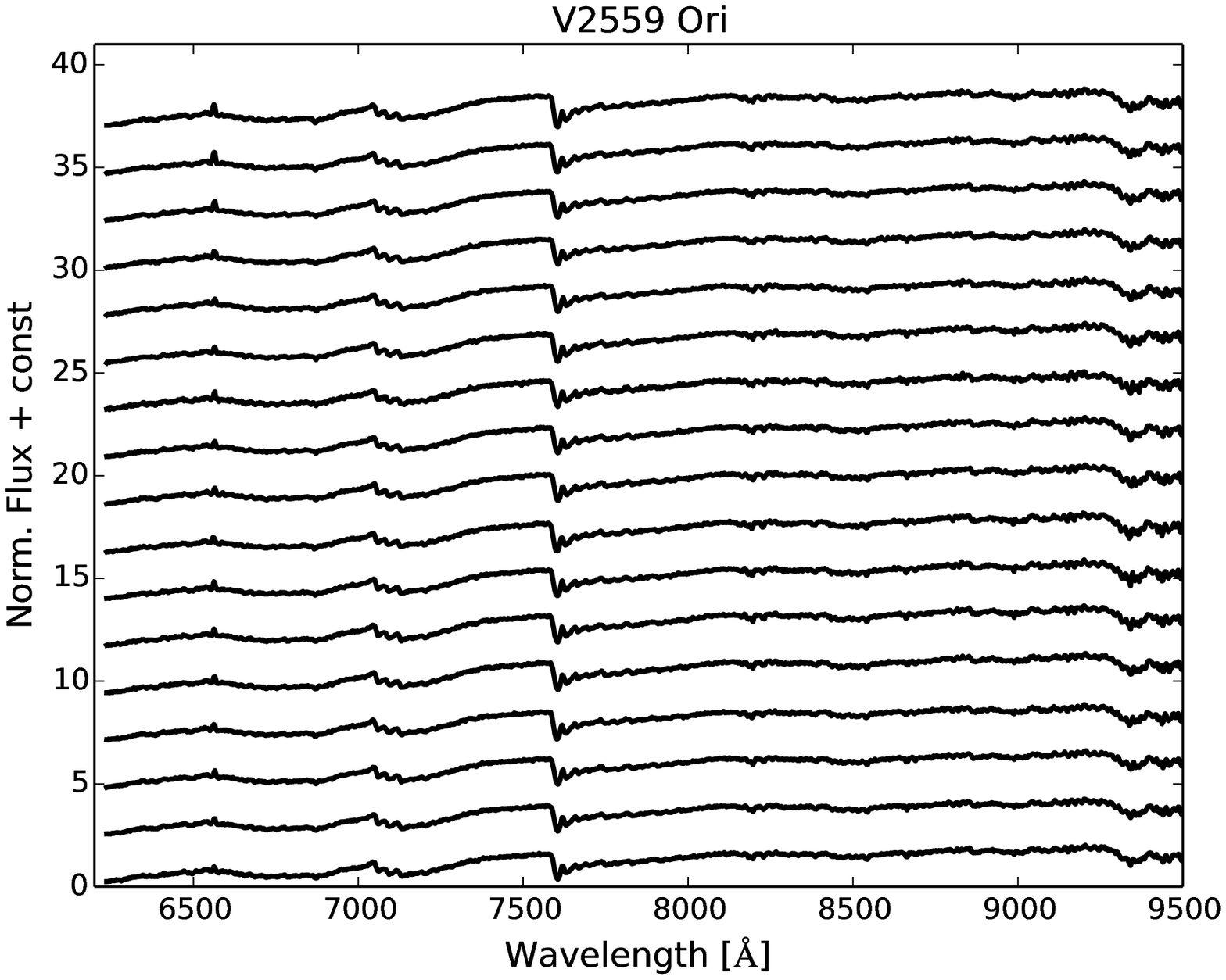}
	   
	\end{tabular}
	\caption{}
\end{figure*}

\renewcommand{\thefigure}{\arabic{figure}}

\section{Spectral analysis}
\label{spectralanalysis}
\subsection{Spectral types}
\label{spectraltypes}

We carry out a spectral classification using the pseudo-continuum PC3 index from \cite{1999AJ....118.2466M}. The PC3 integration limits are 8230-8270$\AA$ for the numerator and 7540-7580$\AA$ for the denominator. The spectral type is then empirically derived to be 
\begin{equation}
	{\rm 	SpT = -6.685 + 11.715 \times (PC3) - 2.024 \times (PC3)^2 }
\end{equation}  

All targets fall between spectral class M2.5 and M5.5, according to this index. Table \ref{specTypeTable} summarises the average spectral type and variations derived for each star over all epochs. The variation is taken as the maximum difference between individual spectral type measurements and the average spectral type. The variations are between 0.2 and 0.4 subtypes. Since the uncertainty in the PC3 coefficients (Eq. 1) is given as 0.28 by \cite{1999AJ....118.2466M}, none of the objects is showing a significant change in spectral type. 

\begin{figure*}
   
	\begin{tabular}{cc}
	    
	   \includegraphics[width=0.5\linewidth]{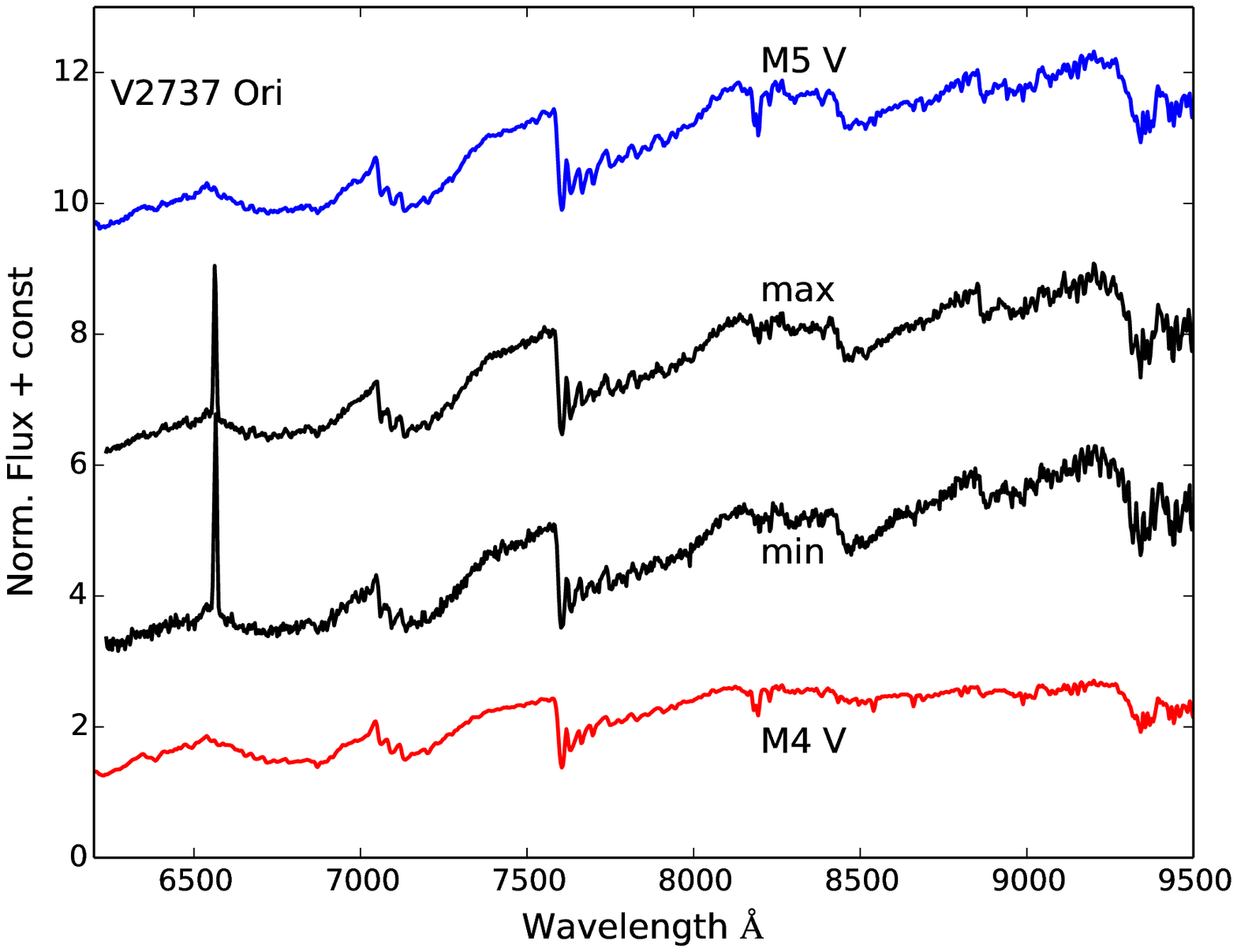}
	   \includegraphics[width=0.5\linewidth]{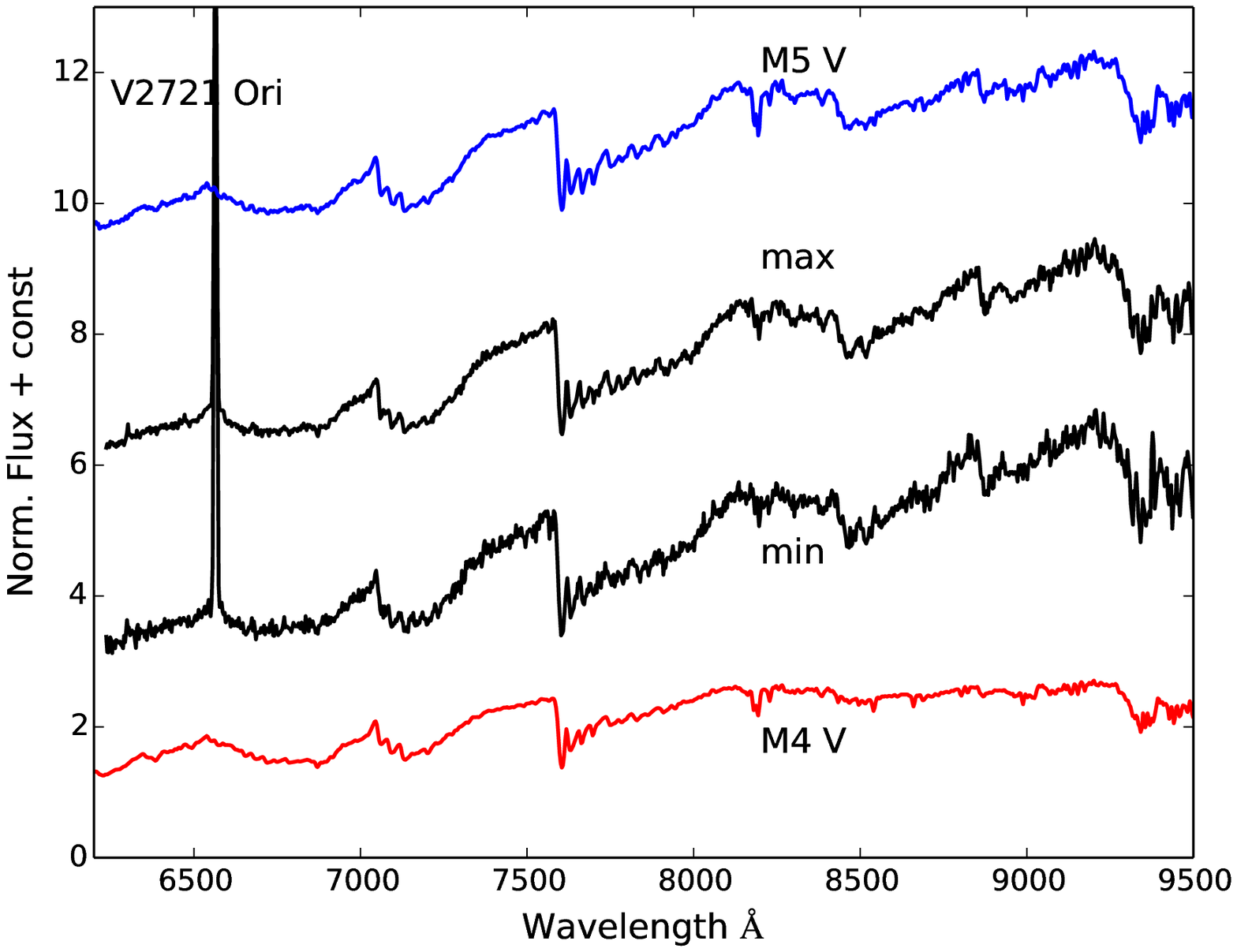} \\
	   
	    \includegraphics[width=0.5\linewidth]{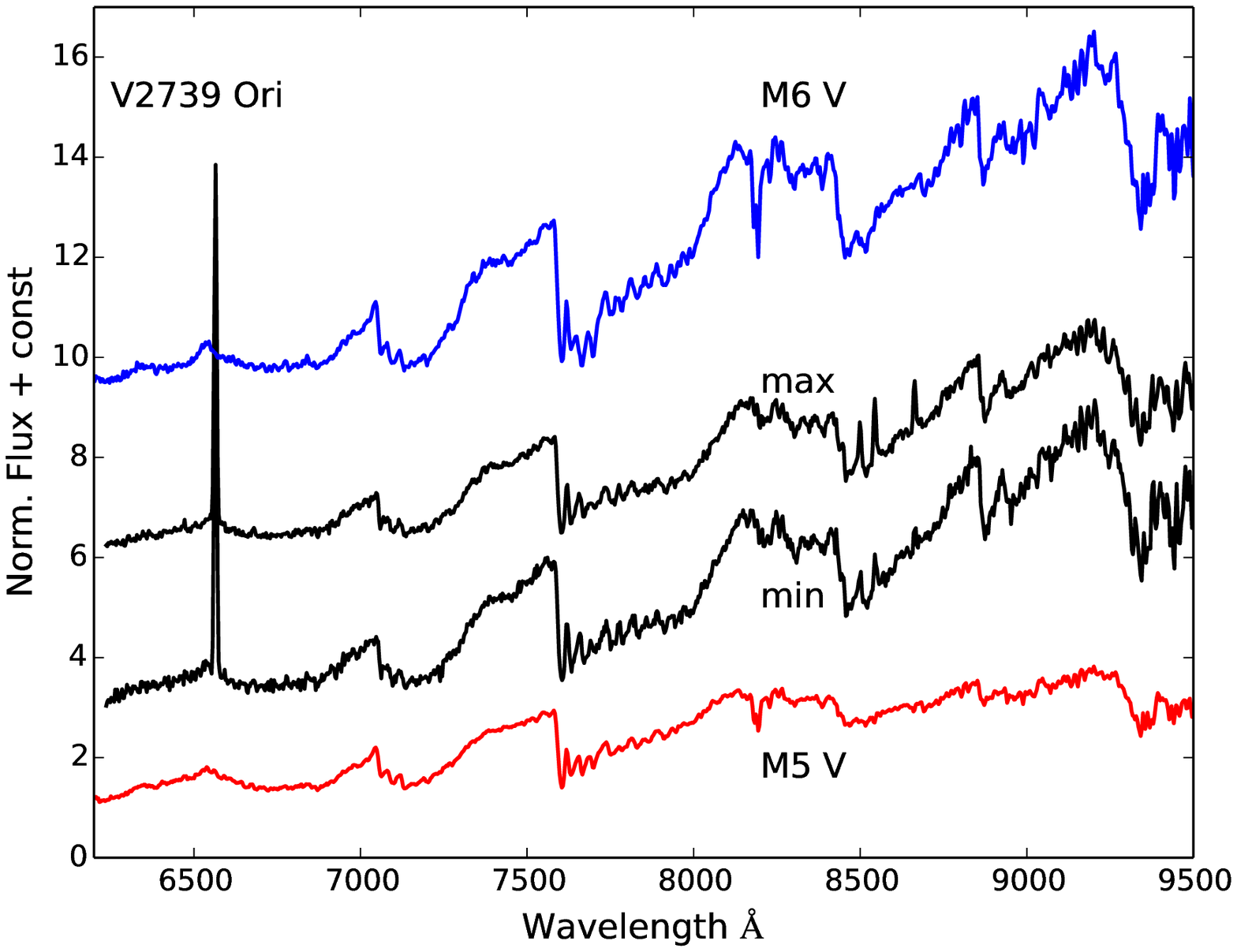}
	   \includegraphics[width=0.5\linewidth]{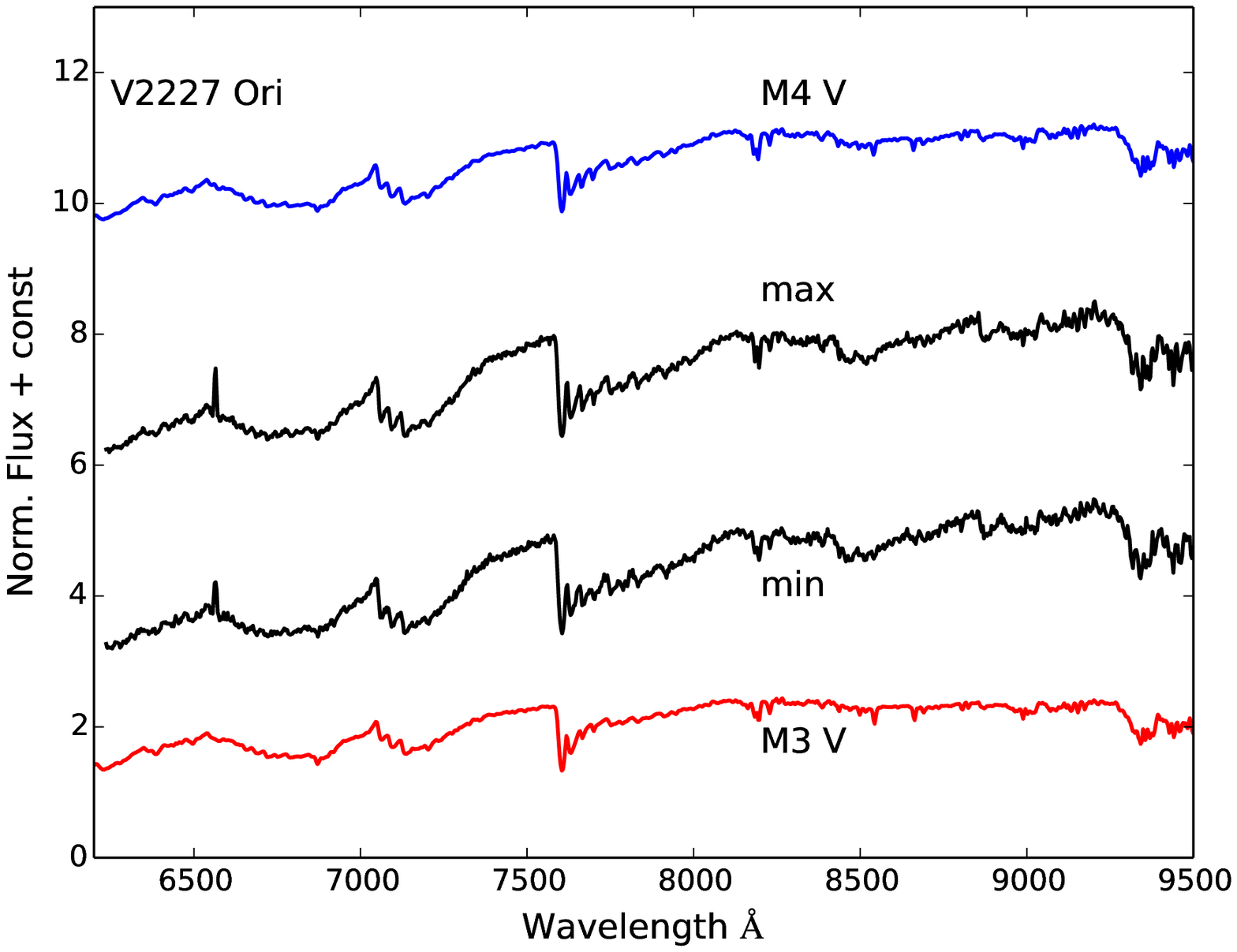} \\	 
	   
	    \includegraphics[width=0.5\linewidth]{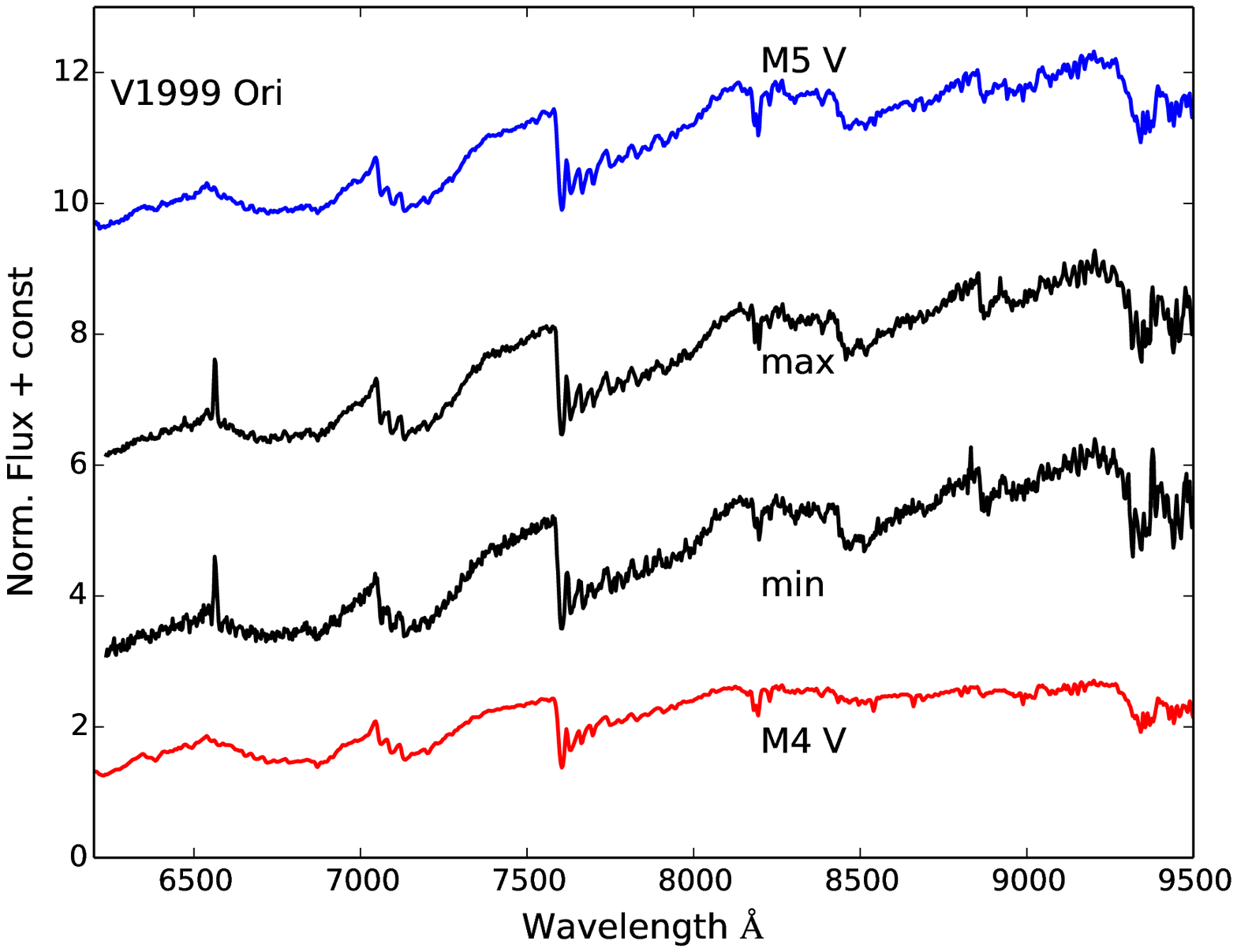}
	   \includegraphics[width=0.5\linewidth]{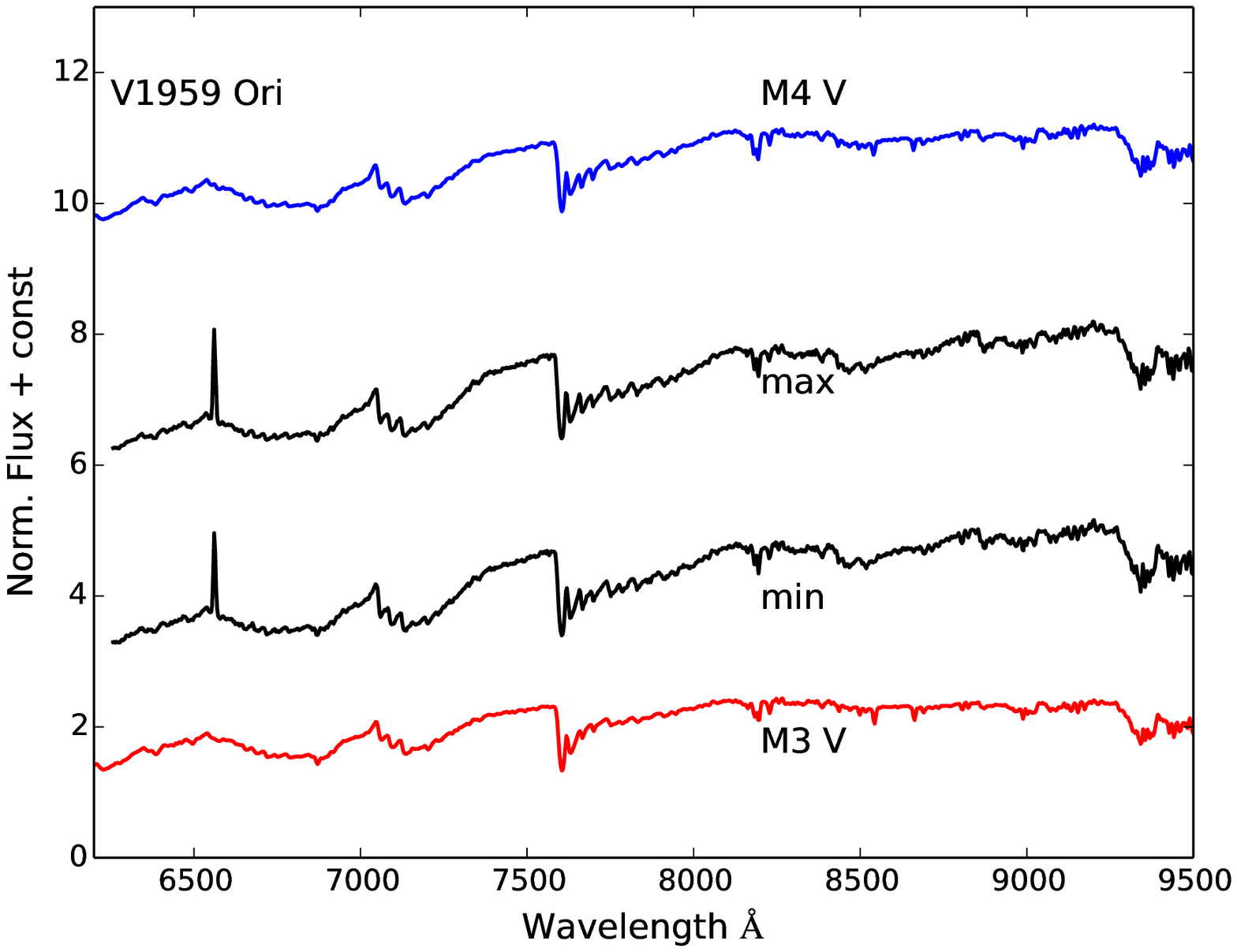} \\  
	  
	\end{tabular}

	\caption{Visual comparison between targets and spectral templates of M dwarfs. The two black spectra correspond to the maximum and minimum epochs in the lightcurves. The coloured spectra represent the templates, normalised at 7680$\AA$.}
	\label{templatesfull}	
	
\end{figure*}

In Figure \ref{templatesfull} we provide further comparison between all targets and closest visual matches from spectral templates (templates are identical to ones employed by \cite{2014ApJ...785..159M}, which they compiled by averaging observed spectra of M dwarfs of similar spectral type, obtained from the dwarf archives\footnote{www.dwarfarchives.org; \citealt{1991ApJS...77..417K,1992PhDT.........7K,1995AJ....109..797K} or unpublished}). Two spectra, corresponding to the epochs of the maximum and minimum in the lightcurve, are plotted for each star. The spectral types of the closest matching templates are summarised in Table \ref{specTypeTable}. The spectral types derived from PC3 and those derived from comparing with templates agree within 1 subtype. We note that neither method is a perfect representation for our targets as our set is composed of very young stars and the templates as well as the stars used to define the PC3 index are evolved field objects. Even so, no measurable spectral type change is seen in either case.

We further make use of eq. 3 in \cite{2014ApJ...785..159M} to calculate an effective temperature for the stars based on their spectral type. No object shows a temperature change of more than 100K which is less than the expected difference for a change of one spectral subtype (170K). What is more, the uncertainty in the temperature determination from spectral type (140K) is higher than any derived change for our targets. We conclude that there are no significant effective temperature changes in our sample during the 4 night run.

\begin{table}

	\caption{ Spectral measurements for our onjects. {\bf Col. 2 and 3:} Average value and variation (see Sect. \ref{spectraltypes}) for the spectral type of each object. Spectral types derived using the PC3 index \citep{1999AJ....118.2466M}. {\bf Col. 4:} Closest visual matches between our spectra and spectral templates. {\bf Col. 5 and 6:} Average value and standard deviation of the measured H$\alpha$ equivalent width for each target (see Sect \ref{halpha}).}
	
	\label{specTypeTable}

	\begin{tabular}{c|c|c|c|c|c}

		\hline	
			Target & $\overline{Sp\,T}$ & SpT var. & Temp. SpT & $\overline{H \alpha \,EW}$ [$\AA$] & $\sigma$ [$\AA$] \\ \hline
			V2737 Ori & M 3.8 & 0.4 & M4-M5 & 30 & 8 \\
			V2721 Ori & M 4.0 & 0.3 & M4-M5 & 97 & 17\\
			V2739 Ori & M 4.9 & 0.3 & M5-M6 & 96 & 17\\
			V2227 Ori & M 3.3 & 0.2 & M3-M4 & 6 & 1\\
			V1999 Ori & M 3.9 & 0.2 & M4-M5 & 14 & 1\\
			V1959 Ori & M 2.9 & 0.2 & M3-M4 & 12 & 2\\
			V2559 Ori & M 3.2 & 0.2 & M3-M4 & 3 & 1\\  \hline

	\end{tabular}
		
\end{table}

\renewcommand{\thefigure}{\arabic{figure} (continued)}
\addtocounter{figure}{-1}

\begin{figure*}
   
	\begin{tabular}{c}
	
	    \includegraphics[width=0.5\linewidth]{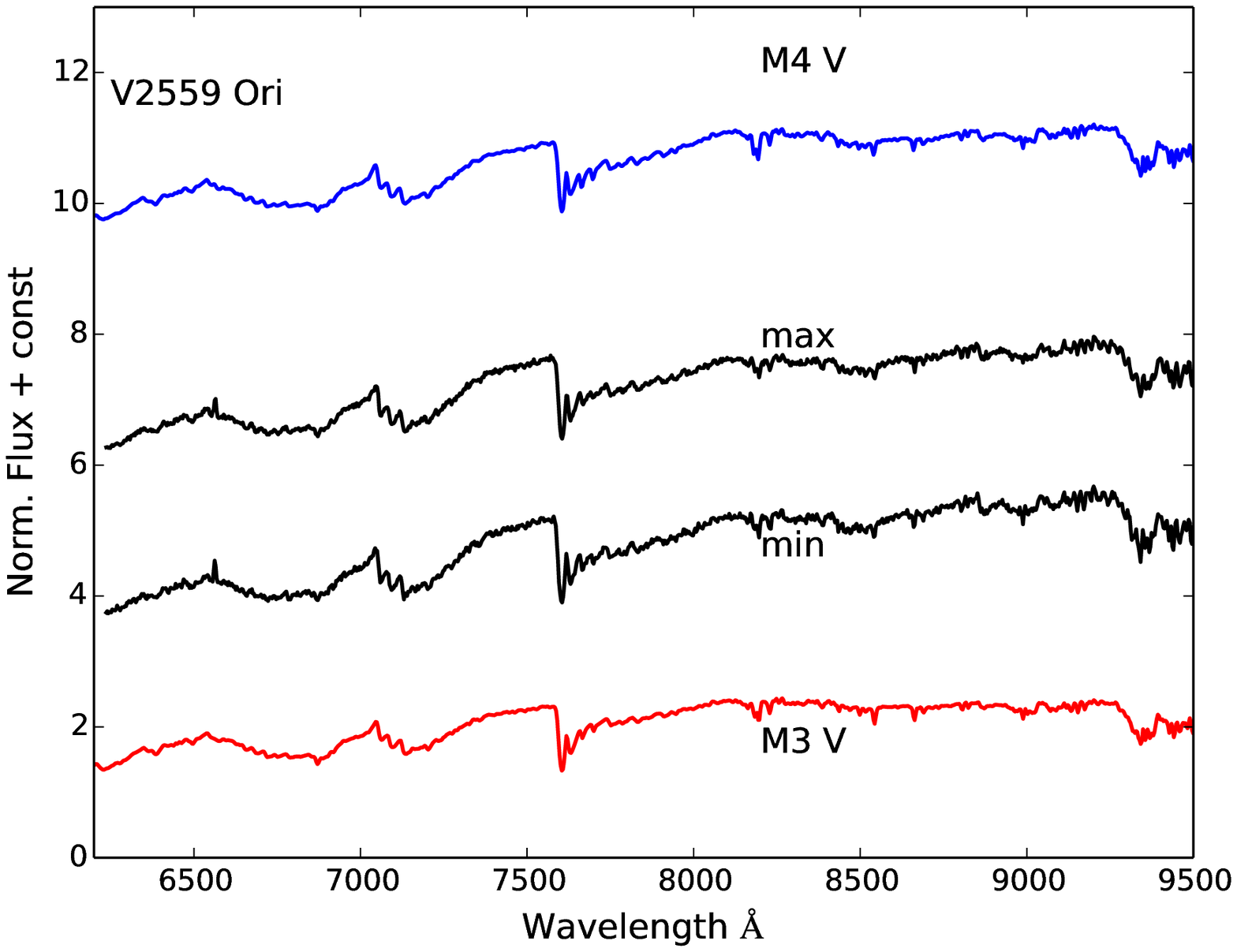}
	   
	\end{tabular}
	
	\caption{}	
	
\end{figure*}

\renewcommand{\thefigure}{\arabic{figure}}

\subsection{H$\alpha$ equivalent widths}
\label{halpha}

H$\alpha$ emission in spectra of young stars can either be caused by magnetic chromospheric activity or by ongoing accretion. While magnetic activity results in narrow, weak H$\alpha$ emission, accretion causes broad and strong lines.
We measure the H$\alpha$ equivalent widths (EW) in each spectrum using 

\begin{equation}
	EW = \sum \bigg(\frac{F_{line}}{F_{cont}} - 1\bigg)\, \Delta\lambda
\end{equation}

where $F_{line}$ is the emission line flux and $F_{cont}$ is the continuum flux.
The continuum is smoothed using optimal averaging with a Gaussian filter of width 15.0$\,\AA$. The line is integrated between 6555 and 6575$\,\AA$.The average EWs and the standard deviation over the time series of spectra are listed in Table \ref{specTypeTable} for each target. The stars near $\sigma$\,Ori show H$\alpha$ EW of over 30 $\AA$ for V2727 and above 95 $\AA$ for V2739 and V2721, in excess of typical levels for chromospheric activity and therefore indicating ongoing accretion \citep{2003AJ....126.2997B}. In contrast, the objects near $\epsilon$\,Ori have EWs of $< 15 \AA$, consistent with chromospheric emission. The error in the EW estimation is 1-2$\,\AA$. The variation in the EW for the $\sigma$\,Ori stars is in the range of 10-20\%, which is strong evidence for a significant variability in these objects, in addition to the I-band photometry (which doesn't cover the H$\alpha$ wavelength). On the other hand, the change in H$\alpha$ in the $\epsilon$\,Ori targets is comparable to the uncertainty in the measurement, suggesting very little or no variability. 

We checked for correlations between I-band variability and H$\alpha$ EW variability for the $\sigma$\,Ori targets. While V2737 may show a hint of a possible correlation, no clear trend can be identified in the sample of only three stars. These plots are included for completeness in Appendix \ref{appB}. The lack of clear correlation would imply that the H$\alpha$ and continuum variations originate from different areas of the stellar systems.

We will also note V2739 is showing CaII triplet emission, a feature associated with accretion. The same feature was clearly detected for V2739 in SE04, as well as V2737 and V2721. However, since our present spectra are low-resolution and only show the CaII triplet for one object, we make no further attempt to quantify this emission.

\subsection{Spectral variations}
\label{spvariations}

\begin{figure*}
   
	\begin{tabular}{cc}
	    
	   \includegraphics[width=0.5\linewidth]{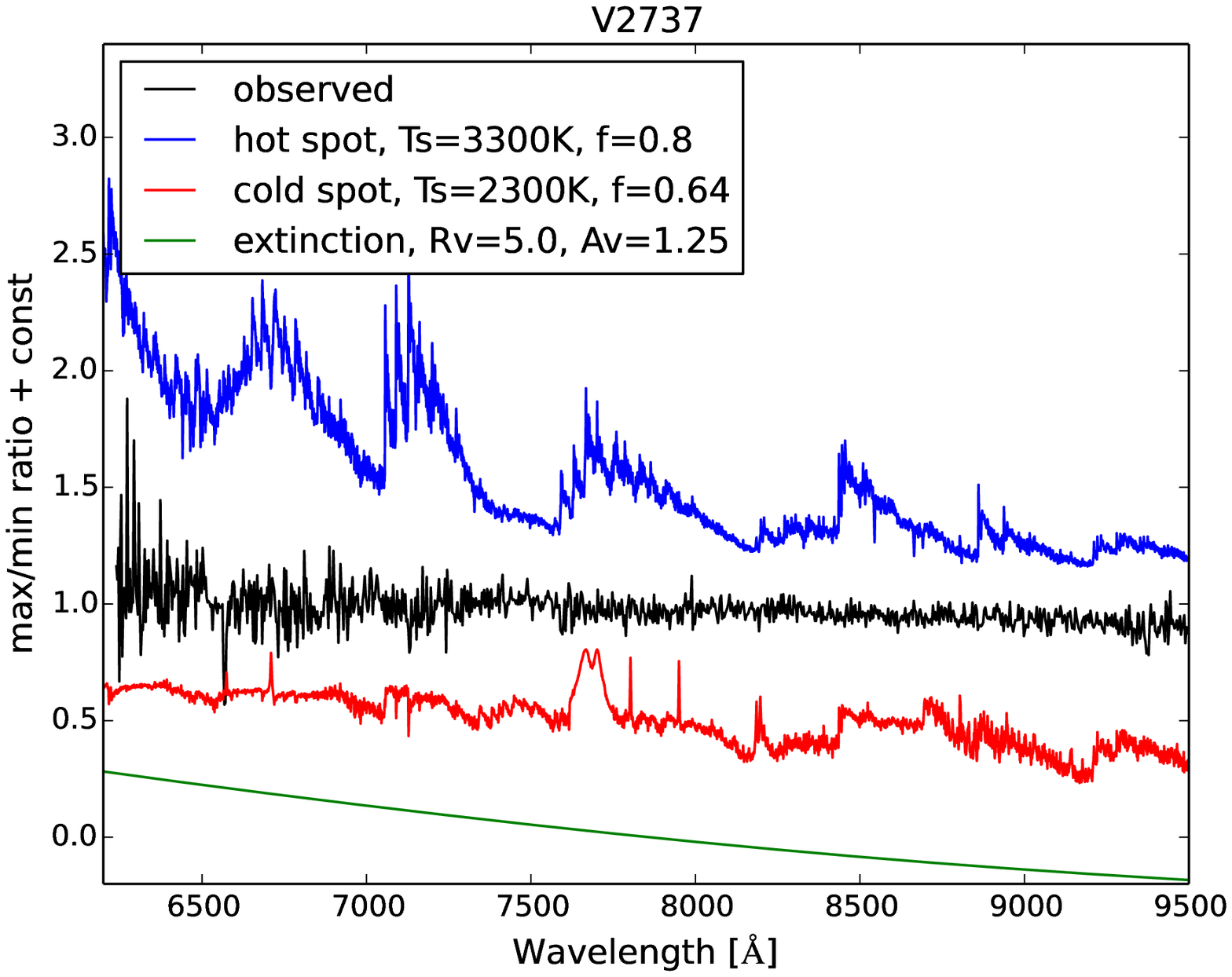}
	   \includegraphics[width=0.5\linewidth]{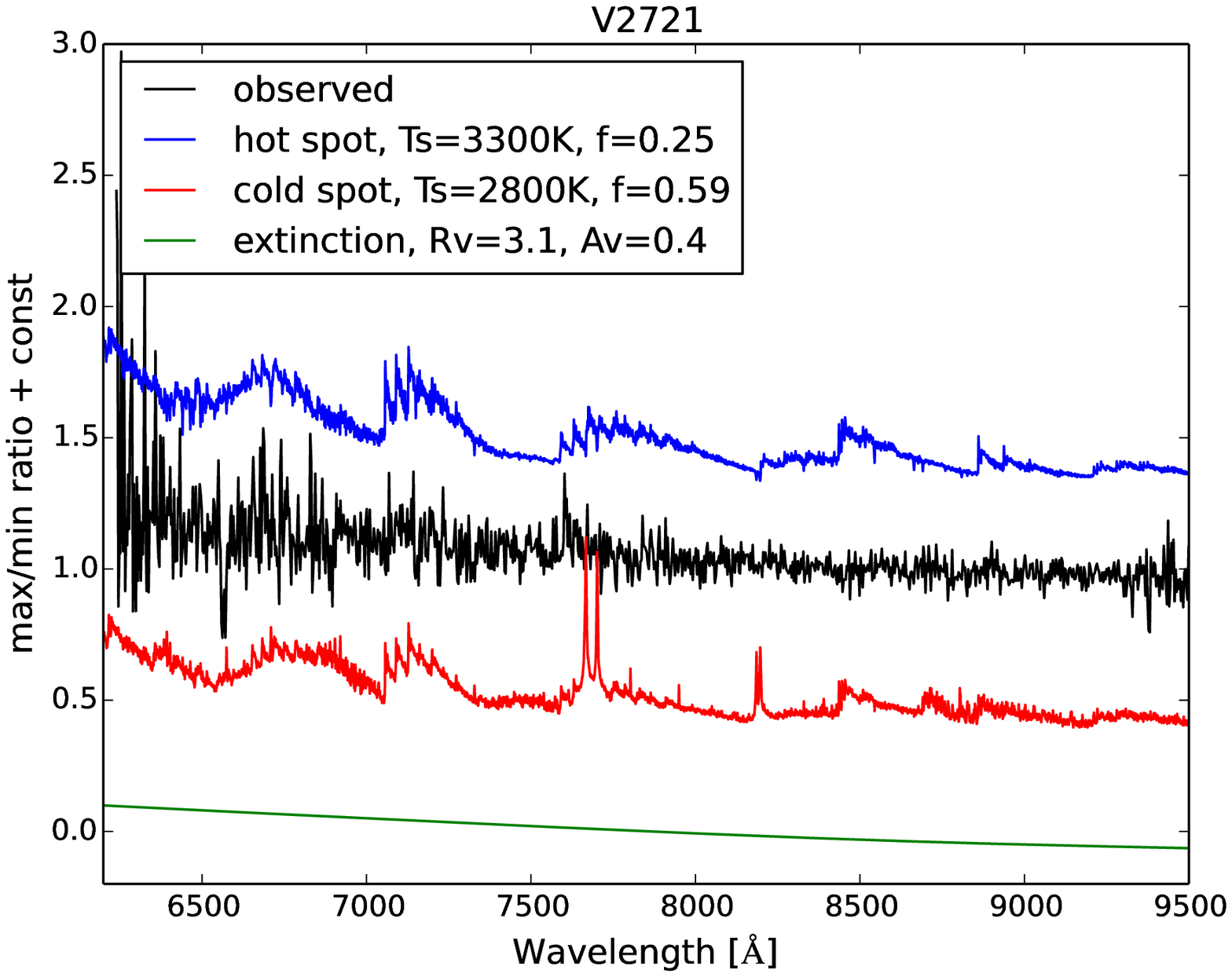} \\
	   
	    \includegraphics[width=0.5\linewidth]{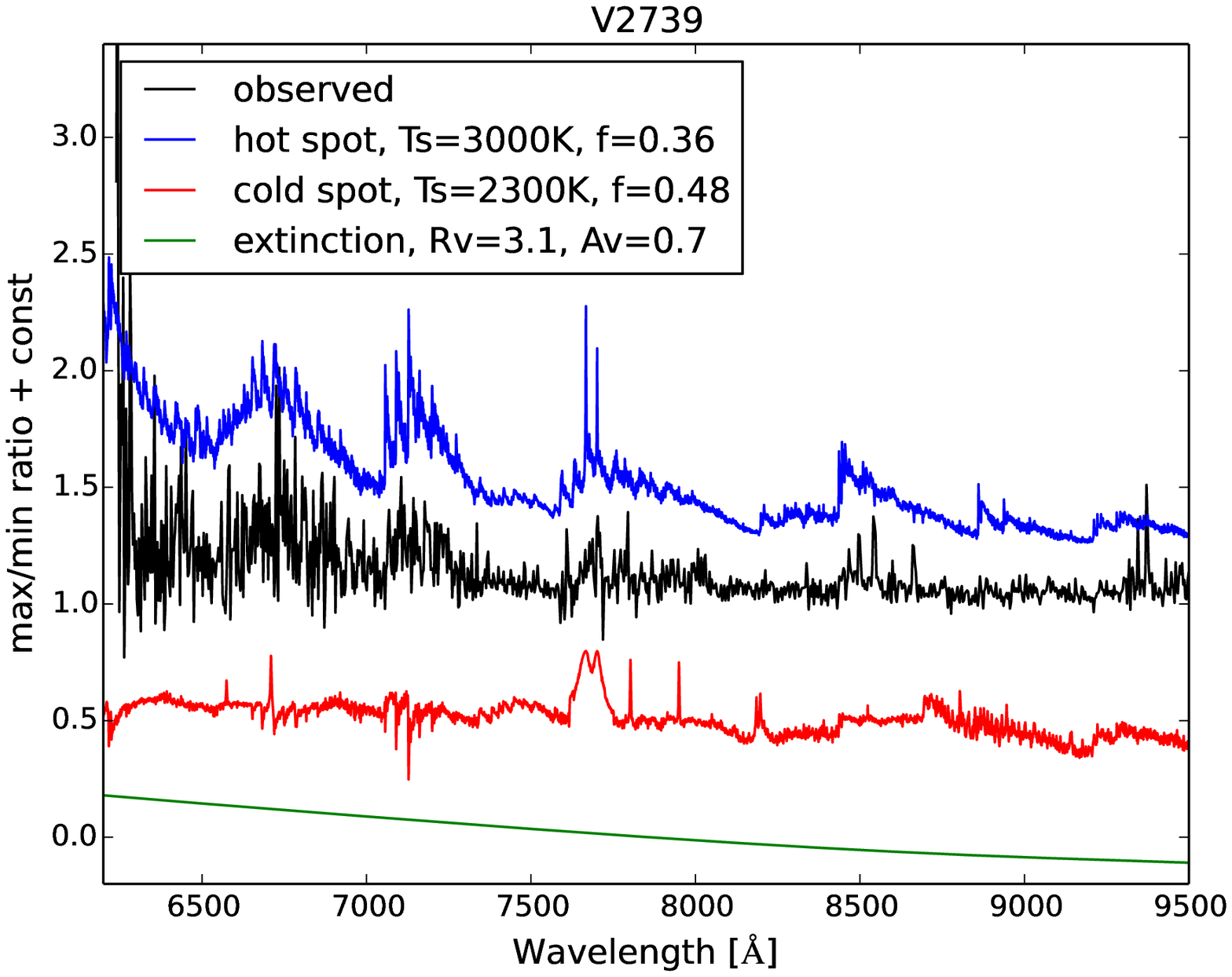}
	   \includegraphics[width=0.5\linewidth]{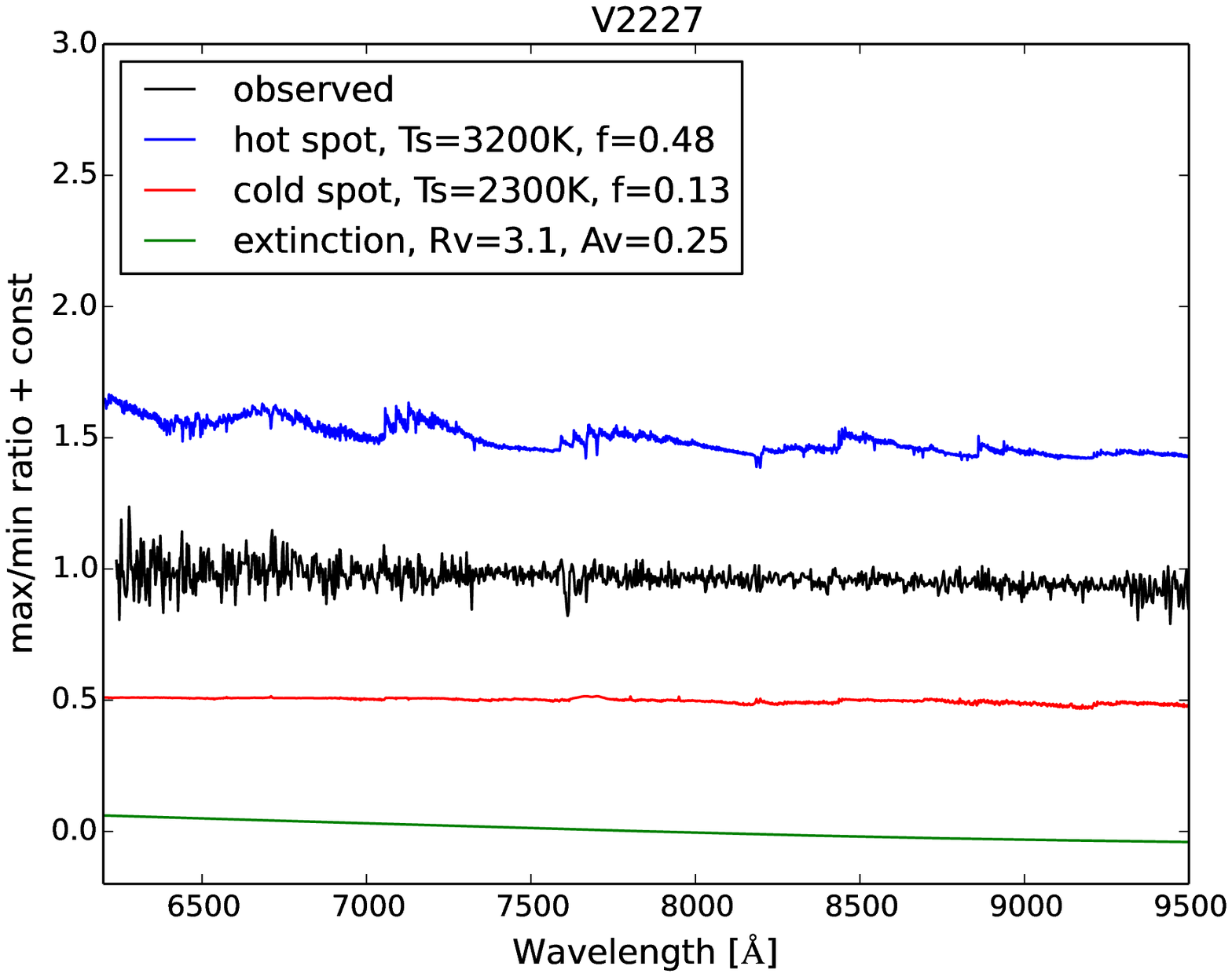} \\	 
	   
	    \includegraphics[width=0.5\linewidth]{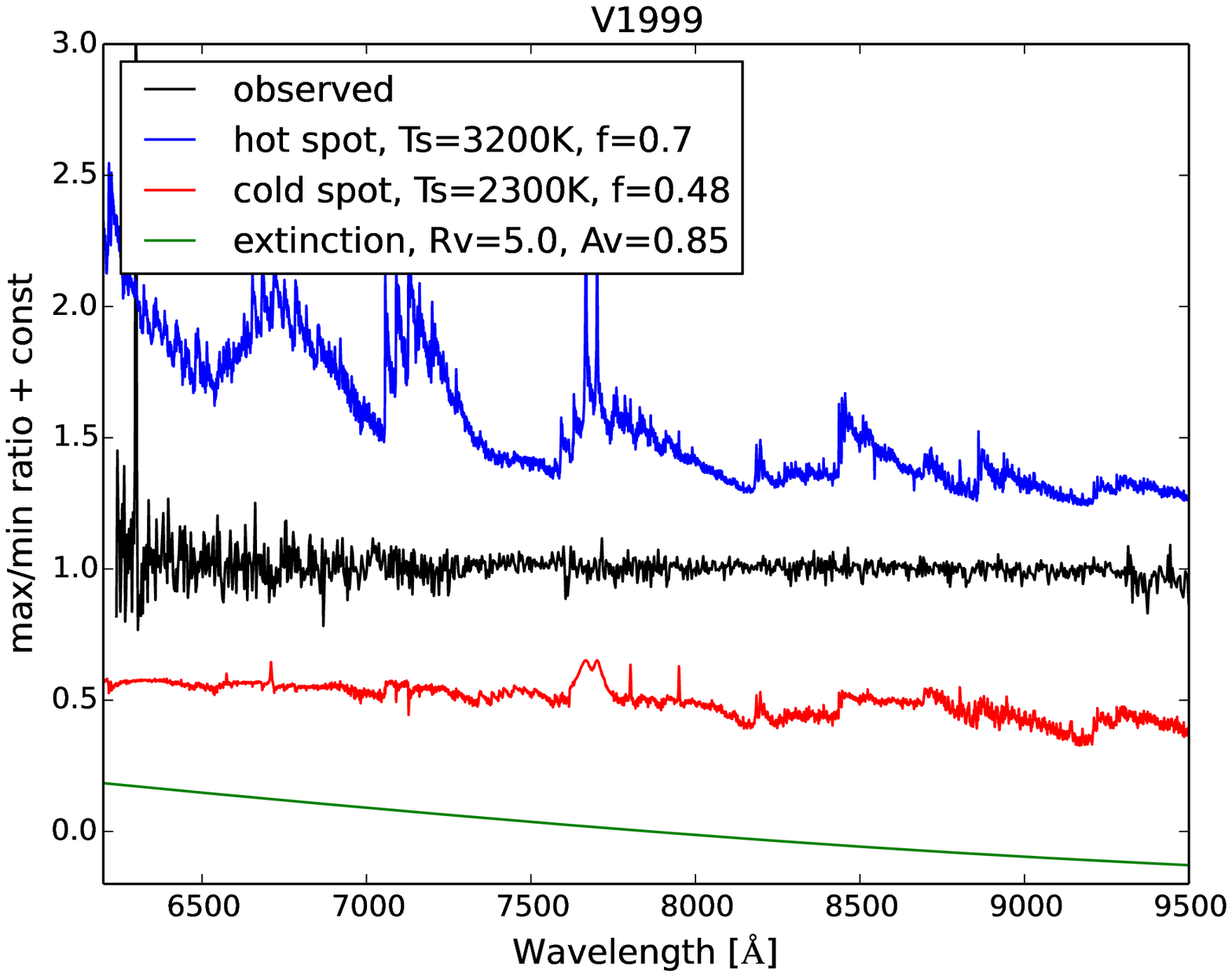}
	   \includegraphics[width=0.5\linewidth]{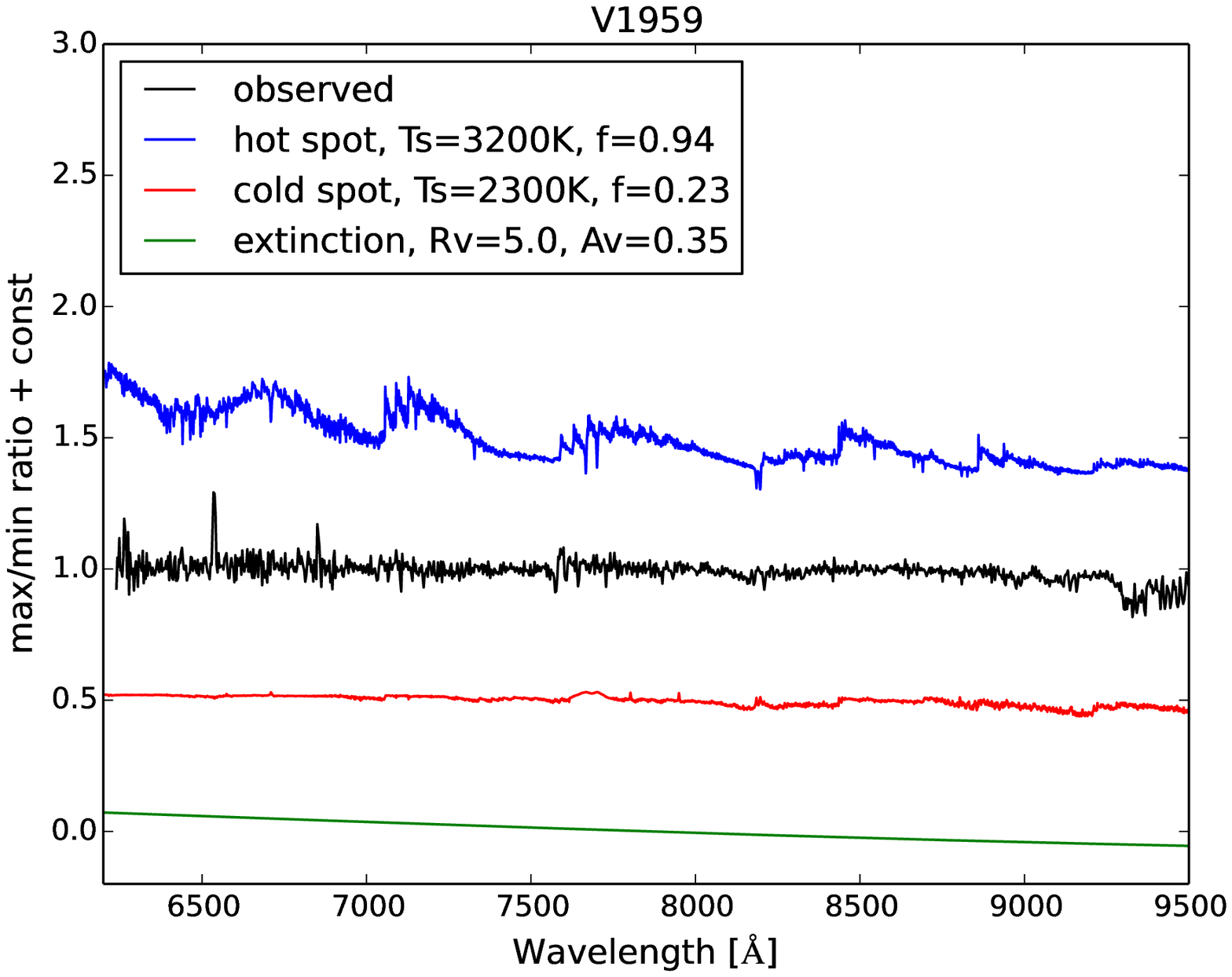} \\  
	  
	\end{tabular}
	
	\caption{Comparison between observed max/min spectrum ratio and best fit hot/cold spot and variable extinction model 
			for each object (see Sect \ref{var_origins}). Note that best fit model does not always imply a good fit was 				obtained. Legend contains best fit parameter values. See text for further details.}
	\label{obs_vs_models}	
	
\end{figure*}

We compare the spectra corresponding to maximum and minimum in the lightcurve for each star.  
The max/min ratio generally varies smoothly with wavelength, with only subtle spectral features. This is again confirming that the spectral shape does not change significantly over the time series (see Sect. \ref{spectraltypes}). In almost all cases we find a slope in the ratio suggesting objects become bluer when brighter and conversely redder when dimmer. Two exceptions are V1999 and V1959 where the ratio is near flat. The observed max/min ratio for each star can be found in Figure \ref{obs_vs_models} (models will be discussed later). 

To further explore changes in the spectra over time, we construct a trailed spectrogram for each object. We calculate an average spectrum for each star, then subtract the average from each individual spectrum and plot the residuals in Figure \ref{specres}. 
In these spectrograms, the dominant source of variability in the $\sigma$\,Ori target spectra (V2737, V2721, V2739) is in the H$\alpha$ emission, which has already been discussed in Sect. \ref{halpha}. A weaker variation can be seen around the head of the TiO band near 7600$\AA$, in all 7 targets, but particularly in V2739.  TiO bands are sensitive to the photospheric temperature of M dwarfs \citep{2004AJ....128.1802O,2015csss...18..665S}; variations can possibly indicate changes in the magnetic spot coverage. However, as the TiO band is affected by a telluric absorption feature, we will not consider it as solid evidence for spectral variability. In addition to the line variations, the spectrograms show slight gradients in the residuals. These trends will be further explored in Sect. \ref{var_origins}.

\renewcommand{\thefigure}{\arabic{figure} (continued)}
\addtocounter{figure}{-1}

\begin{figure*}
   
	\begin{tabular}{c}
	
	    \includegraphics[width=0.5\linewidth]{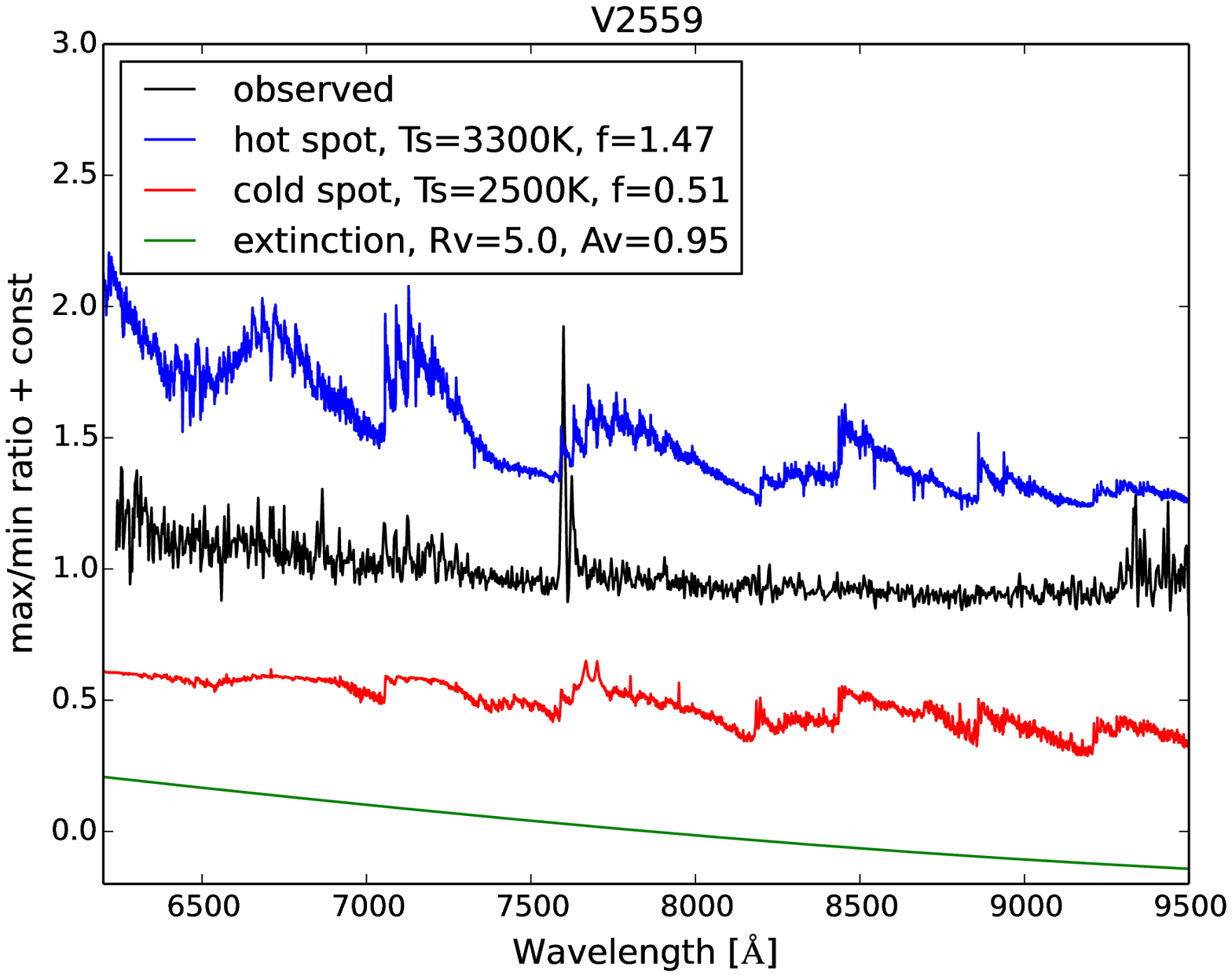}
	   
	\end{tabular}
	\caption{}
\end{figure*}

\renewcommand{\thefigure}{\arabic{figure}}

\begin{figure*}
   
	\begin{tabular}{cc}
	    
	   \includegraphics[width=0.5\linewidth]{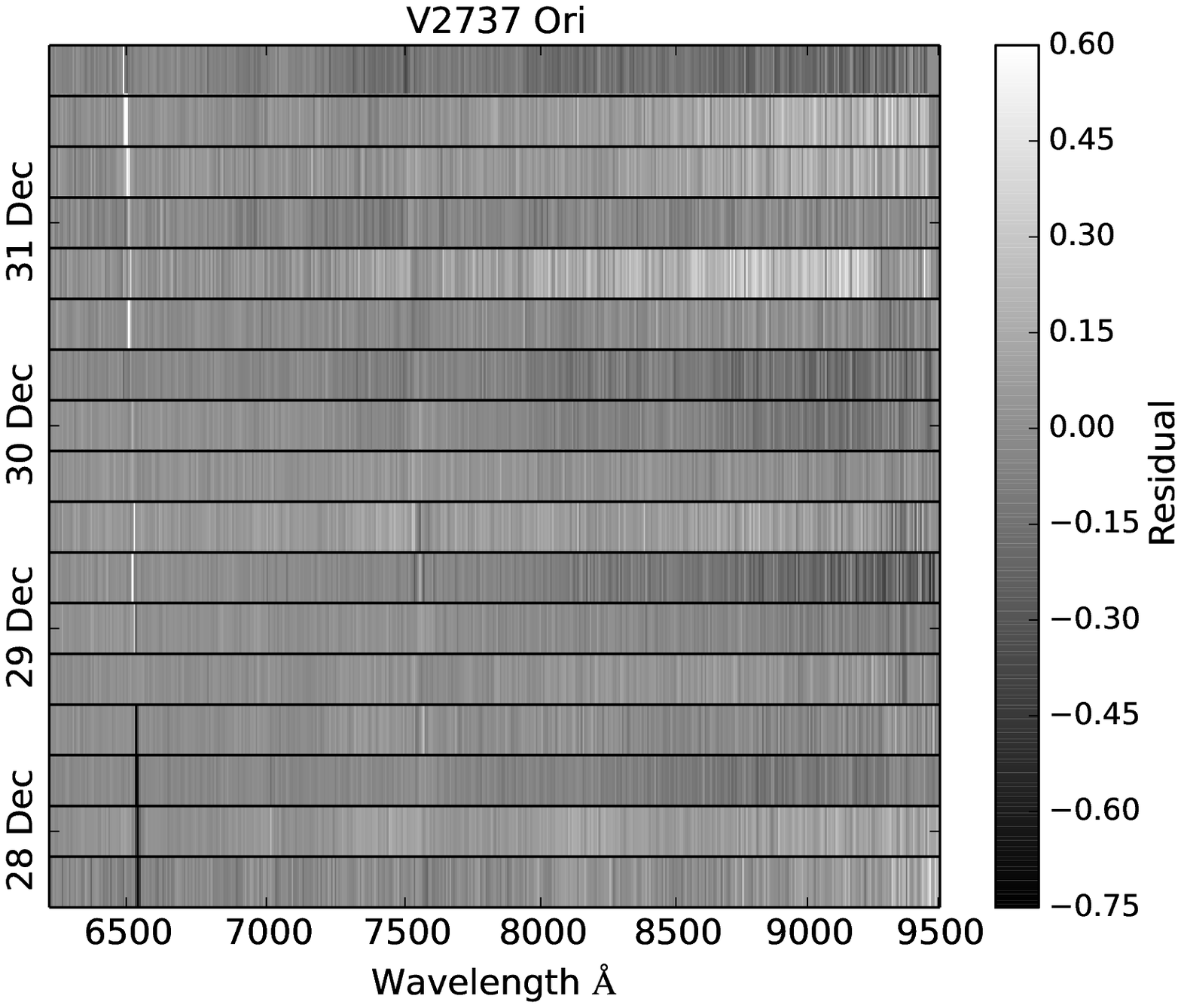}
	   \includegraphics[width=0.5\linewidth]{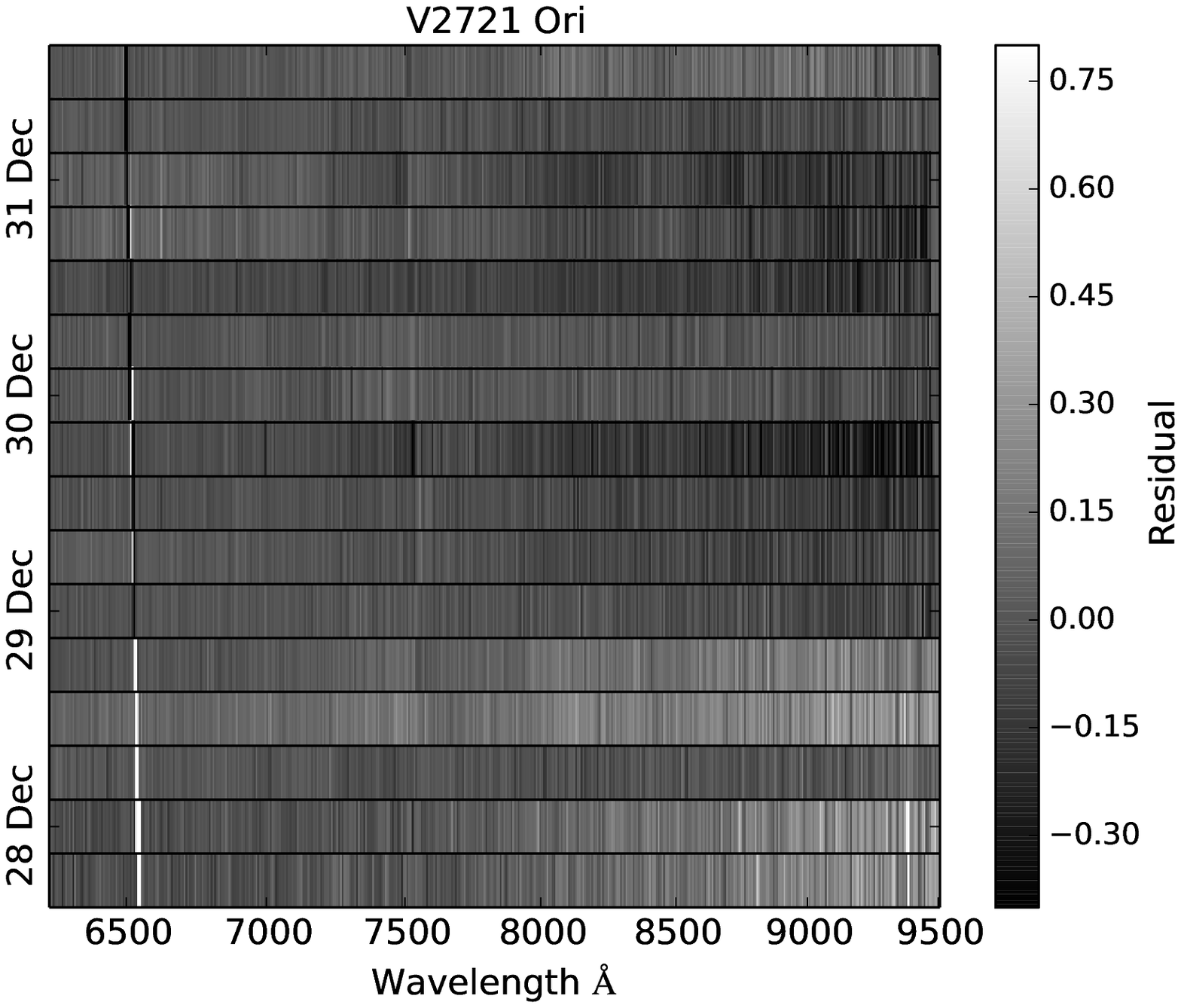} \\
	   
	    \includegraphics[width=0.5\linewidth]{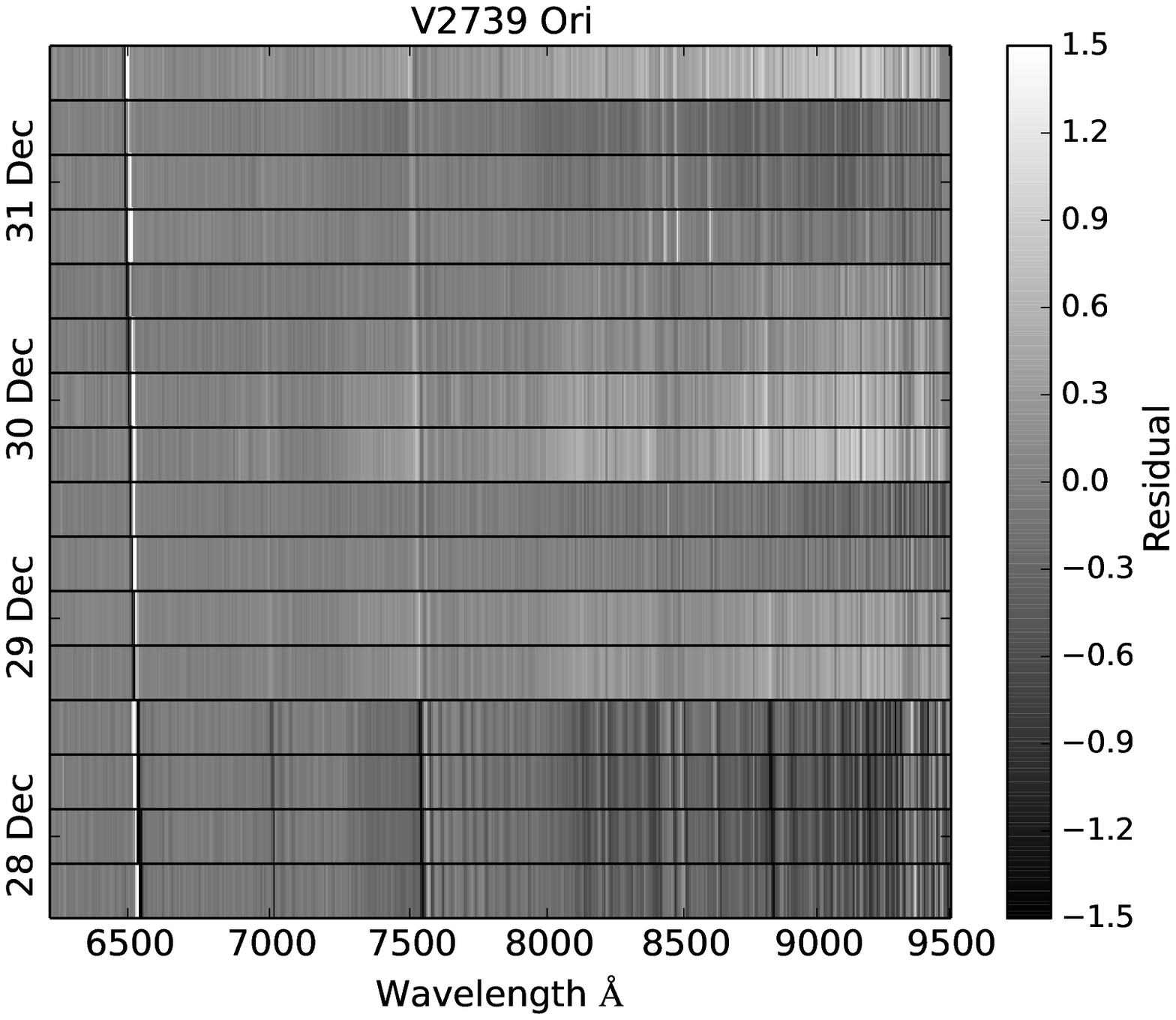}
	   \includegraphics[width=0.5\linewidth]{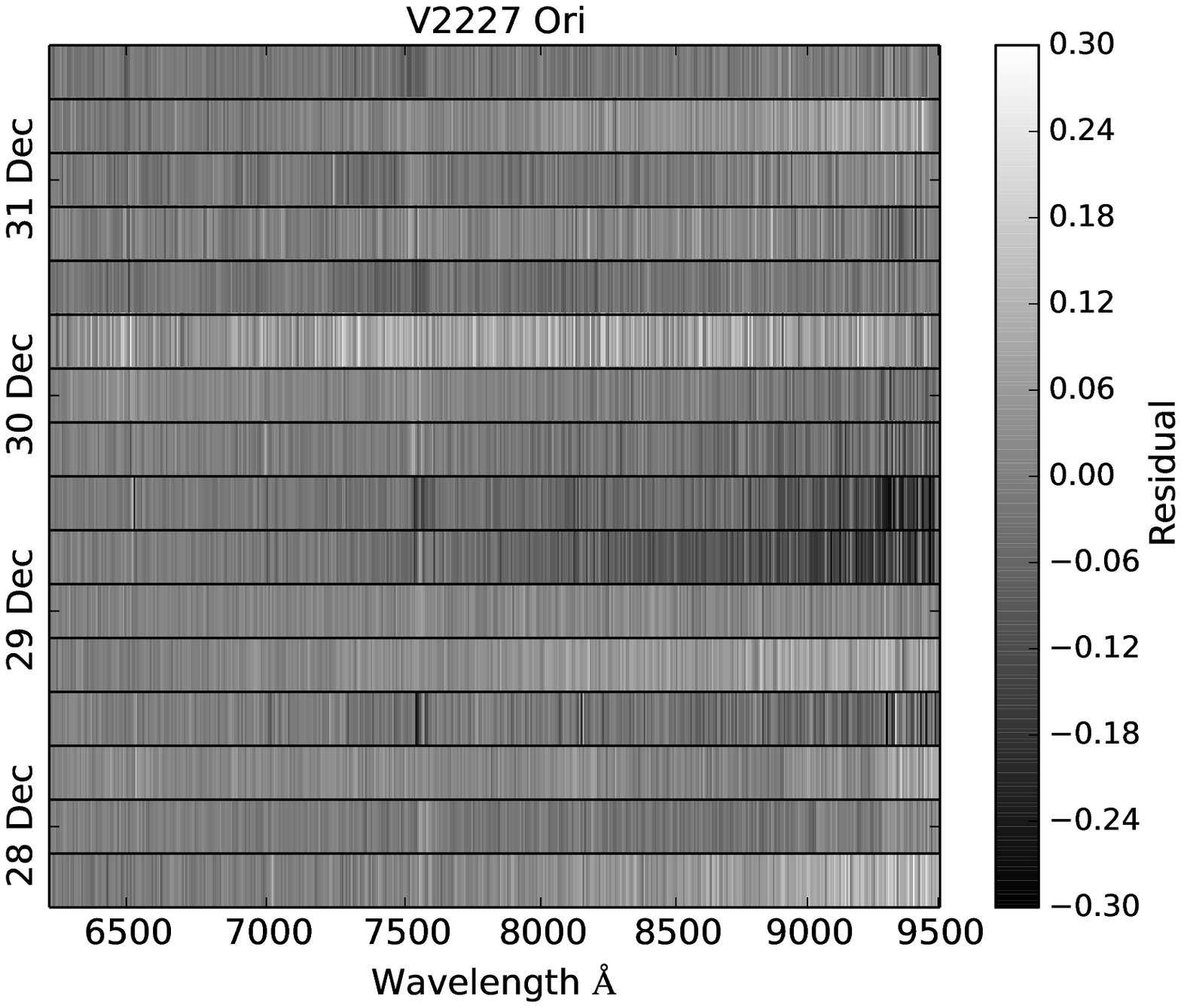} \\	 
	   
	    \includegraphics[width=0.5\linewidth]{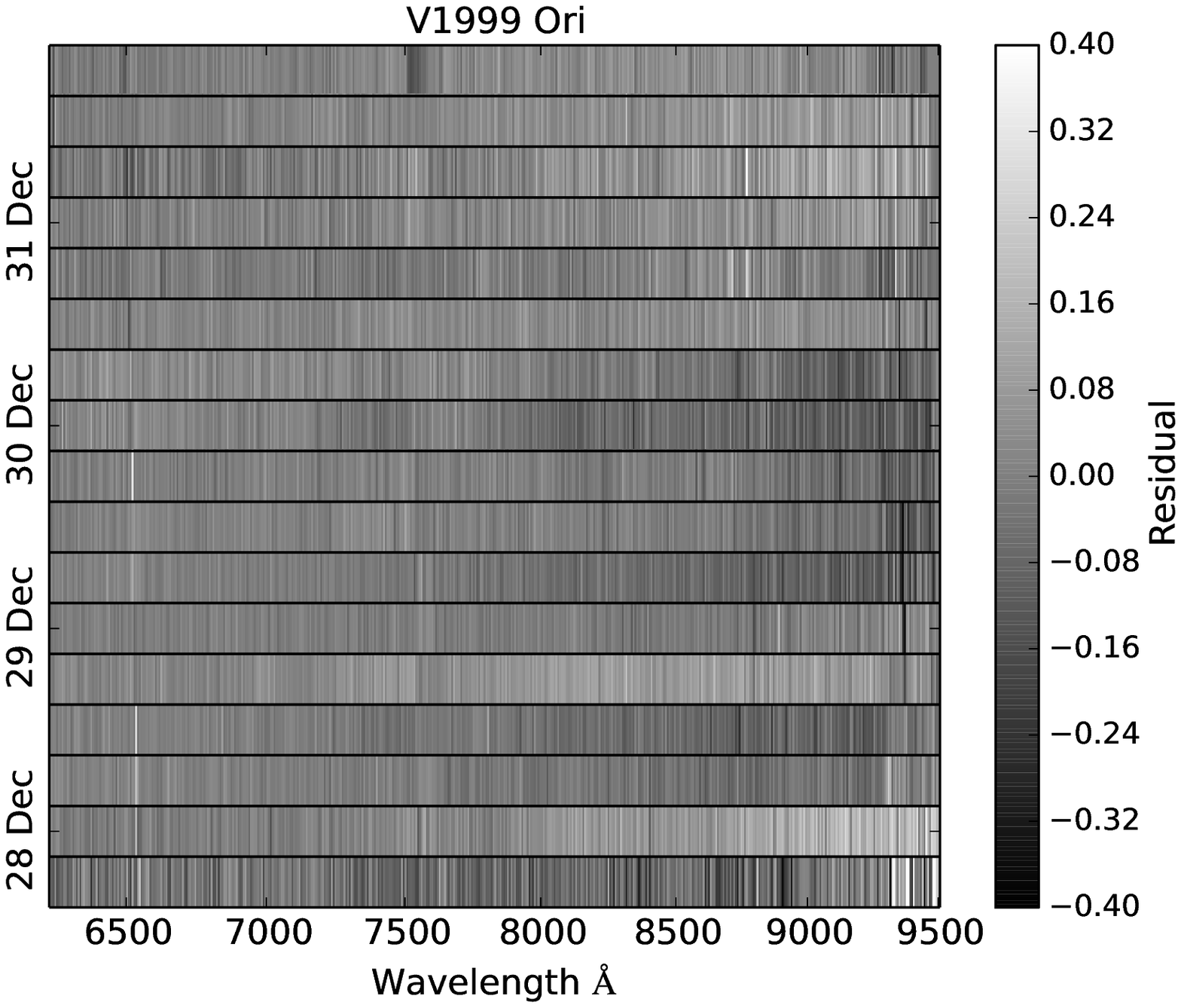}
	   \includegraphics[width=0.5\linewidth]{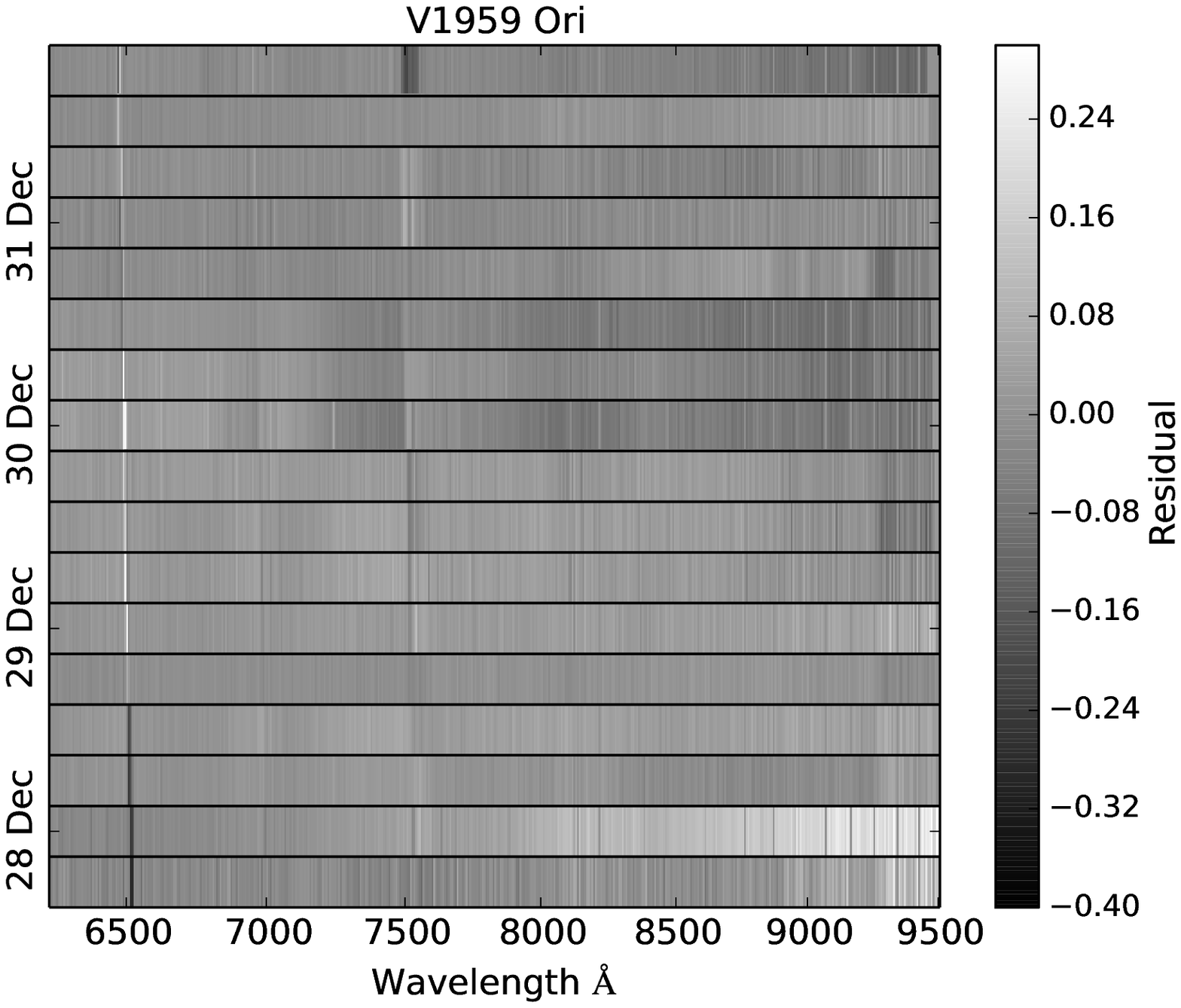} \\  
	  
	\end{tabular}
	
	\caption{Trailed spectrograms for each star. Major variations are seen in the H$\alpha$ line (6563$\AA$); less prominent changes are visible in the head of the TiO band near 7500$\AA$. V2737, V2721, V2739 and V2559 show a colour gradient in the residuals near lightcurve minimum, indicating reddening as targets become fainter.}
	\label{specres}	
	
\end{figure*}

\renewcommand{\thefigure}{\arabic{figure} (continued)}
\addtocounter{figure}{-1}

\begin{figure*}
   
	\begin{tabular}{c}
	
	    \includegraphics[width=0.5\linewidth]{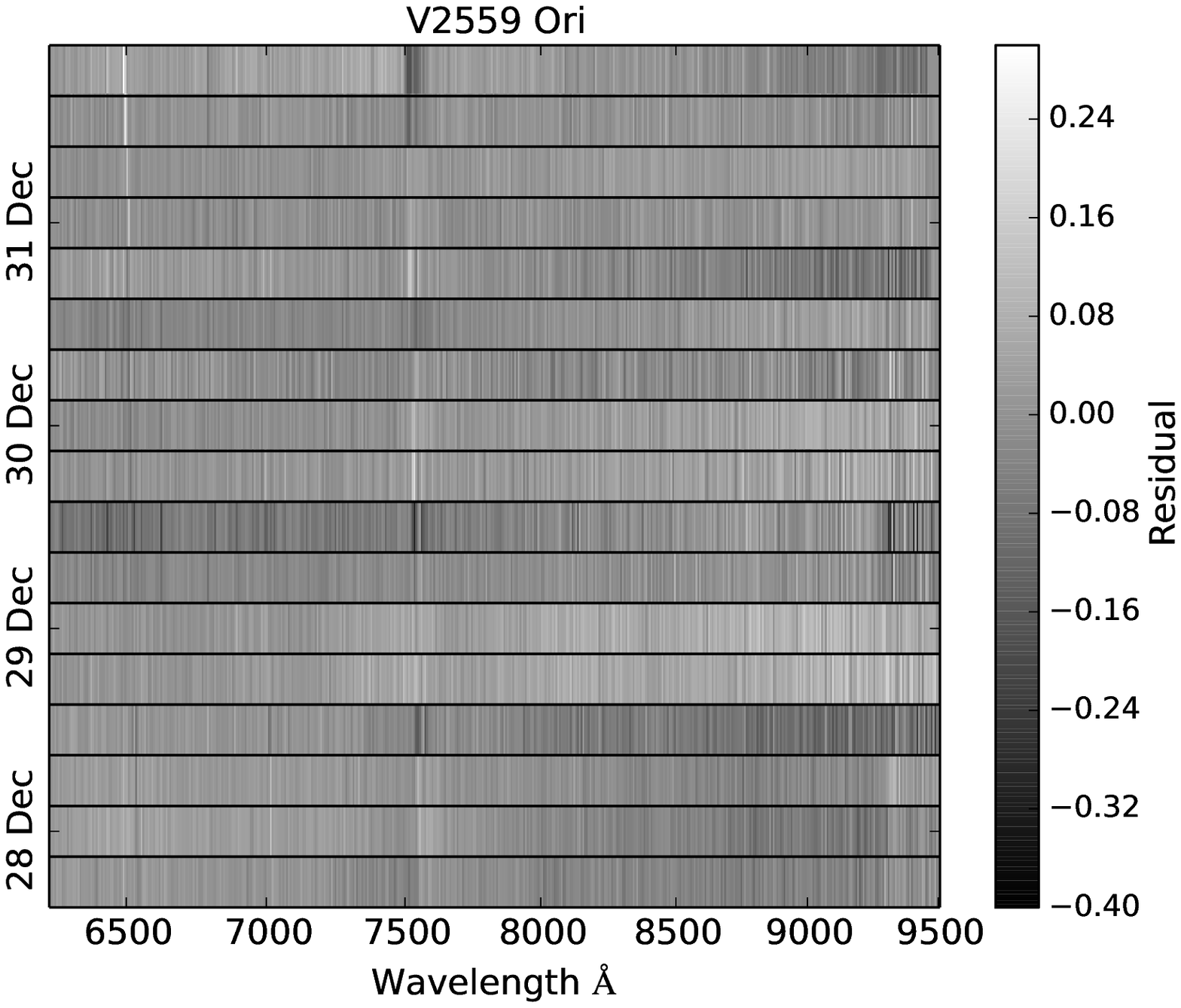}
	   
	\end{tabular}
	\caption{}
\end{figure*}	

\renewcommand{\thefigure}{\arabic{figure}}

\section{Discussion: Origin of the variability}
\label{var_origins}
\subsection{Summary of observational evidence}

The $\sigma$\,Ori stars were originally observed in 2001 (SE04). The I-band lightcurves revealed amplitudes ranging from 0.1 to 1.1 magnitudes. Periodic variations were found on timescales of 0.5-8 days for all but one object, with superimposed irregular components on the scales of hours. V2737 and V2721 were re-visited in a near-infrared campaign in 2006 \citep{2009MNRAS.398..873S}. The stars showed amplitudes of up to 0.9 in the J-band and 0.7 in the K-band. Variability in V2737 was said to be best explained by a hot spot of $T_{eff}$ = 6000K and filling factor of $>=$20$\%$. Changes in V2721 were attributed to inner disk inhomogeneity, estimated to reside between 0.01 and 0.04 AU away from the star. The objects near $\epsilon$\,Ori were first observed in the I-band in 2001. Our targets were identified as highly variable, with quasi-periodic (V1959, V1999 and V2559) or completely irregular signatures (V2227). 

In this paper we present for the first time spectroscopic time series for these sources (in the case of the $\epsilon$\,Ori objects the first spectra altogether). Visual inspection of the lightcurves confirms periodic and aperiodic variations on the order of several days and additional changes on time scales of hours. Thus, the lightcurves in the current study show similar characteristics to the literature datasets. We want to highlight that the shape of the lightcurve of V2559 is remarkably similar to that from the 2001 observations. In contrast to the photometry, we find very little evidence for spectroscopic variability, except for H$\alpha$ (see Sect. \ref{spectralanalysis}).

Table \ref{amplitudes} summarises the derived amplitudes and compares them to previously published peak-to-peak values in the phased lightcurves (SE04, SE05). Amplitudes in this run are comparable to the ones measured before, confirming that the variability is persistent over timescales of at least 5 years (J-band amplitudes of 0.91 and 0.50 mag for V2737 and V2721 respectively were measured in \cite{2009MNRAS.398..873S}). 

The major physical mechanisms discussed in the literature as causes for high level variability of young stars in the optical are hot/cool spots and variable extinction \citep{1993A&A...272..176B,1994AJ....108.1906H}. Both processes affect the stars by making them appear bluer when brighter and conversely redder when dimmer. This trend coincides with the changes in our observed spectra (see Sect. \ref{spvariations}) and we therefore explore these possibilities in more detail below. In Fig. \ref{obs_vs_models} we show the max-min slope for the best fitting models in comparison with observations. 

\subsection{Scenario 1: hot spots}
\label{hotspots}

In this scenario, material from the disk is thought to be channelled onto the star via stellar magnetic field lines. Kinetic energy is converted into heat causing a hot spot to appear at the base of the accretion column on the stellar surface. The spot corotates with the star which should produce periodic brightening. Variations in the accretion rates or flow geometry can then induce irregularities in the observed flux on a variety of time scales, from hours to years.  Photometric studies in multiple bands found that variability in accreting young stars can commonly be explained by hot spots with temperatures a few thousand degrees above photosphere and a filling factor of a few percent \citep{1995A&A...299...89B}. More recently, comparisons between spectra of T Tauri stars and accretion shock models indicate typical filling factors of 1-10\%, with some exceptions at 20-40\% \citep{2013ApJ...767..112I}.

The lack of observed spectral variability poses a general problem for the hot spot scenario. If we choose a smooth spectrum for the spots (e.g., a blackbody), this naturally introduces significant spectral changes. Therefore we choose a spectrum from model atmosphere for the hot spot. Even then, if we assume temperatures significantly above the stellar T$_{\rm{eff}}$, adding a hot spot introduces strong spectral changes again. On the other hand, keeping the temperature close to T$_{\rm{eff}}$ yields large filling factors.

To explore a hot spot scenario, we combine photospheric spectra from model atmospheres ({\sc PHOENIX-ACES-AGSS-COND-2011}\footnote{http://phoenix.astro.physik.uni-goettingen.de}, version 16.01.00B, \cite{2013A&A...553A...6H}) of different effective temperatures for the star and the spot. 
Note that the model spectra do not include the H$\alpha$ emission. The free parameters are the effective temperature and filling factor for the spot. The amplitude produced by the spot is expressed by:

\begin{equation}
	A = -2.5\log{\bigg(\dfrac{F_{star}}{fF_{spot}+(1-f)F_{star}}\bigg)},
\end{equation}

where $F_{star}$ and $F_{spot}$ are the stellar photosphere and hot spot fluxes respectively, and the filling factor $f$ is the fractional area of the stellar surface covered by spots. We calculate amplitudes for a variety of parameter values: 3000K $<= T_{spot} <= $12000K and 0.01$ <=f<= $1.0.

For the comparison with the observations, we assume that the lightcurve minimum represents the unspotted photosphere, whereas at maximum we observe the flux from photosphere plus hot spot. In a first step, we select the models with ($T_{spot}, f$) pairs that reproduce the observed photometric amplitudes for every target (Table \ref{amplitudes}). In a second step we compare the ratio of the models at minimum and maximum with the observed values (see Sect. \ref{spvariations}). A first important finding is that spot temperatures exceeding the photospheric temperature by more than $\sim 500$\,K produce strong spectral features in the max/min ratio. This is inconsistent with the observed max/min ratios, as discussed in Sect. \ref{spvariations}. Thus, to explain the photometric amplitudes {\it and} the lack of spectral changes with this model, spot temperatures have to be relatively close to the effective temperature of the stars. 

We find a plausible solution only for two objects in $\sigma$\,Ori -- V2721 and V2739. Incidentally, these are also the two objects with the highest EW in H$\alpha$, i.e. presumably the highest accretion rate, and their lightcurves appear partly periodic, which fits in this scenario. 

For these two objects, our solutions indicate spot temperatures close to the photosphere (300\,K above effective temperature), and large filling factors of 0.2-0.4. This is in contrast to measurements done for more massive accreting young stars, which in most cases find spot temperatures of over 1000\,K above the photosphere and filling factors of only a few percent \citep{1993A&A...272..176B,1995A&A...299...89B,2013ApJ...767..112I}. It would imply that accretion in very low mass stars is at least on some cases not strongly funneled by magnetic fields.
Even for these two sources, hot spots cannot be the only source of the variations. In the red part of the spectrum ($>7500\AA$) our simple model is inadequate and does not reproduce the smooth slope. 

We note that V2721 was previously examined with J- and K-band two filter lightcurves in  \cite{2009MNRAS.398..873S}. In their analysis, the color variability in the near-infrared was found to be consistent with variable extinction. In the current case, variable extinction underpredicts the observed slope change of the spectrum at minimum (see below). 

\subsection{Scenario 2: cool spots}

Cool spots result from the interaction between the stellar magnetic fields and gas in the atmosphere. M dwarfs like our targets are known to be magnetically active and therefore exhibit cool spots. As in the hot spot case, the cool spots corotate with the star but unlike the previous scenario the variations are expected to be strictly periodic as  cool spots are stable on the time-scale of our observations. 

The modelling of the cool spots is identical to the hot spots, except that we now assume that the maximum in the lightcurve corresponds to the unspotted photosphere. For the spot effective temperature we adopt a range from 2300 to 3000\,K, in line with typical spot temperature for M dwarfs inferred from Doppler Imaging \citep{2009IAUS..259..363S}.
Again, the observed photometric amplitude for every star can be reproduced by a number of model parameter combinations. As in Sect. \ref{hotspots}, we break the degeneracy by comparing with the observed max/min ratio of the spectral slope. 

In order to explain the high photometric amplitudes observed for most of our objects, very high filling factors $>0.5$ have to be adopted, which is not physical. For two objects (V2227 and V1959) we find a solution with a plausible filling factor of 0.1-0.2, but these have eclipse-shaped or partly irregular lightcurves, which cannot be explained by cool spots alone. Thus, while cool spots may contribute to the observed variations in some objects, they cannot be the dominant cause.

\subsection{Scenario 3: variable extinction}

When stars are surrounded by circumstellar material, there might be variable exctinction along the line of sight, causing optical and infrared variability \citep{2003ApJ...594L..47D,2009AstL...35..114G,2015AJ....149..130S,2015A&A...577A..11M}. If the material along the line of sight is optically thick, this could also lead to an eclipse of the central object. Variable extinction and eclipses would be associated with inhomogenities in the circumstellar material, which could be caused by a number of physical processes. Depending on the temporal stability of these features, the variability may appear periodic or irregular. Variable extinction causes brightness changes that are smoothly varying with wavelength, in contrast to the spot models explored above and similar to what is seen in our observations.

To investigate the impact of variable extinction on our targets, we use the average $R_V$-dependent extinction law defined by \cite{1989ApJ...345..245C}:

\begin{equation}
	< A(\lambda) / A(V) > = a(x) + b(x)/R_V  ,
\end{equation}

where $a$ and $b$ are functions of wavelength (see \cite{1989ApJ...345..245C}, eq. 2 and 3). $A(V)$ and $R_V$ are free parameters. We vary $A(V)$ from 0.0 to 1.5 and $R_V$ from 3.1 to 5.0. We assume here that the lightcurve maximum corresponds to the object seen at zero extinction. Hence, we use a model atmosphere to represent the star at maximum brightness and apply extinction to match the minimum brightness. We then compare with the max/min slope to pick the best fitting combination of $A(V)$ and $R_V$. This approach assumes ISM-like extinction caused by very small grains ($<1\,\mu m$). Larger grain sizes, as they are expected to be present in disks, would produce smaller colour variations.

We are able to find parameter values that can reproduce the photometric amplitudes. However, these solutions cause too much spectral variation compared with the observations, i.e. the slope in max/min ratio is too large. The ratio in the observed spectra deviates by $\pm 0-10$\%
from 1.0, whereas the simple model requires $\pm 10-20$\% to match the amplitudes. The only object where we find a plausible solution is V2559, with $R_V = 5.0$ and $A(V) = 0.95$. For the other 6 objects, we can exclude variable extinction by ISM-type grains as dominant cause of variability. 

Variability extinction by large dust grains, however, remains an attractive scenario, particularly for the objects in $\epsilon$\,Ori without accretion and with eclipse-shaped lightcurves (V2227, V1999, V1959). For these three plus V2559 obscuration by circumstellar material might be the only viable explanation, the only scenario that can explain both the shape of the lightcurve and the lack of spectral variability. The obvious problem is that two of these objects do not show mid-infrared excess emission and thus there is no evidence for the presence of a disk. One plausible scenario may be collisions of massive planetesimals that rapidly generate a significant mass of small dust grains as suggested by \cite{2015ApJ...805...77M}. These four objects will be further explored in the future.

In that case, the timescales of the variations give constraints about the location of the material that is responsible for the additional exctinction. The timespan between minimum and maximum in our lightcurves is in the range of 1-4\,d, which corresponds to a distance from the star of 0.01-0.05\,AU, assuming Keplerian rotation. This is consistent with the expected location of the inner edge of the dust disk (if a disk is in fact present). 



\subsection{Notes on V2737}

None of our models produce a satisfactory explanation for this object. In \cite{2009MNRAS.398..873S} the star was found to have a slope which was too small to be consistent with variable extinction. The variability was attributed to hot spots of temperature 6000\,K and filling factor of over 20\%. Our current results show that variable extinction overpredicts the change in spectral slope, in agreement with the previous study. However, we find that hot spots with a temperature even a few hundred degrees higher than the photosphere produces strong spectral features, which are not seen in the observations. The cold spot model comes closest to reproducing the slope but it also generates features that are too strong and requires a very large filling factor. This object clearly requires more sophisticated modelling which is beyond the scope of this paper.

\subsection{Summary}

We have explored several scenarios in order to explain our observations. We find that the sources of variability are complex; no single source can fully account for the observed photometric and spectroscopic variability implying multiple processes contribute to various degrees. A summary of observed and model I-band amplitudes and max/min spectrum ratio slopes is provided in Table \ref{modpartable}. We find two results that are surprising and in contrast to the literature.

1) In the cases where "hot" spots are a plausible explanation (V2721, V2739) our models rule out the parameter space that is normally found for T Tauri stars (high temperatures, low filling factors). Instead, we find evidence for large filling factors and temperatures only slightly above photosphere. The 
variable continuum does not show a strong correlation with H$\alpha$. This indicates that the continuum and H$\alpha$ emission possibly originate from different spatial zones.

2) We find that even objects with very little or no measurable excess emission at mid-infrared
wavelength ($\epsilon$\,Ori) show evidence for being surrounded by clouds/clumps of interstellar material causing variable extinction.

\section{Conclusions}
	
We present a spectro-photometric variability study for seven very low mass stars in clusters in Orion, using four
nights of data from the FORS2 instrument at the ESO/VLT in December 2003. Our targets were previously known to be strongly variable.
In the following we summarise our main findings.	

\begin{itemize}
\item{Combining our data with literature lightcurves from 2001 and 2006, the photometric variability persist over 
at least 5 years with no major changes.}
\item{Three of our targets, located in the cluster $\sigma$\,Ori, show strong mid-infrared color excess, strong H$\alpha$
emission indicative of accretion, and large-amplitude partially periodic lightcurves, comparable to typical behaviour of 
classical T Tauri stars.}
\item{In contrast, the remaining targets, located near $\epsilon$\,Ori and presumably slightly older, have weak H$\alpha$
emission, weak or no mid-infrared excess, but still show significant and partially irregular variability.}
\item{While the I-band magnitudes vary by 0.1-0.8\,mag in our sample, the spectra remain remarkably constant in all objects.
The spectral types remain constant within 0.5 subtypes and the inferred effective temperature within 100\,K. 
The only significant spectral variations are found in H$\alpha$ (for the accreting objects).}
\item{Four out of 7 objects exhibit a slight gradient ($\pm 10\%$) in the ratio between spectra taken at maximum and minimum 
flux, in the sense that they become redder when fainter. For the remaining two this ratio is flat.}
\item{We calculate simplified models for three possible variability causes, hot spots, cold spots, and variable extinction.
For two accreting objects in $\sigma$\,Ori, the spectral changes are reasonably well explained by hot spots. This scenario
would also fit with the strong H$\alpha$ emission and the shape of the lightcurves. However, we have to assume temperature
close to the photosphere ($+300$\,K) and high filling factors (0.2-0.4), which are not the typical hot spot parameters found
for more massive stars and from magnetospheric accretion. It might indicate a change in the geometry towards more
homogenuous accretion in the very low mass regime.}
\item{The objects in $\epsilon$\,Ori are best explained by variable extinction caused by large grains. This is also in line
with the lack of accretion and with the eclipse-like shape of the lightcurves. The obvious problem with this explanation is
the lack of infrared excess in two of these sources out to 10$\,\mu m$.}
\end{itemize}
	
This study demonstrates that spectral information is essential for interpreting the variability in young stellar
objects. The combination of spectra and time-domain information is a useful tool to explore the immediate environment of
young stellar objects down to substellar masses.

\section{Acknowledgements}

We thank KM for kindly providing us with spectral templates of M dwarfs. 
The authors acknowledge support from the Science \& Technology Facilities Council
through grants no.ST/K502339/1 and ST/M001296/1. 
This work is based on observations collected at the European Organisation for Astronomical Research in the Southern Hemisphere under ESO programme 072.C$-$0197(A).
AS would like to thank the team at the VLT/FORS2 for their assistance during the observing run.

\bibliographystyle{mnras}
\bibliography{inb}

\clearpage 
\appendix
\section{Target Spectral Energy Distributions}
\label{appA}
SEDs based of 2MASS, Spitzer IRAC and MIPS (24 $\mu m$) and WISE data, where available. Overplotted are black body SEDs with effective temperature matching those chosen for the models fitting (Sect \ref{hotspots}). Some WISE datapoints represent upper limit estimates for the objects (see figure caption). 

\begin{figure*}
   
	\begin{tabular}{cc}
	    
	   \includegraphics[width=0.5\linewidth]{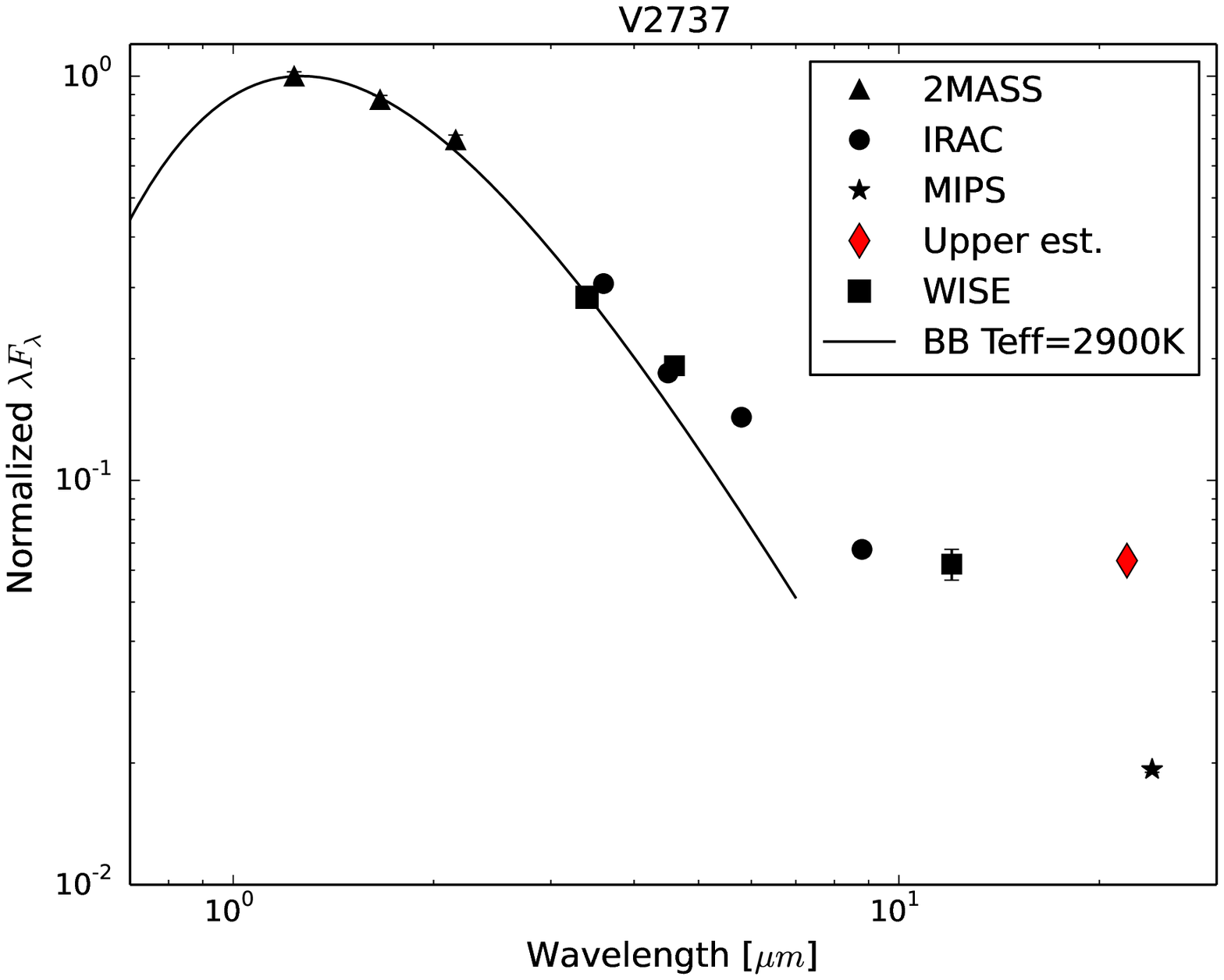}
	   \includegraphics[width=0.5\linewidth]{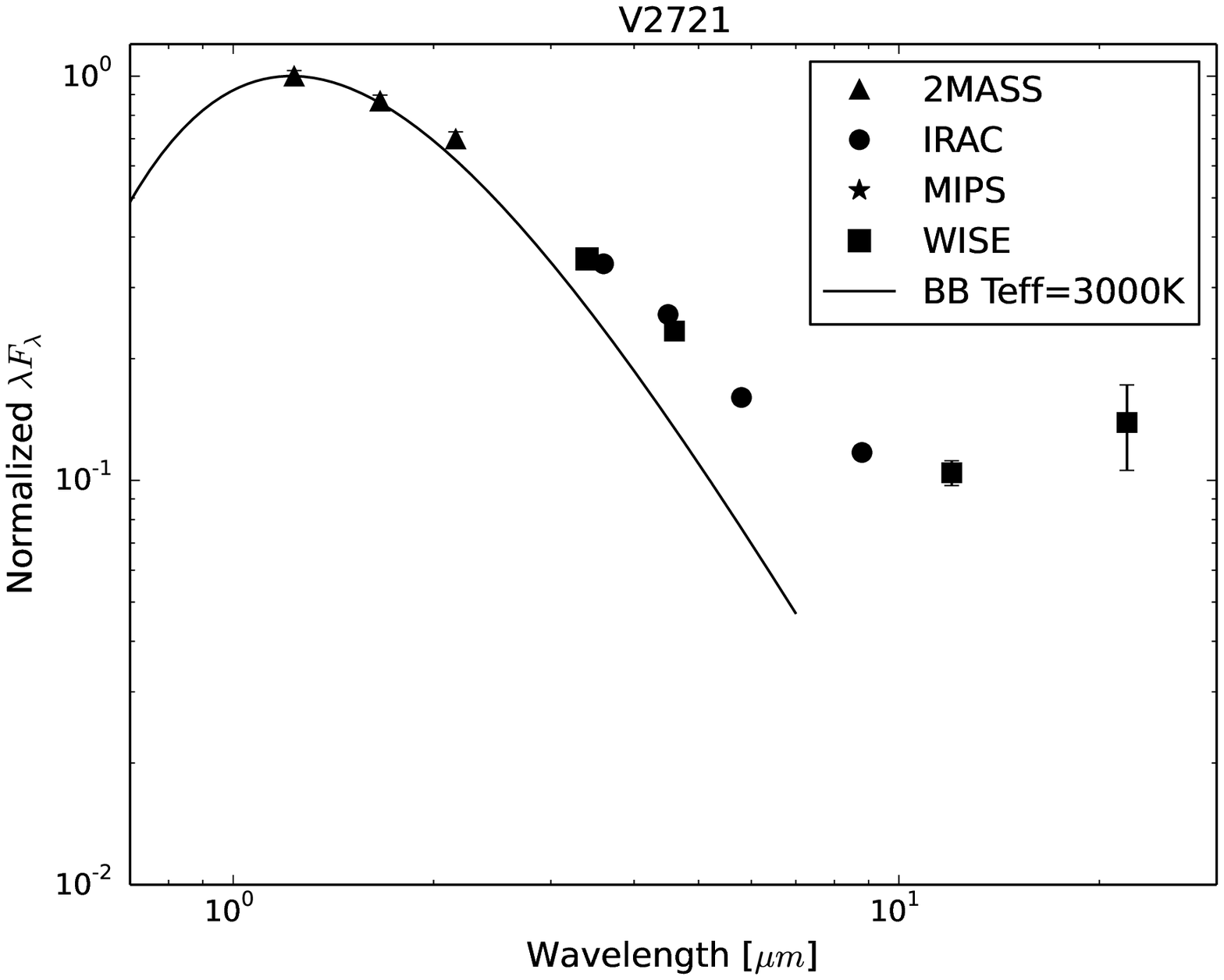} \\
	   
	    \includegraphics[width=0.5\linewidth]{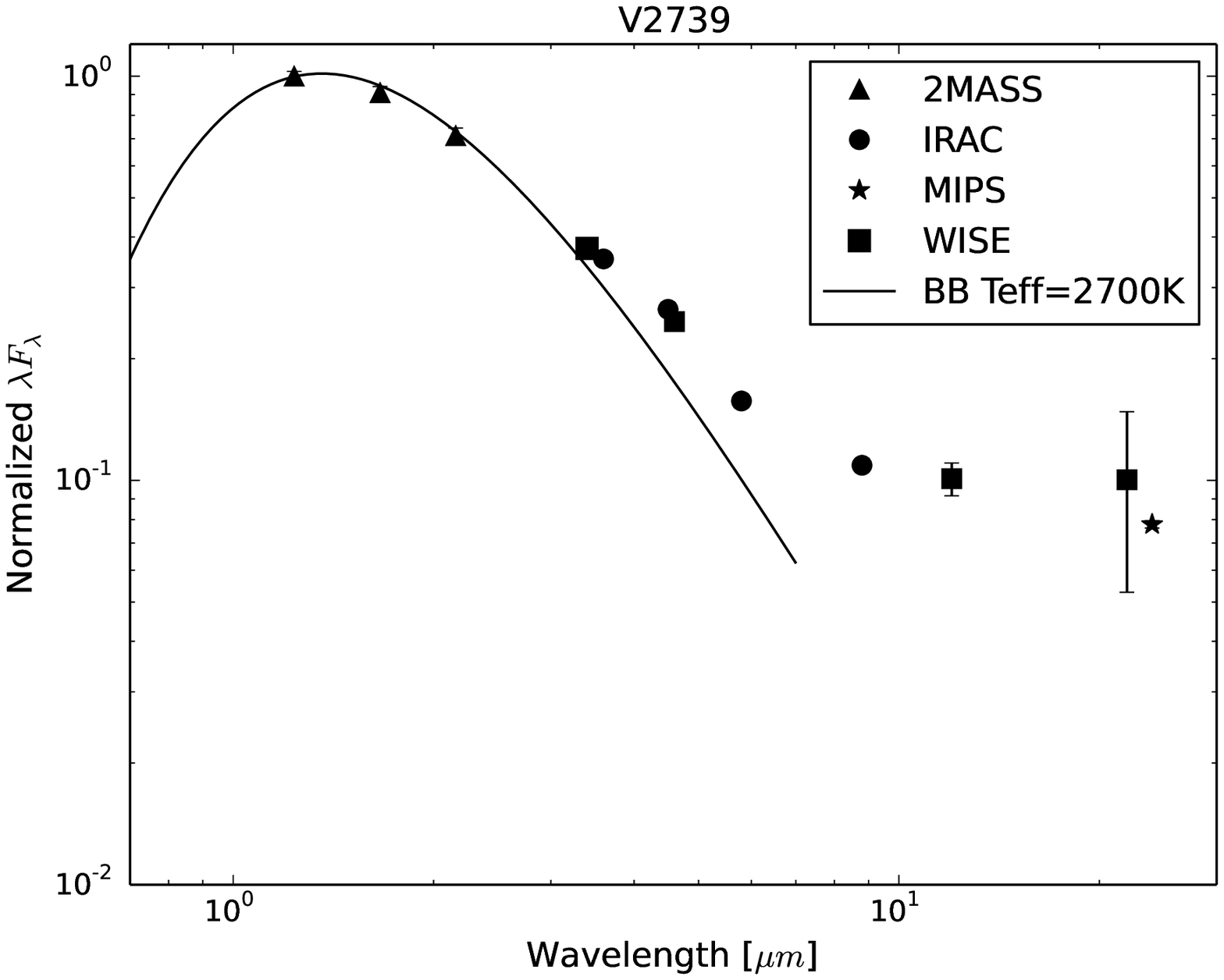}
	   \includegraphics[width=0.5\linewidth]{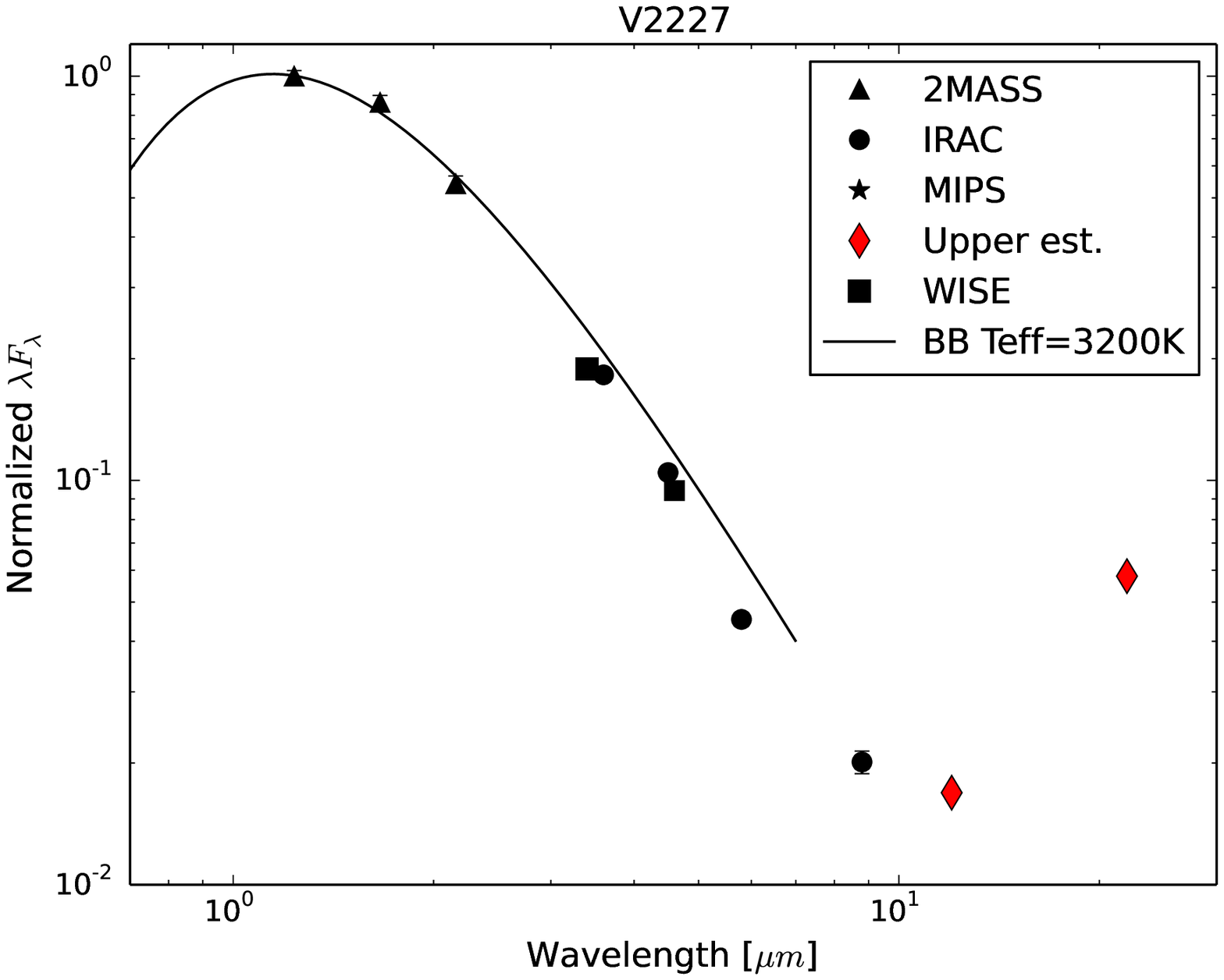} \\	 
	   
	    \includegraphics[width=0.5\linewidth]{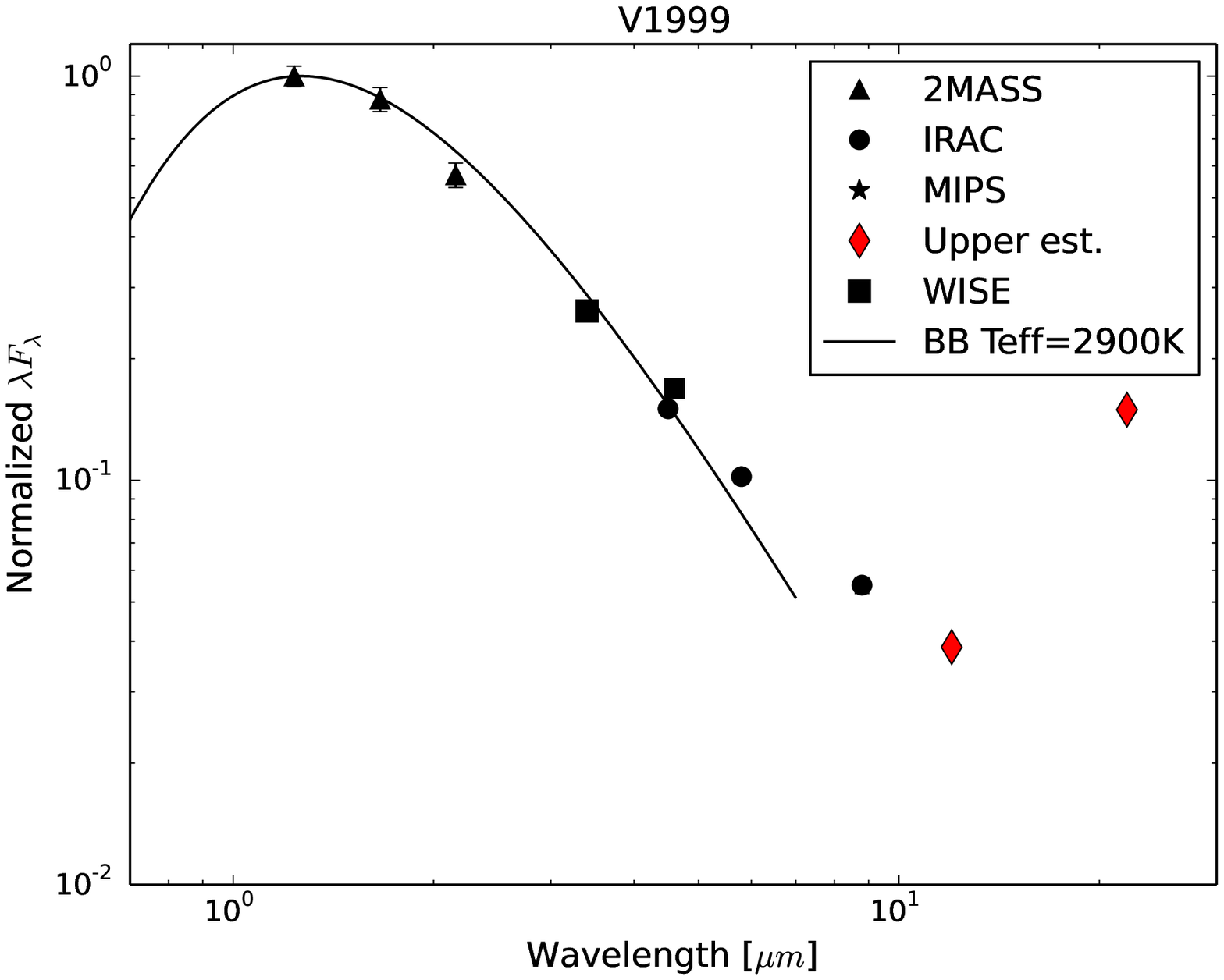}
	   \includegraphics[width=0.5\linewidth]{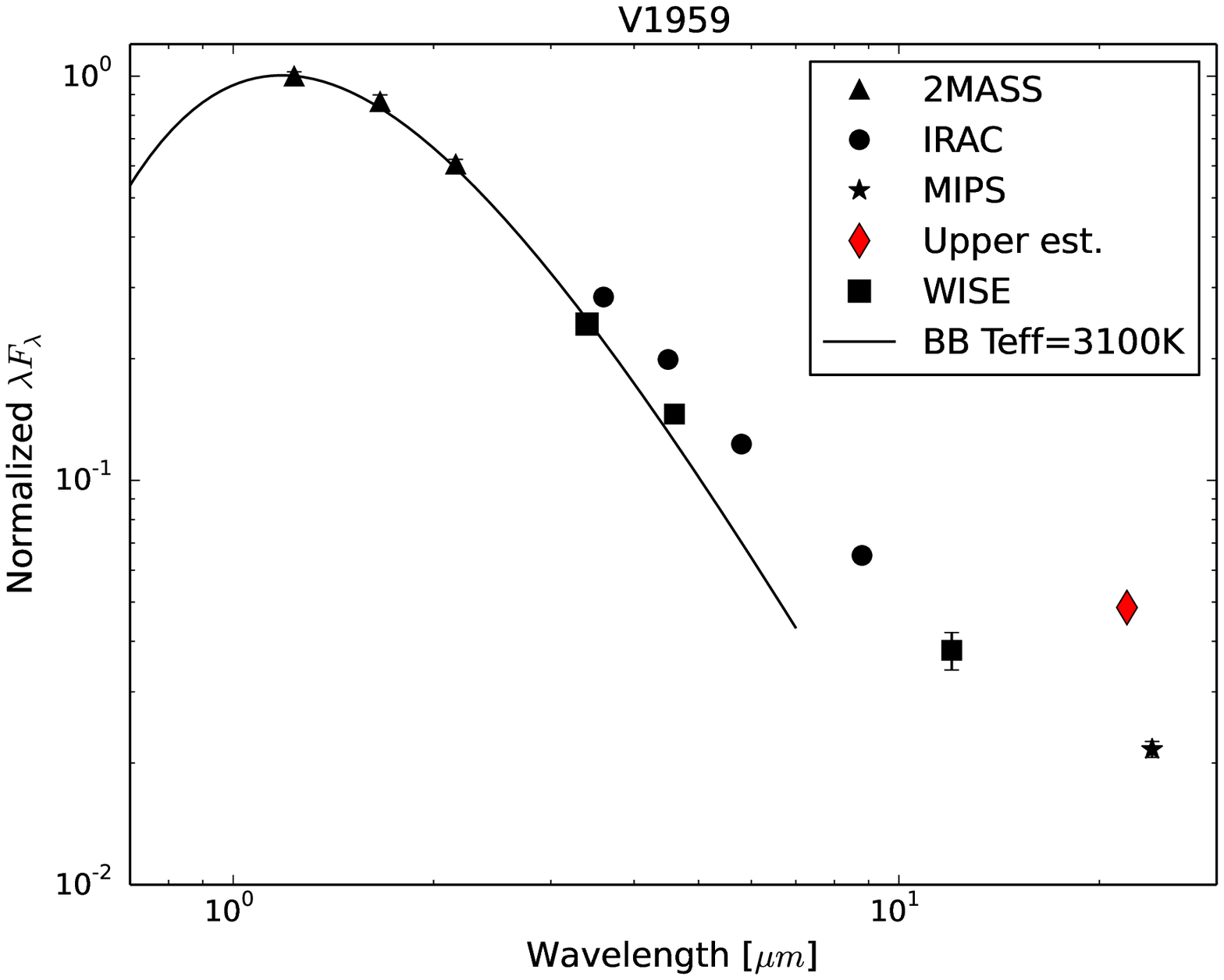} \\  
	  
	\end{tabular}
	
	\caption{Spectral energy distributions for the stars in our sample. The smooth curves represent a black body with effective temperature equal to our best estimates for the objects. The $\sigma$\,Ori (V2737, V2721 and V2739) show a clear colour excess indicative of the presence of a disk. V1999 and V1959 have a smaller albeit clear excess, while V2227 and V2559 and more consistent with a photosphere. Red data points represent upper limit estimates.}
	\label{seds}
	
\end{figure*}

\renewcommand{\thefigure}{\arabic{figure} (continued)}
\addtocounter{figure}{-1}

\begin{figure*}
   
	\begin{tabular}{c}
	
	    \includegraphics[width=0.5\linewidth]{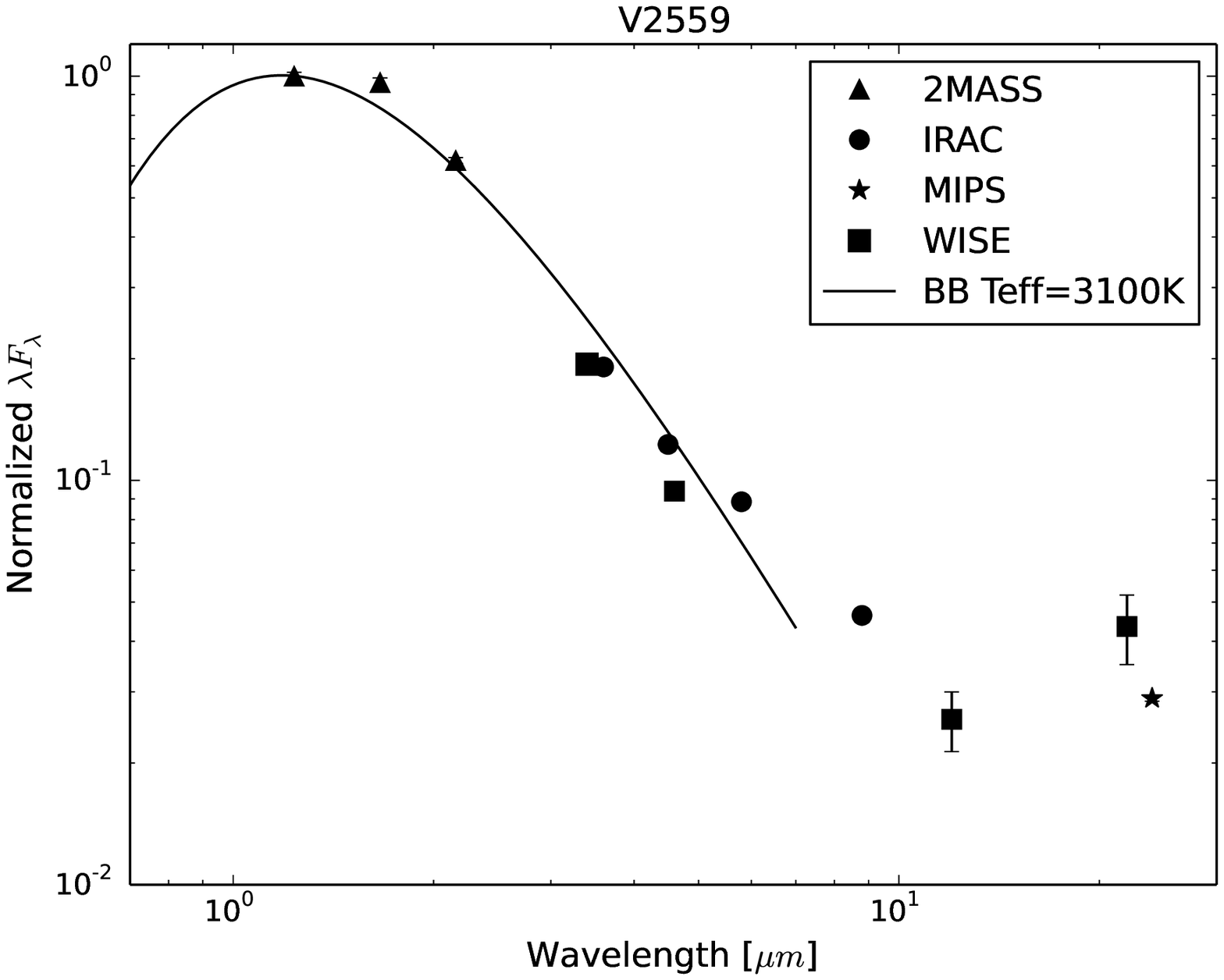}
	   
	\end{tabular}

\caption{}
\end{figure*}
\renewcommand{\thefigure}{\arabic{figure}}

\section{H${\alpha}$ plots and best fit model slopes and I-band amplitudes}
\label{appB}
H$\alpha$ equivalent width (EW) (see Sect. \ref{halpha}) as a function of time and lightcurve magnitude. No trend in correlation with I-band lightcurves are found.

   
	    
	   
	   
	  
	
	

   
	
	   
	


\begin{figure*}
   
	\begin{tabular}{cc}
	    
	   \includegraphics[width=0.5\linewidth]{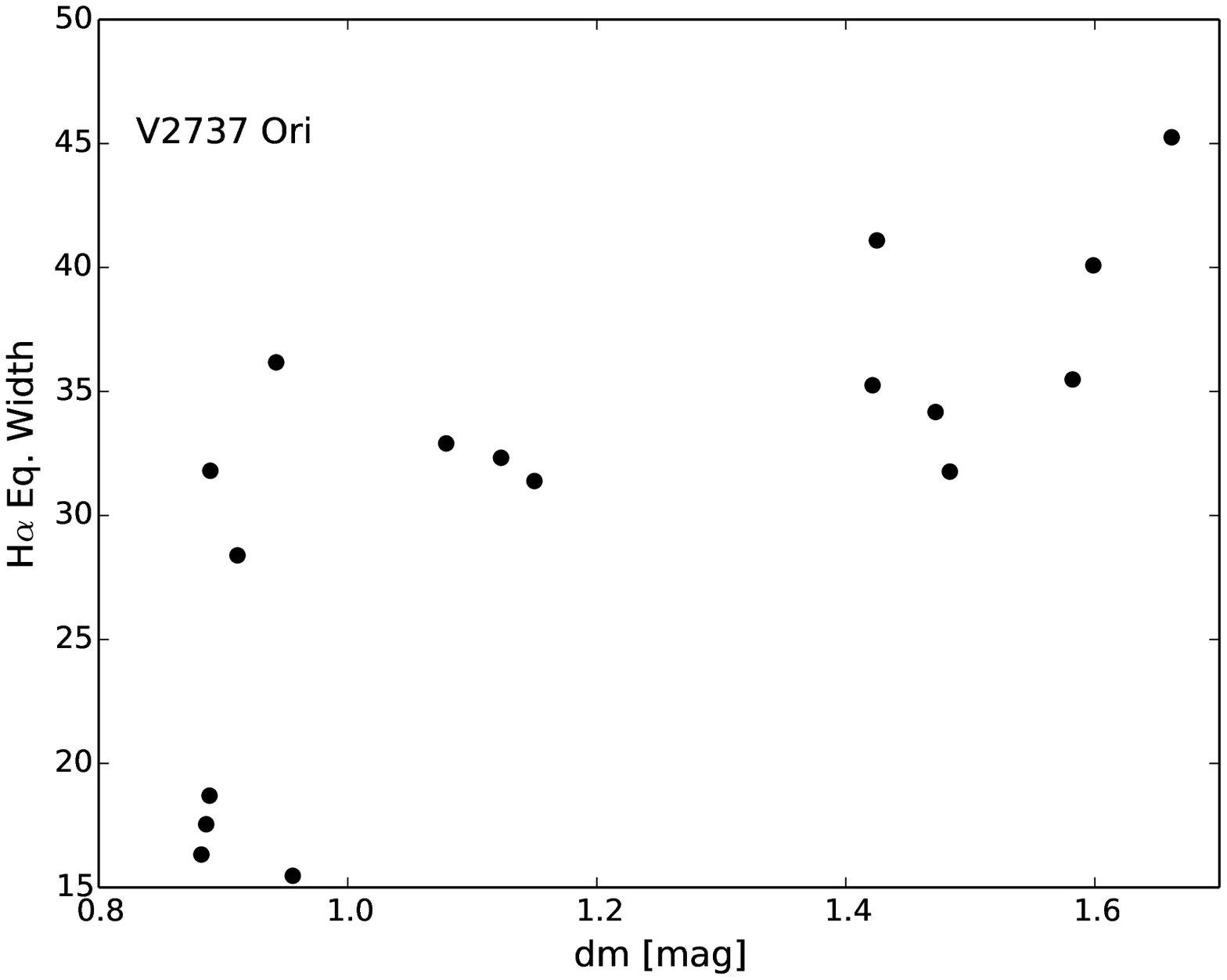}
	   \includegraphics[width=0.5\linewidth]{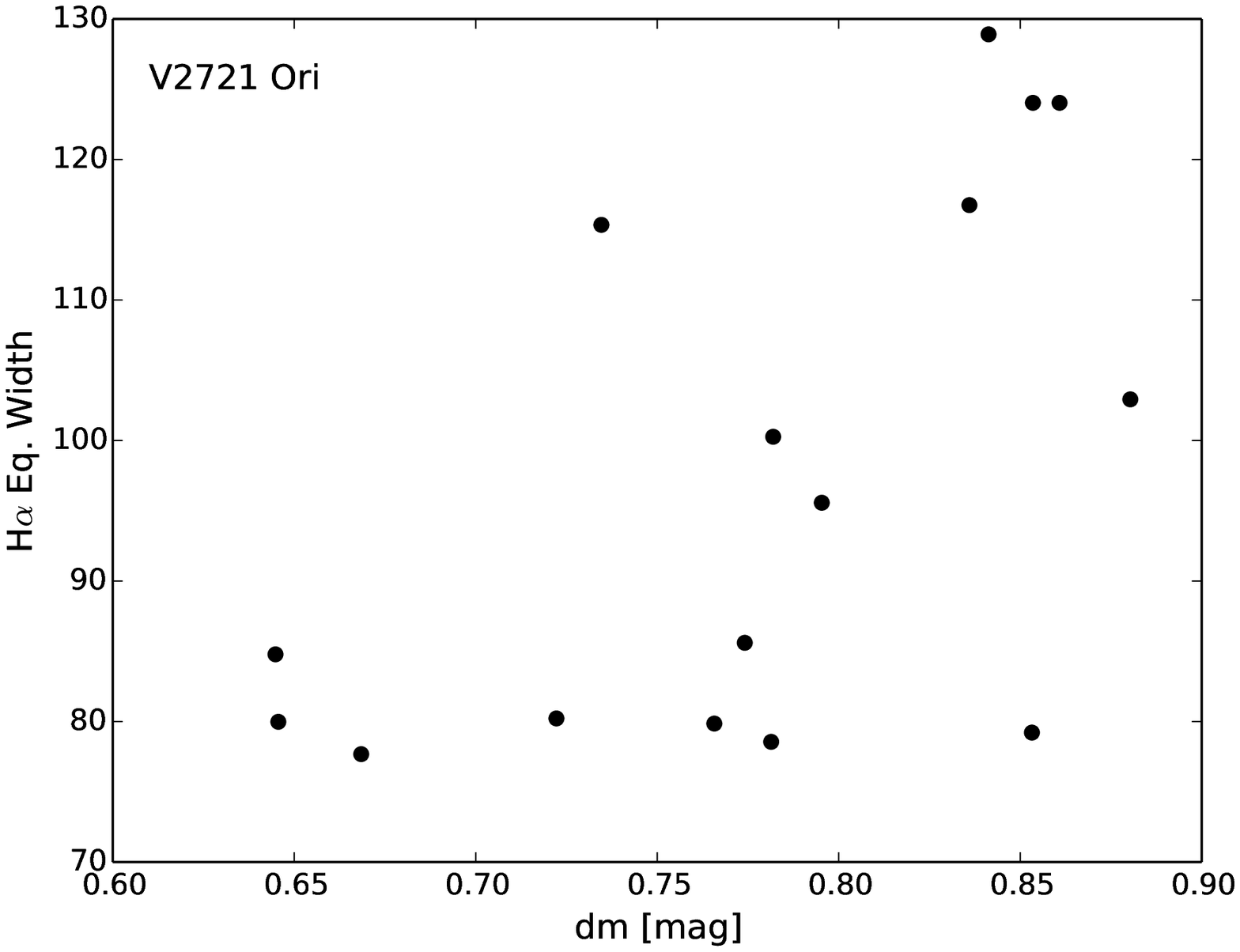} \\
	   
	    \includegraphics[width=0.5\linewidth]{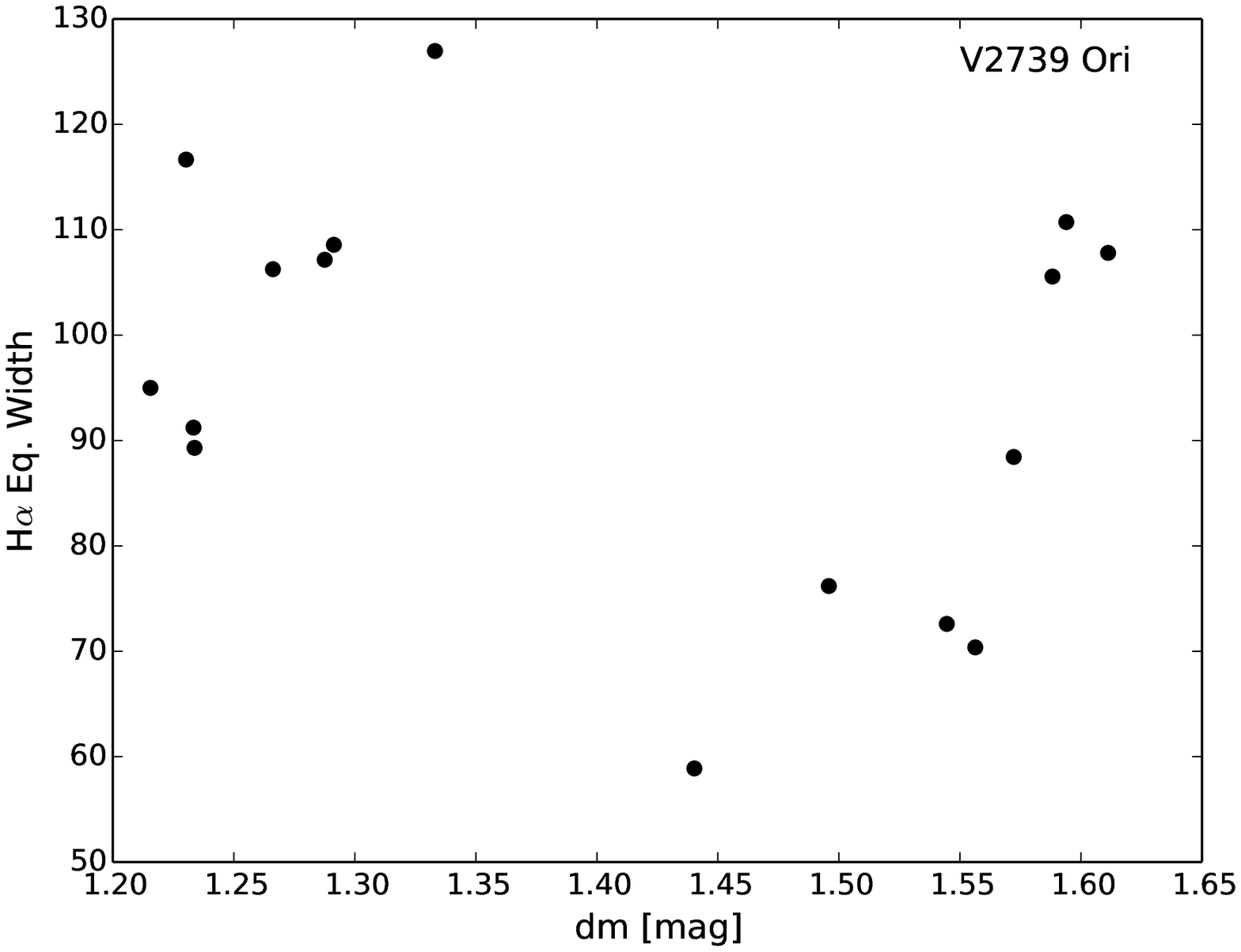}

	\end{tabular}
	
	 \caption{H$\alpha$ EW as a function of lightcurve differential magnitude. Again, plots show no solid correlation of H$\alpha$ EW and magnitude. }
	\label{HavsMag}	
	
\end{figure*}

\begin{table*}

	\caption{Summury of observed and best fit model I-band amplitudes (A) and max/min ratio slopes (m) and parameter values for each object (see Sect. \ref{var_origins}, Fig. \ref{obs_vs_models}). The slopes represent the best linear fit to the ratio data.}
	\label{modpartable}
	\begin{tabular}{ccccccccc}
		
		\hline
		& V2737 & V2721 & V2739 & V2227 & V1999 & V1959 & V2559 \\ \hline
		observed A & 0.79 & 0.24 & 0.40 & 0.13 & 0.54 & 0.24 & 0.61	\\ \hline
		hot spot A & 0.80 & 0.24 & 0.40 & 0.13 & 0.54 & 0.24 & 0.61	\\ \hline
		cold spot A & 0.79 & 0.24 & 0.42 & 0.13 & 0.54 & 0.24 & 0.61 	\\ \hline
		extinction A & 0.81 & 0.23 & 0.40 & 0.13 & 0.55 & 0.24 &  0.62	\\ \hline
		observed m & -3.82e-5 & -8.05e-5 & -8.90e-5 & -2.17e-5 & -1.91e-5 & -2.25e-5 & -6.95e-5	\\ \hline
		hot spot m & -3.27e-4 & -1.12e-4 & -1.98e-4 & -5.24e-5 & -2.30e-4 & -6.69e-4 & -2.03e-4	\\ \hline
		cold spot m & -8.84e-5 & -7.95e-5 & -4.70e-5 & -8.23e-6 & -5.20e-5 & -1.61e-5 & -7.32e-5	\\ \hline
		extinction m & -1.45e-4 & -5.19e-5 & -9.19e-5 & -3.22e-5 & -9.75e-5 & -3.96e-5 & -1.09e-4 	\\ \hline
		Best fit parameters &&&&&&& \\ \hline
		hot spot &&&&&&& \\
		temperature (K) & 3300 & 3300 & 3000 & 3200 & 3200 & 3200 & 3300 \\
		filling factor   & 0.80 & 0.25 & 0.36 & 0.48 & 0.70 & 0.94 & 1.47 \\ \hline
		cold spot &&&&&&& \\
		temperature (K) & 2300 & 2800 & 2300 & 2300 & 2300 & 2300 & 2500 \\
		filling factor   & 0.64 & 0.59 & 0.48 & 0.13 & 0.48 & 0.23 & 0.51 \\ \hline
		extinction &  &&&&&& \\
		Rv & 5.0 & 3.1 & 3.1 & 3.1 & 5.0 & 5.0 & 5.0 \\
		Av   & 1.25 & 0.4 & 0.7 & 0.25 & 0.85 & 0.35 & 0.95 \\ \hline
		
	\end{tabular}

\end{table*}

   
	
	   
	


\end{document}